%% file: Ch7_gal.tex
\documentclass[12pt]{article}

\usepackage[dvips]{graphicx}

\title{Introduction to galactic dynamos}

\author{Anvar Shukurov}

\input{dyn_book}

\newcommand{\cs}{c_\mathrm{s}}  
\newcommand{\Remcr}{\Rem_\mathrm{cr}}  
\newcommand{\uu}{u}  
\newcommand{\UU}{U}  

\newcommand{\HI}{{\rm H\,\scriptstyle I}}  
\newcommand{\RM}{{\rm RM}} 
\newcommand{\ncr}{n_{\rm cr}}  
\newcommand{\nel}{n_{\rm e}}   

\newcommand\col[2]{\left(\begin{array}{c}#1\\#2\end{array}\right)}

\newcommand\deriv[2]{\frac{\partial #1}{\partial #2}} 
\newcommand\la{\;
  \raise0.3ex\hbox{$<$\kern-0.75em\raise-1.1ex\hbox{$\sim$}}\;\hskip-2pt }
\newcommand\ga{\;
  \raise0.3ex\hbox{$>$\kern-0.75em\raise-1.1ex\hbox{$\sim$}}\;\hskip-2pt }
\newcommand\sfrac[2]{{\textstyle{\frac{#1}{#2}}}} 
\newcommand\vect[1]{{\mbox{\bf #1}}}
\newcommand{\kg}{\,{\rm kg}}
\newcommand{\kms}{\,{\rm km\,s^{-1}}}
\newcommand{\g}{\,{\rm g}}
\newcommand{\cm}{\,{\rm cm}}
\newcommand{\cmcube}{\,{\rm cm^{-3}}}
\newcommand{\G}{\,{\rm G}}
\newcommand{\gcmcube}{\,{\rm g}\,{\rm cm^{-3}}}
\newcommand{\m}{\,{\rm m}}

\newcommand{\mkG}{\,\mu{\rm G}}

\newcommand{\erg}{\,{\rm erg}}
\newcommand{\K}{\,{\rm K}}
\newcommand{\kpc}{\,{\rm kpc}}
\newcommand{\p}{\,{\rm pc}}
\newcommand{\yr}{\,{\rm yr}}
\newcommand{\s}{\,{\rm s}}

\newcommand{\radm}{\,{\rm rad\,m^{-2}}}

\begin{document}

\maketitle

\begin{figure}
\mbox{}
   \includegraphics[width=0.55\hsize]{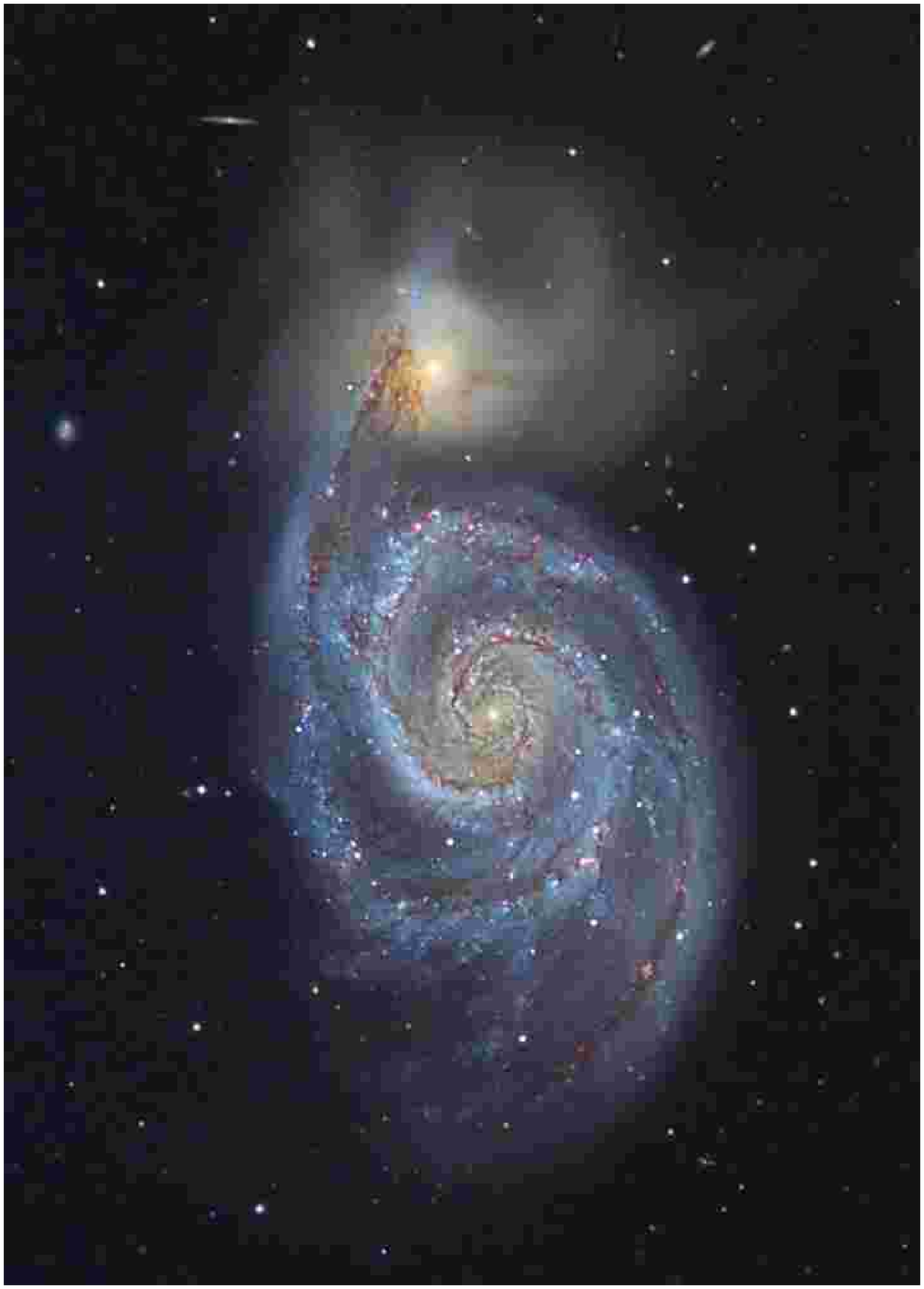}
\includegraphics[width=0.434\hsize]{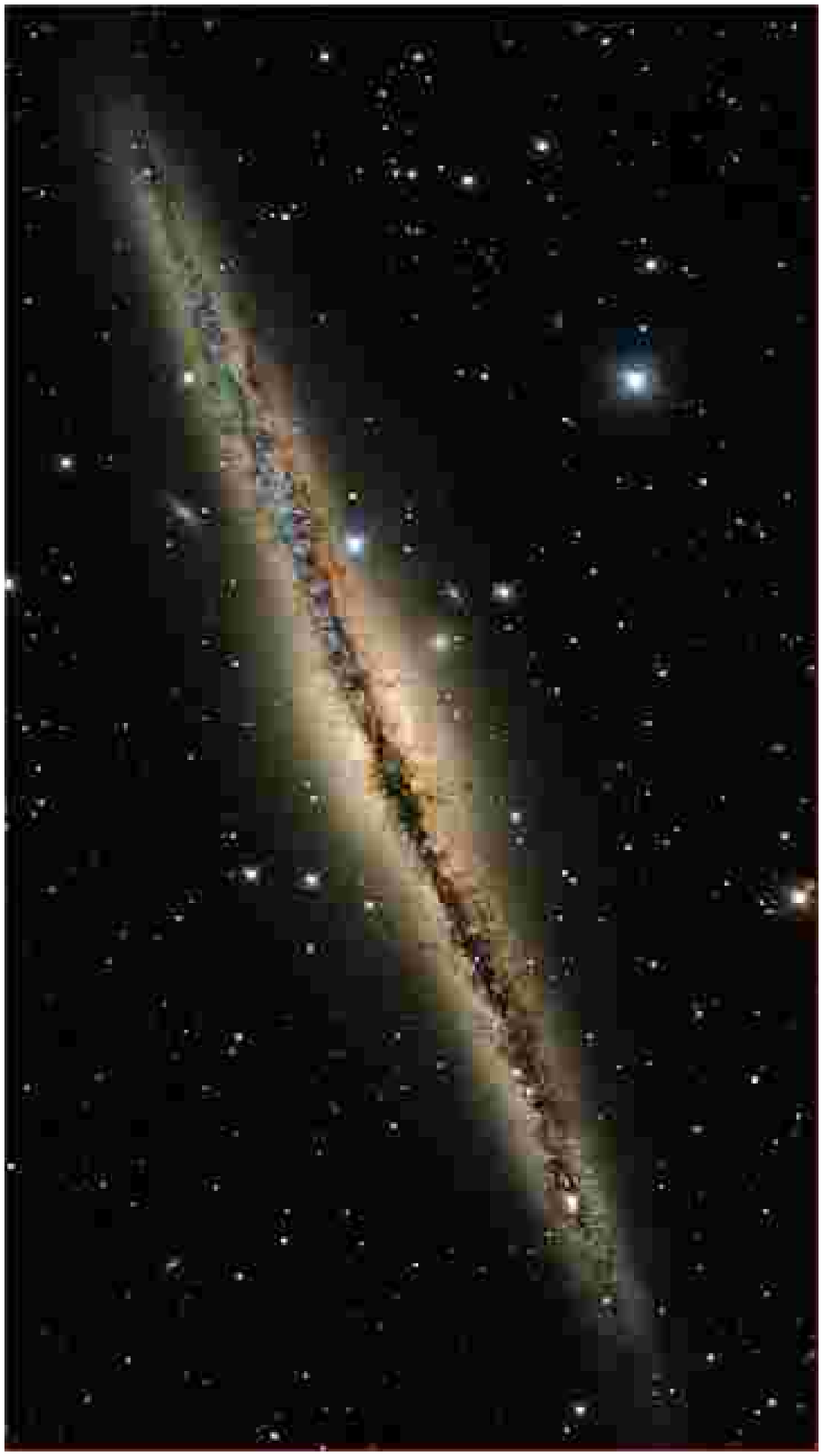}
\caption{\label{ch7f1}Optical images of two nearby spiral galaxies. Left: M51,
the Whirlpool galaxy (with a satellite galaxy at the top). Right: NGC~891
(Courtesy of the Canada--France--Hawaii Telescope/J.-C.~Cuillandre/Coelum). M51
is one of nearby galaxies (distance 9.6\,Mpc) notable for its prominent
spiral pattern. M51 is the first galaxy where a well ordered, large-scale
magnetic field was detected (Segalovitz \etal, 1976) and studied in fine
detail. NGC~891 is at about the same distance as M51,
but seen nearly edge-on, so the thinness of the galactic disc is evident.
The dark strip along the galactic disc and filaments extended away from the
galactic plane are due to obscuration by interstellar dust. The filaments trace
gas outflow from the disc into the halo.}
\end{figure}

\section{Introduction}
Galaxies are attractive objects to study. The magnetism of their natural
beauty adds to the fascinating diversity of physical processes that occur
over an enormous range of scales from the global dimension of order
10\,kpc\footnote{A length unit appropriate to galaxies is
$1\kpc\approx3.1\times10^{19}\m\approx3262~\mbox{light years}$. The distance
of the Sun from the centre of the Milky Way is $s_\odot\approx8.5\kpc$.} down
to the viscous turbulent scales of 1000\,km and less.
The visual image of a galaxy (see Fig.~\ref{ch7f1}) is
dominated by the optical light mostly produced by stars that contribute most
of the visible galactic mass ($2\times10^{11}M_\odot$
for the Milky Way, where $M_\odot=2\times10^{30}\kg$ is the mass of the
Sun). A few percent of the galactic mass is due to the interstellar gas that
resides in the gravitational field produced by stars and dark matter. Spiral
galaxies are flat (Fig.~\ref{ch7f1}) because the stars and gas rapidly rotate.
The gas is ionized by the UV and X-ray radiation and by cosmic rays; the
degree of ionization of diffuse gas ranges from 30\% to 100\% in various
phases --- see Sect.~\ref{tmps}). Interstellar gas is involved in turbulent
motions that can be detected because the associated Doppler shifts broaden
spectral lines emitted by the gas beyond their width expected from thermal
motions alone. The effective mean free path of interstellar gas particles is
small enough to justify a fluid description under a broad range of conditions.
Altogether, interstellar gas can be reasonably described as an electrically
conducting, rotating, stratified turbulent fluid --- and thus a site of MHD
processes discussed elsewhere in this volume, including various types of
dynamo action.

The energy density of interstellar magnetic fields is observed to be
comparable to the kinetic energy density of interstellar turbulence and cosmic
ray energy density, and apparently exceeds the thermal energy density of
interstellar gas (Cox, 1990). Therefore, interstellar gas, magnetic field and
cosmic rays form a complex, nonlinear physical systems whose behaviour is
equally affected by each of the three components. The system is so complex
that magnetic fields and cosmic rays --- the components that are more
difficult to observe and model --- are often neglected. Such a simplification
is perhaps justifiable at very large scales of order 10\,kpc, where the
motions of interstellar gas (mainly the overall rotation) are governed by
gravity: systematic motions at a speed in excess of 10--$30\kms$ are too
strong to be affected by interstellar magnetic fields. However, motions at
smaller scales (comparable to and less than the turbulent scale,
$\ell\approx0.1\kpc$) are strongly influenced by magnetic fields. In
particular, interstellar turbulence is in fact an MHD turbulence. In this
respect, the interstellar environment does not differ much from stellar and
planetary interiors.

Until recently, interstellar magnetic fields had been a rather isolated area
of galactic astrophysics. The reason for that was twofold. Firstly, magnetic
fields are difficult to observe and model. Secondly, they were understood too
poorly to provide useful insight into the physics of
interstellar gas and galaxies in general. The widespread attitude of galactic
astrophysicists to interstellar magnetic fields was succinctly described by
Woltjer (1967):
\begin{quote}
The argument in the past has frequently been a process of elimination: one
observed certain phenomena, and one investigated what part of the phenomena
could be explained; then the unexplained part was taken to show the effects of
the magnetic field. It is clear in this case that, the larger one's ignorance,
the stronger the magnetic field.
\end{quote}
The attitude hardly changed in 20 subsequent years, when Cox (1990) observed
that
\begin{quote}
As usual in astrophysics, the way out of a difficulty is to invoke the poorly
understood magnetic field. \ldots  One tends to ignore the field so
long as one can get away with it.
\end{quote}
The situation has changed dramatically over the last 10--15 years.
Theory and observations of galactic magnetic fields are now advanced enough to
provide useful constraints on the kinematics and dynamics of interstellar gas,
and the importance and r\^ole of galactic magnetic fields are better
appreciated.

In this chapter, we review in Sect.~\ref{ISM} those aspects of galactic
astrophysics that are relevant to magnetic fields, and briefly summarize in
Sect.~\ref{Obs} our observational knowledge of magnetic fields in spiral
galaxies. Section~\ref{Origin} is an exposition of the current ideas on the
origin of galactic magnetic fields, including the dynamo theory. The
confrontation of theory with observations is the subject of Sect.~\ref{OEDASG}
where we summarize the advantages and difficulties of various theories and
argue that the mean-field dynamo theory remains the best contender. Magnetic
fields in elliptical galaxies are briefly discussed in Sect.~\ref{Elliptic}.

\section{Interstellar medium in spiral galaxies}        \label{ISM}

\subsection{Turbulence and multi-phase structure}       \label{tmps}
The interstellar medium (ISM) is much more inhomogeneous and active than
stellar and planetary interiors. The reason for that is ongoing star formation
where massive young stars evolve rapidly (in about $10^6\yr$) and then explode
as supernova stars (SN) releasing large amounts of energy
($E_\mathrm{SN}\sim10^{51}\erg$ per event). These explosions control the
structure of the ISM.

SN remnants are filled with hot, overpressured gas and first expand
supersonically; at this stage the gas surrounding the blast wave is not
perturbed. However, a pressure disturbance starts propagating faster than the
SN shell as soon as the expansion velocity becomes comparable to or lower than
the speed of sound in the surrounding gas --- at this stage the expanding SN
remnant drives motions in the surrounding gas, and its energy is partially
converted into the kinetic energy of the ISM. When pressure inside an SN
remnant reduces to values comparable to that in the surrounding gas, the
remnant disintegrates and merges with the ISM. Since SN occur at (almost)
random times and positions, the result is a random force that drives random
motions in the ISM that eventually become turbulent. The size of an SN remnant
when it has reached pressure balance determines the energy-range turbulent
scale,
\[
\ell\approx0.05\mbox{--}0.1\kpc.
\]
A useful review of supernova dynamics can be found, e.g., in Lozinskaya
(1992), and the spectral properties of interstellar turbulence are discussed
by Armstrong \etal\ (1995). Among numerous reviews of the multi-phase ISM we
mention that of Cox (1990) and a recent text of Dopita \& Sutherland (2003).

About $f=0.07$ of the SN energy is converted into the
ISM's kinetic energy. With the SN frequency of
$\nu_\mathrm{SN}\sim(30\yr)^{-1}$ in the Milky Way (i.e., one SN per $30\yr$),
the kinetic energy supply rate per unit mass is $\dot e_\mathrm{SN}=
f\nu_\mathrm{SM}E_\mathrm{SN}M_\mathrm{gas}^{-1}
\sim10^{-2}\erg\g^{-1}\s^{-1}$, where
$M_\mathrm{gas}=4\times10^9\,M_\odot$ is the total mass of
gas in the galaxy. This energy supply can drive turbulent motions at a speed
$\uu_0$ such that $2\uu_0^3/\ell=\dot e_\mathrm{SN}$ (where the factor 2
allows for equal contributions of kinetic and magnetic turbulent energies),
which yields
\[
\uu_0\sim10\mbox{--}30\kms,
\]
a value similar to the speed of sound at a temperature $T=10^4\K$ or higher.
The corresponding turbulent diffusivity follows as
\begin{equation}        \label{etat}
\turb{\eta}\sim\sfrac13\ell\uu_0
\approx(0.5\mbox{--}3)\times10^{26}\cm^2\s^{-1}\;.
\end{equation}

Supernovae are the main source of turbulence in the ISM. Stellar winds is
another significant source, contributing about 25\% of the total energy supply
(e.g., \S{VI.3} in Ruzmaikin \etal, 1988).

The time interval between supernova shocks passing through a given point is
about (McKee \& Ostriker, 1977; Cox, 1990)
\[
\tau=(0.5\mbox{--}5)\times10^6\yr.
\]
After this period of time, the velocity field at a given position
completely renovates to become independent of its previous form. Therefore,
this time can be identified with the correlation time of interstellar
turbulence. The renovation time is 2--20 times shorter than the `eddy
turnover' time $\ell/\uu_0\sim10^7\yr$. This means that the short-correlated
(or $\delta$-correlated) approximation, so important in turbulence and dynamo
theory (e.g., Zeldovich \etal, 1990; Brandenburg \& Subramanian, 2004), can be
quite accurate in application to the ISM ---  this is a unique
feature of interstellar turbulence. Note that the standard estimate (\ref{etat})
is valid if the correlation time is $\ell/\uu_0$. If the renovation time was
used instead, the result would be
$\turb{\eta}\sim\ell^2/\tau\sim10^{27}\cm^2\s^{-1}$, a value an order of
magnitude larger than the standard estimate.

\begin{figure}
\centerline{\includegraphics[width=0.487\hsize]{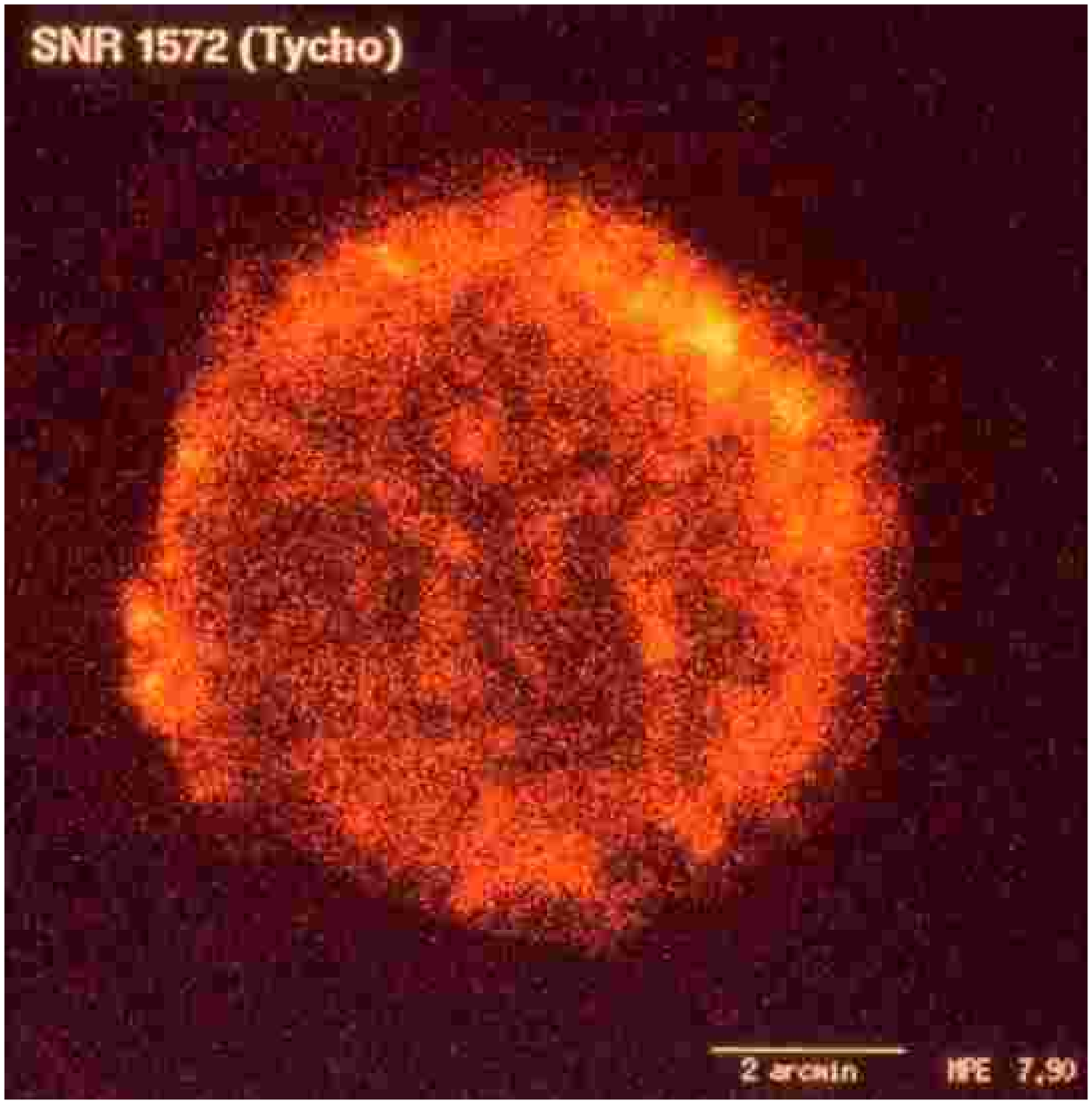}
\includegraphics[width=0.493\hsize]{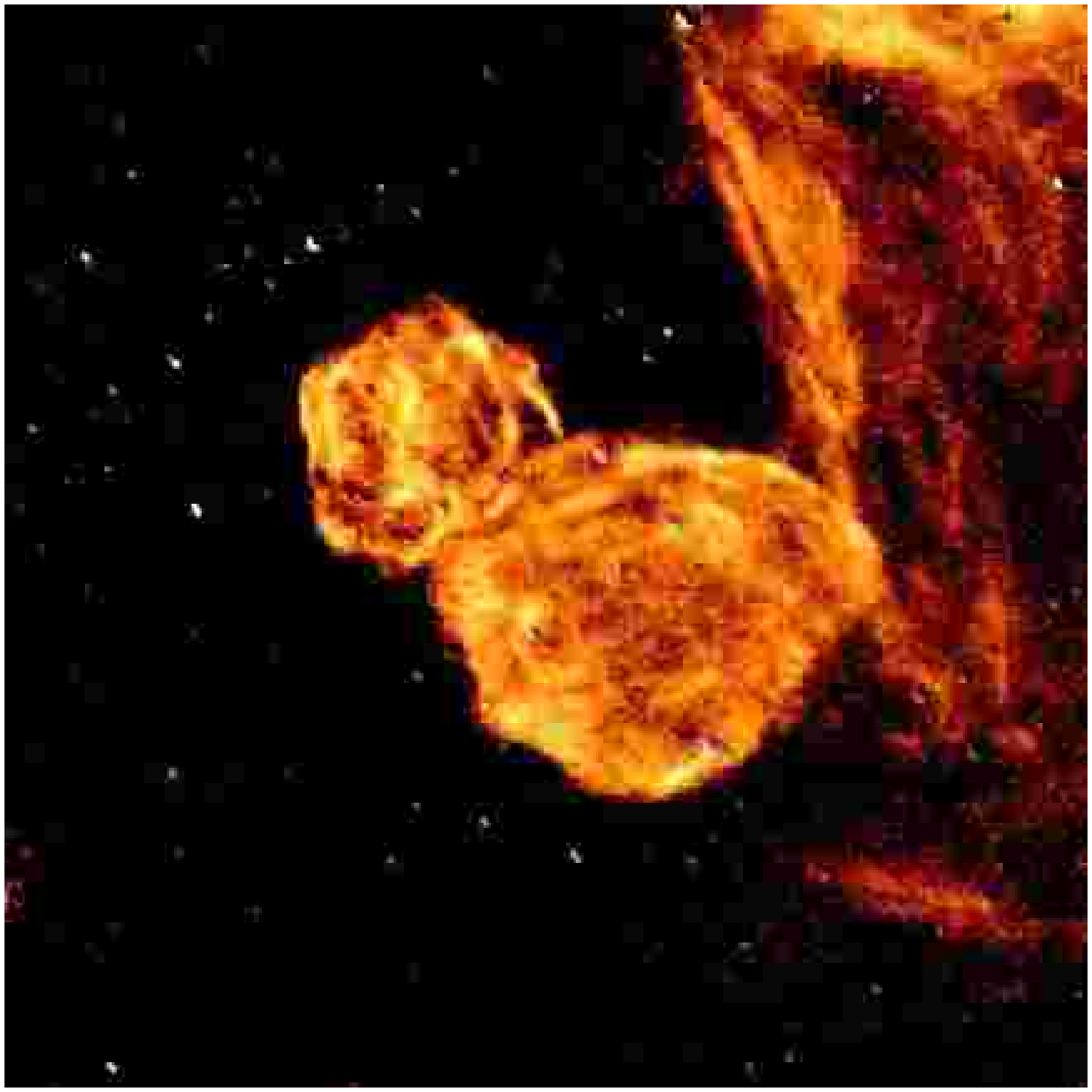}}
\caption{\label{SNe}SN remnants are expanding bubbles of hot gas
that emits thermal X-rays. This is illustrated by the X-ray image of
Tycho's supernova remnant (left panel; courtesy of the ROSAT Mission and the
Max-Planck-Institut f\"ur extraterrestrische Physik) whose parent star's
explosion in 1572 was recorded by the famous Danish astronomer Tycho Brahe.
The hot gas cools only slowly, and SN remnants often merge. The right panel
shows a false-colour optical (H$\alpha$) image of two SN remnants DEM~L316 in
the Large Magellanic Cloud which appear to be colliding (Williams \etal,
1997; image produced by the Magellanic Cloud Emission-Line Survey, reprinted
with permission).}
\end{figure}

\begin{table}
\caption{\label{phases}The multi-phase ISM. The origin and parameters of the
most important phases of interstellar gas: $n$, the mid-plane number density in
hydrogen atoms per cm$^3$; $T$, the temperature in K; $\cs$, the speed of sound
in km\,s$^{-1}$; $h$, the scale height in kpc; and $f_V$, the volume filling
factor in the disc of the Milky Way, in per cent.}
\begin{center}
\begin{tabular}{llccccc}
\hline
\hline
Phase           &Origin         &$n$        &$T$     &$\cs$  &$h$   &$f_V$\\
\hline
Warm    &                       &0.1        &$10^4$  &10     &0.5   &60--80\\
Hot             &Supernovae     &$10^{-3}$  &$10^6$  &100    &3     &20--40\\
Hydrogen clouds &Compression    &20         &$10^2$  &1      &0.1   &2\\
Molecular clouds&Self-gravity,  &$10^3$     &10      &0.3    &0.075 &0.1\\
                &thermal instability\\
\hline
\hline
\end{tabular}
\end{center}
\end{table}

Another important result of supernova activity is a large
amount of gas heated to $t=10^6\K$ (Fig.~\ref{SNe}). The gas is so tenuous
that the collision rate of the gas particles is low, and so its radiative
cooling time is very long and exceeds $\tau$: the hot bubbles produced by
supernovae can merge before they cool (Fig.~\ref{SNe}). A result is a network
of hot tunnels that form the hot component of the ISM. Altogether, the
interstellar gas is found in several distinct states, known as `phases' (this
usage may be misleading as most of them are not proper thermodynamic phases)
whose parameters are presented in Table~\ref{phases}. Some of the parameters
(especially the volume filling factors) are not known confidently, so
estimates of Table~\ref{phases} should be approached with healthy skepticism.
The warm diffuse gas can be considered as a background against which the ISM
dynamics evolves; this is the primary phase that occupies a connected
(percolating) region in the disc, whereas the hot gas may or may not fill a
connected region. The warm gas is ionized by the stellar ultraviolet radiation
and cosmic rays; its degree of ionization is about 30\% at the Galactic
midplane. The hot gas is so hot that it is fully ionized by gas particle
collisions.

The locations of SN stars are not entirely random: 70\% of them cluster in
regions of intense star formation (known as OB associations as they contain
large numbers of young, bright stars of spectral classes O and B) where gas
density is larger than on average in the galaxy. Collective energy input from
a few tens (typically, 50) SN within a region about 0.5--1\kpc\ in size
produces a superbubble that can break through the galactic disc (Tenorio-Tagle
\& Bodenheimer, 1988). This removes the hot gas into the galactic halo and
significantly reduces its filling factor in the disc (from about 70\% to
10--20\%). This also gives rise to a systematic outflow of the hot gas to
large heights where the gas eventually cools, condenses and returns to the
disc after about $10^9\yr$ in the form of cold, dense clouds of neutral
hydrogen (Wakker \& van Woerden, 1997). This convection-type flow is known as
the galactic fountain (Shapiro \& Field, 1976), and it can plausibly support
a mean-field dynamo of its own (Sokoloff \& Shukurov, 1990). Another
aspect of its r\^ole in galactic dynamos is discussed in Sect.~\ref{hel}. The
vertical velocity of the hot gas at the base of the fountain flow is
100--$200\kms$ (e.g., Kahn \& Brett, 1993; Korpi \etal, 1999a,b).

\subsection{Galactic rotation}\label{Grot}
Spiral galaxies have conspicuous flat components because they rotate rapidly
enough. The Sun moves in the Milky Way at a velocity of about
$u_\odot=s_\odot\Omega_\odot=220\kms$, to complete one orbit of a radius
$s_\odot\approx8.5\kpc$ in $2\pi/\Omega_\odot=2.4\times10^8\yr$. These values
are representative for spiral galaxies in general. The Rossby number is
estimated as
\[
\Ros=\frac{\uu_0}{\ell\Omega_\odot}\sim 4\;.
\]
The vertical distribution of the gas is controlled, to the first approximation,
by hydrostatic equilibrium in the gravity field produced by stars and dark
matter, with pressure comprising thermal, turbulent, magnetic and cosmic
ray components in roughly equal proportion (e.g., Boulares \& Cox, 1990;
Fletcher \& Shukurov, 2001). The semi-thickness of the warm gas layer
is about $h=0.5\kpc$, i.e., the aspect ratio of the gas disc is
\begin{equation}        \label{aspect}
\eps=\frac{h}{s_\odot}\sim0.06\;.
\end{equation}
Since the gravity force decreases with radius $s$ together with the stellar
mass density, $h$ grows with $s$ at $s\ga10\kpc$ (see \S{VI.2} in Ruzmaikin
\etal, 1988, for a review).

However, the hot gas has larger speed of sound and turbulent velocity, and its
Rossby number can be as large as 10 given that its turbulent scale is about
$0.3\kpc$ (see Poezd \etal, 1993). Hence, the hot gas fills a quasi-spherical
volume, where its pressure scale height of order $5\kpc$ is comparable to
the disc radius.

$\Ros=1$ at a scale $0.4\kpc$ in the warm gas, which is similar to the
scale height of the gas layer. This implies that rotation
significantly affects turbulent gas motions, making them helical on average. A
convenient estimate of the associated $\alpha$-effect can be obtained from
Krause's formula,
\begin{equation}        \label{alKrause}
\alpha_0\sim\frac{\ell^2\Omega}{h}\approx0.5\kms\;,
\end{equation}
where $\Omega$ is the angular velocity, and the numerical estimate refers to
the Solar neighbourhood of the Milky Way. Thus, $\alpha_0\sim0.05\uu_0$ near
the Sun and increases in the inner Galaxy together with $\Omega$.

\begin{figure}
\centerline{\includegraphics[width=0.49\textwidth]{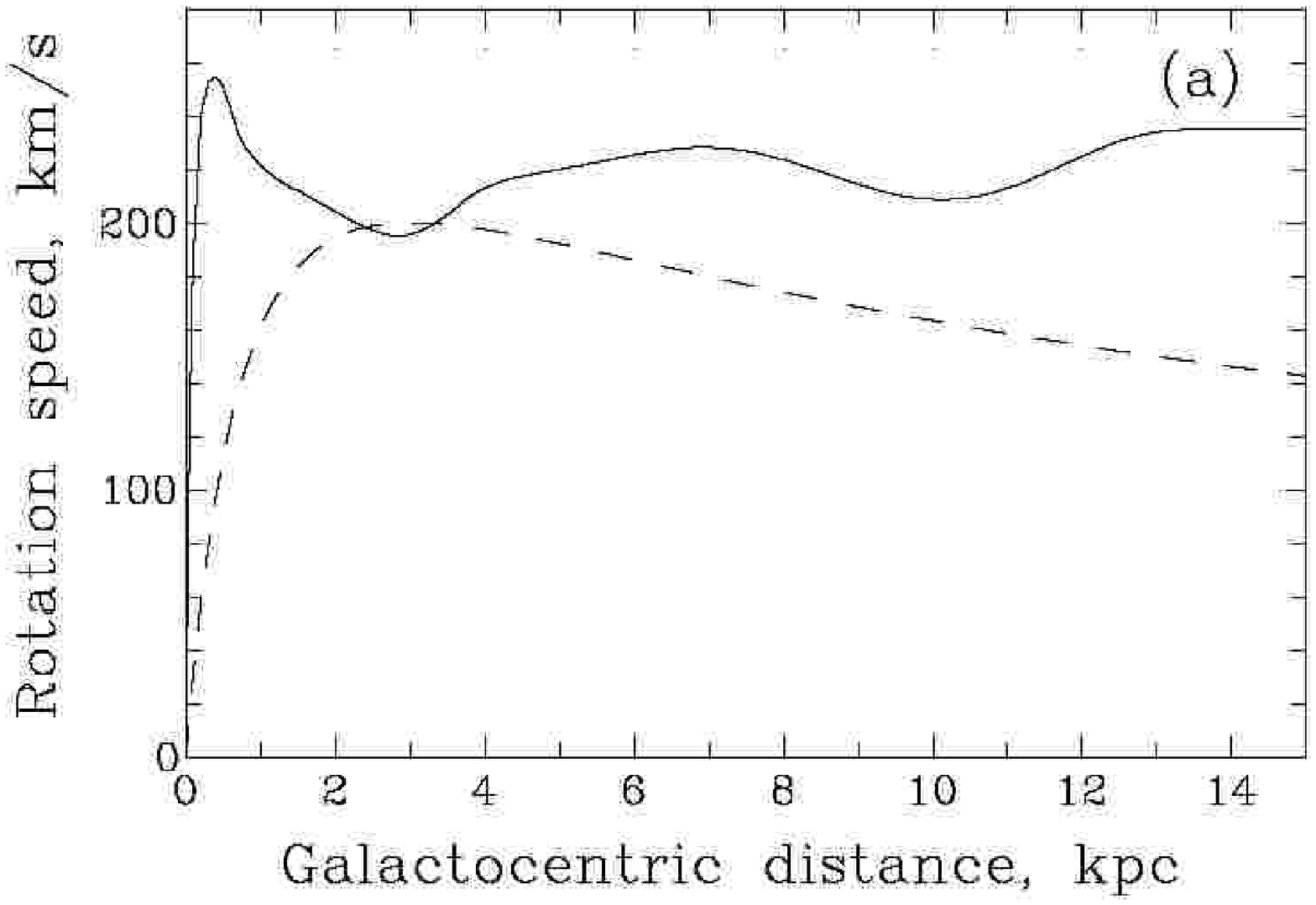}
\includegraphics[width=0.49\textwidth]{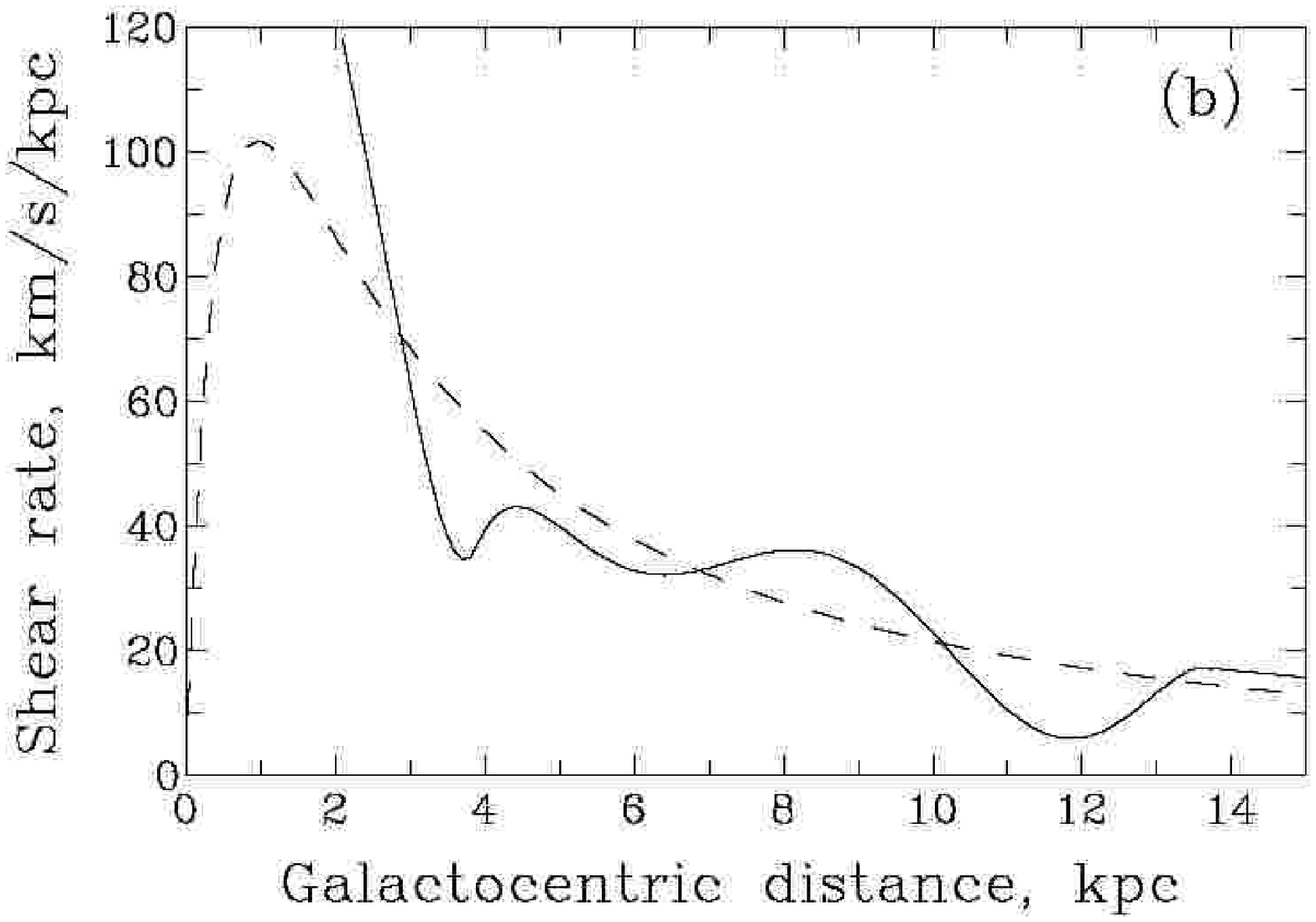}}
\caption{\label{rotcurve}{\bf(a)}: The rotation speed $s\Omega(s)$ in the
galactic midplane versus galactocentric radius $s$ in the Milky Way (solid)
(Clemens, 1985), and the generic Schmidt's rotation curve with
$\UU_0=200\kms$, $s_0=3\kpc$ and $n=1$ (dashed). {\bf(b)}: The corresponding
rotation shear rates (taken with minus sign), $-s\partial\Omega/\partial s$.}
\end{figure}

The spatial distribution of galactic rotation is known for thousands galaxies
(Sofue \& Rubin, 2001) from systematic Doppler shifts of various
spectral lines emitted by stars and gas. In this respect, galaxies are
 much better explored than any star or planet (including the Sun and the
Earth) where reliable data on the angular velocity in the interior are much
less detailed and reliable or even unavailable. The radial profile of the
galactic rotational velocity is called the rotation curve. Rotation curves of
most galaxies are flat beyond a certain distance from the axis, so
$\Omega\propto s^{-1}$ is a good approximation for $s\ga5\kpc$. The rotation
curve of a generic galaxy, known as the Schmidt rotation curve and shown in
Fig.~\ref{rotcurve}, has the form
\[
s\Omega(s)=\UU_0\frac{s}{s_0}
\left[\sfrac13+\sfrac23\left(\frac{s}{s_0}\right)^n\right]^{-3/2n}\;,
\]
where the parameters vary between various galaxies in the range
$s_0\sim5$--$20\kpc$, $\UU_0\sim200\kms$ and $n\sim0.7$--1.
This rotation curve is not flat at large radii, but it provides an acceptable
approximation at moderate distances from galactic centre where magnetic field
generation is most intense. Some galaxies have more complicated rotation curves.
Notably, the Milky Way and M31 are among them --- see Figs~\ref{rotcurve} and
\ref{M31pitch}. The complexity of the rotation curves is explained by a
complicated distribution of the gravitating (stellar and dark) mass in those
galaxies. It is evident from Fig.~\ref{rotcurve}b that the rotation shear is
strong at all radii even for the Schmidt rotation curve, and so the rotation
in the inner part of a spiral galaxy cannot be approximated by the solid-body
law, even if the shape of some rotating curves tempts to do so.

The vertical variation of the rotation velocity is only poorly known. In a
uniform gravitating disc of infinite radial extent the angular velocity of
rotation would be constant in $z$. Then it is natural to expect that $\Omega$
should decrease along $z$ at a scale comparable to the radial scale length of
the gravitating mass in the disc, typically $s_*=3$--$5\kpc$. Recent
observations of gas motions in galactic halos have confirmed such a decrease
(Fraternali \etal, 2003). In the absence of detailed models, an approximation
$\Omega\propto\mathrm{exp}\,(-z/s_*)$ seems to be appropriate.

\section{Magnetic fields observed in galaxies}          \label{Obs}
Estimates of magnetic field strength in the diffuse interstellar medium
of the Milky Way and other galaxies are most efficiently obtained from the
intensity and Faraday rotation of synchrotron emission. Other methods are only
sensitive to relatively strong magnetic fields that occur in dense clouds
(Zeeman splitting) or are difficult to quantify (optical polarization of star
light by dust grains). The total $I$ and polarized $P$ synchrotron intensities
and the Faraday rotation measure $\RM$ are weighted integrals of magnetic
field over the path length $L$ from the source to the observer, so they
provide a measure of the average magnetic field in the emitting or
magneto-active volume:
\begin{eqnarray}
I&=&K\int_L\ncr B_\perp^2\,ds\;,\nonumber\\
P&=&K\int_L\ncr\mean{B}_\perp^2\,ds\;,       \label{ints}\\
\RM&=&K_1\int_L\nel B_\parallel\,ds\;\nonumber,
\end{eqnarray}
where $\ncr$ and $\nel$ are the number densities of relativistic and thermal
electrons,
$\bfB$ is the total magnetic field comprising a regular
$\mean{\bfB}$ and random $\bfb$ parts,
$\bfB=\mean{\bfB}+\bfb$ with $\langle\bfB\rangle=\mean{\bfB},\
\langle\bfb\rangle=0$ and $\langle{B^2}\rangle=B^2+\langle b^2\rangle$,
angular brackets denote averaging, subscripts $\perp$ and $\parallel$
refer to magnetic field components perpendicular and parallel to the line of
sight, and $K$ and
$K_1=e^3/(2\pi\m_\mathrm{e}^2 c^4)=0.81\radm\cm^3\mkG^{-1}\p^{-1}$ are certain
dimensional constants (with $e$ amd $m_\mathrm{e}$ the electron charge and
mass and $c$ the speed of light). The degree of polarization $p$ is related to
the ratio $\langle b^2\rangle/\mean{B}^2$,
\begin{equation}                \label{polar}
p\equiv\frac{P}{I} \approx
p_0 \frac{\mean{B}_\perp^2}{B_\perp^2}
=p_0\frac{\mean{B}_\perp^2}{\mean{B}_\perp^2+\sfrac23\langle b^2\rangle}\;,
\end{equation}
where
the random field $\bfb$ has been assumed to be isotropic in the last equality,
$\ncr$ is assumed to be a constant, and $p_0\approx0.75$ weakly depends on the
spectral index of the emission. This widely used relation is only approximate.
In particular, it does not allow for any anisotropy of the random magnetic
field, for the dependence of $n_\mathrm{cr}$ on $B$, and for depolarization
effects; some generalizations are discussed by Sokoloff \etal\ (1998).

The orientation of the apparent large-scale magnetic field in the sky plane is
given by the observed $B$-vector of the polarized synchrotron emission.
Due to Faraday rotation, the true orientation can differ by an angle of
$\RM\lambda^2$, which amounts to $10^\circ$--$20^\circ$ at a wavelength
$\lambda=6\cm$. The special importance of the Faraday rotation measure, $\RM$,
is that this observable is sensitive to the direction of $\bfB$ (the sign of
$\mean{B}_\parallel$) and this allows one to determine not only the
orientation of $\mean{\bfB}$ but also its direction. Thus, analysis of Faraday
rotation measures can reveal the three-dimensional structure of the magnetic
vector field (Berkhuijsen \etal, 1997; Beck \etal, 1996).

Since $\ncr$ is difficult to measure, it is often assumed that magnetic
field and cosmic rays are in pressure equilibrium or energy equipartition;
this allows to express $\ncr$ in terms of $B$.
The physical basis of this assumption is the fact that
cosmic rays (charged particles of relativistic energies) are confined by
magnetic fields. An additional assumption involved is that the energy density
of relativistic electrons responsible for synchrotron emission (energy of
several GeV per particle) is one percent of the proton energy density in the
same energy interval, as measured near the Earth.

The cosmic ray number density$\ncr$  in the Milky Way can be determined
independently from $\gamma$-ray emission produced when cosmic ray particles
interact with the interstellar gas. Then magnetic field strength can be
obtained without assuming equipartition (Strong \etal, 2000); the results are
generally consistent with the equipartition values. However, Eq.~(\ref{polar})
is not consistent with the equipartition or pressure balance between cosmic
rays and magnetic fields as it assumes that $\ncr=\mbox{const}$. Therefore,
$\mean{B}$ obtained from Eq.~(\ref{polar}) can be inaccurate (Beck \etal,
2003).

The mean thermal electron density $\nel$ in the ISM can be obtained from the
emission measure of the interstellar gas, an observable defined as ${\rm
EM}\propto\int_L\nel^2\,ds$, but this involves the poorly known filling factor
of interstellar clouds. In the Milky Way, the dispersion measures of pulsars,
${\rm DM}=\int_L\nel\,ds$ provide information about the mean thermal electron
density, but the accuracy is limited by our uncertain knowledge of distances
to pulsars. Estimates of the strength of the regular magnetic field in the
Milky Way are often obtained from the Faraday rotation measures of pulsars
simply as

\begin{equation}        \label{obsB}
B_\parallel=\frac{\RM}{K_1\,\rm DM}\;.
\end{equation}
This estimate is meaningful if magnetic field and thermal electron density are
statistically uncorrelated. If the fluctuations in magnetic field and thermal
electron density are correlated with each other, they will contribute
positively to $\RM$ and Eq.~(\ref{obsB}) will yield overestimated
$\mean{B}_\parallel$. In the case of anticorrelated fluctuations, their
contribution is negative and Eq.~(\ref{obsB}) is an underestimate. As shown by
Beck \etal\ (2003), physically reasonable assumptions about the statistical
relation between magnetic field strength and electron density can lead to
Eq.~(\ref{obsB}) being in error by a factor of 2--3.

\begin{figure}
\centerline{\includegraphics[width=0.8\textwidth]{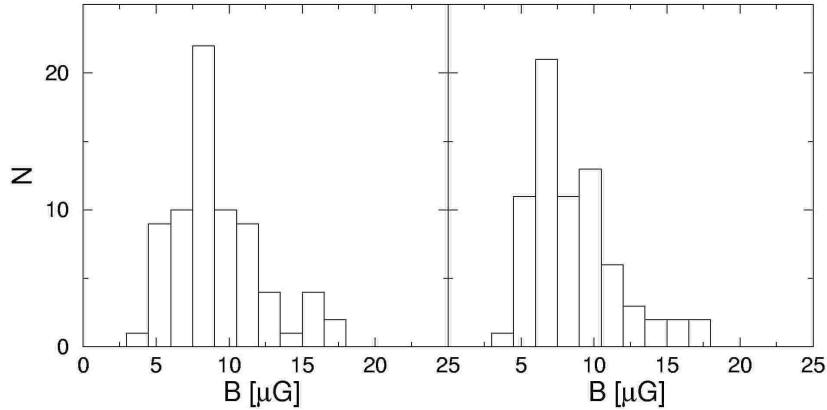}}
\caption{\label{equipartition}The distributions of the strength of the total
magnetic field in a sample of spiral galaxies obtained from the
observed synchrotron intensity $I$ using energy equipartition between magnetic
fields and cosmic rays (p.~109 in Niklas, 1995) under slightly different
assumptions. The estimates of the left-hand panel were derived from
integrating the observed synchrotron intensity in the range corresponding to
the relativistic electron energies from 300\,MeV to infinity, and in the
right-hand panel the integration was over a frequency range 10\,MHz--10\,GHz.
Results presented in the left-hand panel are better justified physically
(\S{2.1} in Beck \etal, 1996; \S{III.A.1} in Widrow, 2002).}

\end{figure}

The observable quantities (\ref{ints}) have provided extensive data on
magnetic field strengths in both the Milky Way and external galaxies
(Ruzmaikin \etal., 1988; Beck \etal, 1996; Beck, 2000, 2001). The average
total field strengths in nearby spiral galaxies obtained from total
synchrotron intensity $I$ range from $B\approx4\mkG$ in the galaxy M31 to
about $15\mkG$ in M51, with the mean for the sample of 74 galaxies of
$B=9\mkG$ (Beck, 2000). Figure~\ref{equipartition} shows the distribution of
magnetic field strength in a sample of spiral galaxies. The typical degree of
polarization of synchrotron emission from galaxies at short radio wavelengths
is $p=10$--$20\%$, so Eq.~(\ref{polar}) gives $\mean{B}/B=0.4$--0.5; these are
always lower limits due to the limited resolution of the observations, and
$\mean{B}/B=0.6$--0.7 is a more plausible estimate. Most existing polarization
surveys of synchrotron emission from the Milky Way, having much better spatial
resolution, suffer from Faraday depolarization effects and missing large-scale
emission and cannot provide reliable values for $p$. The total equipartition
magnetic field in the Solar neighbourhood is estimated as $B=6\pm2\mkG$ from
the synchrotron intensity of the diffuse Galactic radio background
(E.~M.~Berkhuijsen, in Beck, 2001). Combined with $\mean{B}/B=0.65$, this
yields a strength of the local regular field of $\mean{B}=4\pm1\mkG$. Hence,
the typical strength of the local Galactic random magnetic fields,
$b=(B^2-\mean{B}^2)^{1/2}=5\pm2\mkG$, exceeds that of the regular field by a
factor $b/\mean{B}=1.3\pm0.6$. $\RM$ data yield similar values for this ratio
(\S{IV.4} in Ruzmaikin \etal, 1988).

Meanwhile, the values of $\mean{B}$ in the Milky Way obtained from Faraday
rotation measures seem to be systematically lower than the above values (see
Beck \etal, 2003, and references therein). $\RM$ of pulsars and extragalactic
radio sources yield $\mean{B}=1$--$2\mkG$ in the Solar vicinity, a value about
twice smaller than that inferred from the synchrotron intensity and
polarization. There can be several reasons for the discrepancy between the
estimates of the regular magnetic field strength from Faraday rotation and
synchrotron intensity. Both methods suffer from systematic errors due to our
uncertain knowledge of thermal and relativistic electron densities, so one
cannot be sure if the difference is significant. Nevertheless, the discrepancy
seems to be worrying enough to consider carefully its possible reasons.

The discrepancy can be explained, at least in part, if the methods described
above sample different volumes. The observation depth of total synchrotron
emission, starlight polarization and of Faraday rotation measures are all of
the order of a few kpc. Polarized emission, however, may emerge from more
nearby regions. However, a more fundamental reason for the discrepancy can be
partial correlation between fluctuations in magnetic field and electron
density. Such a correlation can arise from statistical pressure balance where
regions with larger gas density have weaker magnetic field, and vice versa.
As discussed by Beck \etal\ (2003), the term $\langle b_\parallel\nel\rangle$
then differs from zero and contributes to the observed $\RM$ leading to
underestimated $\mean{B}$. In a similar manner, correlation between $B$ and the
cosmic ray number density biases the estimates of magnetic field from
synchrotron intensity and polarization (see also Sokoloff \etal, 1998).
Altogether, $\mean{B}=4\mkG$ and $b=5\mkG$ seem to be acceptable estimates of
magnetic field strengths near the Sun. The geometry and three-dimensional
structure of the magnetic fields observed in spiral galaxies are further
discussed in Sect.~\ref{OEDASG}.

\section{The origin of galactic magnetic fields}        \label{Origin}
There are two basic approaches to the origin of global magnetic
structures in spiral galaxies --- one of them asserts that the observed
structures represent a primordial magnetic field twisted by
differential rotation, and the other that they are due to ongoing
dynamo action within the galaxy. The simplicity of the former theory is
appealing, but it fails to explain the strength, geometry and apparent
lifetime of galactic magnetic fields (Ruzmaikin \etal, 1988; Beck \etal,
1996; Kulsrud, 1999; Widrow, 2002; see Sect.~\ref{OEDASG} below).
Furthermore, there are no mechanisms known to produce cosmological magnetic
fields of required strength and scale (Beck \etal, 1996), although Kulsrud
\etal\ (1997) argue that suitable magnetic field can be produced in
protogalaxies. Dynamo models appear to be much better consistent with the
observational and theoretical knowledge of interstellar gas, and all models of
magnetic fields in specific galaxies, known to the author, have been
formulated in terms of dynamo theory. It seems to be very plausible that
galactic magnetic fields are generated by some kind of dynamo action, i.e.,
that they are produced {\em in situ.\/} The most promising is the mean-field
turbulent dynamo.

\subsection{Mean-field models of the galactic dynamo}\label{MFD}
As discussed in Sect.~\ref{Grot}, the discs of spiral galaxies are thin.
This provides a natural small parameter, the disc aspect ratio,
Eq.~(\ref{aspect}). This greatly facilitates modelling of many global
phenomena in galaxies, including large-scale magnetic fields. Parker (1971)
and Vainshtein \& Ruzmaikin (1971, 1972) were the first to suggest mean-field
dynamo models for spiral galaxies. These were local models discussed in
Sect.~\ref{local}, where only derivatives across the disc (in $z$) are
retained. The theory has been extended to two and more dimensions and applied
to specific galaxies (see Ruzmaikin \etal, 1988, Beck \etal, 1996, and
Widrow, 2002, and references therein). Rigorous asymptotic solutions for the
$\alpha\omega$-dynamo in a thin disc were developed by Soward (1978, 1992a,b)
and further discussed by Priklonsky \etal\ (2000) and Willis \etal\ (2003).
Reviews of these results can be found in Ruzmaikin \etal\ (1988), Beck
\etal\ (1996), Kulsrud (1999) and Soward (2003).

In this section we present asymptotic solutions of the mean-field dynamo
equations in a thin disc surrounded by vacuum.
We first consider axially symmetric solutions of the kinematic problem, and
then discuss generalizations to non-axisymmetric modes and to nonlinear
regimes. Cylindrical coordinates $(s,\phi,z)$ with the origin the galactic
centre and the $z$-axis parallel to the galactic angular velocity are used
throughout this chapter. In this section we use dimensionless variables, with
$s$ and $z$ measured in the units of the characteristic disc radius and disc
half-thickness (e.g., $s_0=s_\odot\approx8.5\kpc$ and $h_0=0.5\kpc$),
respectively. Then the dimensionless radial and axial distances are both of
order unity within the disc as they are measured in different units in order to
make the disc thinness explicit. The corresponding time unit is the turbulent
magnetic diffusion time across the disc,
$h_0^2/\turb{\eta}\approx7.5\times10^8\yr$.

It is convenient to introduce a unit rotational shear rate $G_0$,
\[
G=s\deriv{\Omega}{s}\equiv G_0 g(s,z)\;,
\]
with $g(s,z)$ its dimensionless value and $G_0=-\Omega_\odot$ for a flat
rotation curve, $\Omega\propto s^{-1}$, and adopt the characteristic magnitude
of the $\alpha$-coefficient near the Sun as given by Eq.~(\ref{alKrause}).

\subsubsection{Kinematic, axially symmetric solutions}  \label{KAS}
The three components of an axially symmetric magnetic field can be expressed
in terms of the azimuthal components of the large-scale magnetic field
$\mean{B}_\phi$ and vector potential $\mean{A}_\phi$,
\[
 \mean{\vect{B}}=\left(-\frac{\partial\mean{A}_\phi}{\partial z},\
\mean{B}_\phi,\ \frac{1}{s}\,\deriv{}{s}(s\mean{A}_\phi)\right). \label{BA}
\]
The dimensionless governing equations, resulting from
standard mean-field dynamo equations have the form
\begin{eqnarray}
\deriv{\mean{B}_\phi}{t} &=&
      -R_\omega g\deriv{\mean{A}_\phi}{z}+\deriv{^2\mean{B}_\phi}{z^2}
      +\eps^2\deriv{}{s}\left[\frac{1}{s}\deriv{}{s}(s\mean{B}_\phi)\right],
\label{meanB}\\
\deriv{\mean{A}_\phi}{t} &=&
      R_\alpha\alpha\mean{B}_\phi+\deriv{^2\mean{A}_\phi}{z^2}
      +\eps^2\deriv{}{s}\left[\frac{1}{s}\deriv{}{s}(s\mean{A}_\phi)\right],
\label{meanA}
 \end{eqnarray}
where
\begin{equation}        \label{Dnumbers}
R_\omega=\frac{G_0 h_0^2}{\turb{\eta}}\;,\qquad
R_\alpha=\frac{\alpha_0 h_0}{\turb{\eta}}\;,
\end{equation}
are the turbulent magnetic Reynolds numbers that characterize the intensity of
induction effects due to differential rotation and the mean helicity of
turbulence, respectively. We have neglected the vertical shear
$\partial\Omega/\partial z$ which can easily be restored, and assumed for
simplicity that $\turb{\eta}=\mbox{const}$. A term containing $\alpha$ has
been neglected in Eq.~(\ref{meanB}) for the sake of simplicity (but can easily
be restored), so the equations are written in the
$\alpha\omega$-approximation.

The kinematic, axially symmetric asymptotic solution in a thin disc has the
form
\[
\col{\mean{B}_\phi}{\mean{A}_\phi}=\exp^{\Gamma t}
\left[Q(\eps^{-1/3}s)\col{{\cal B}(z;s)}{{\cal A}(z;s)}+\ldots\right],
\label{asymptexp}
\]
where $\Gamma$ is the growth rate, $({\cal B},{\cal A})$ represent the
suitably normalized local solution (obtained for fixed $s$), and $Q$ is the
amplitude of the solution which can be identified with the field strength at a
given radius.

\subsubsection{The local solution}      \label{local}
The local solution (with $s$ fixed) arises in the lowest order in $\eps$.  Its
governing equations, obtained from Eqs~(\ref{meanB}) and (\ref{meanA}) by
putting $\eps=0$, contain only derivatives with respect to $z$, with
coefficients depend on $s$ as a parameter (hence, the notation of the
arguments of $b$ and $a$ with semicolon separating $z$ and $s$):
\begin{eqnarray}
\gamma(s){\cal B}&=&-R_\omega g(s)\deriv{{\cal A}}{z}
        +\deriv{^2{\cal B}}{z^2}\;.\label{localb}\\
\gamma(s){\cal A}&=&R_\alpha \alpha(s,z){\cal B}
        +\deriv{^2{\cal A}}{z^2}\;,\label{locala}
\end{eqnarray}
were $\gamma(s)$ is the local growth rate. The boundary conditions often
applied at the disc surface $z=\pm h(s)$ correspond to vacuum outside the
disc. For axisymmetric fields and to the lowest order in $\eps$ they are (see
below)
\begin{equation}                \label{localbc}
{\cal B}=0\quad \mbox{and}\quad
        \deriv{{\cal A}}{z}=0\quad\mbox{at } z=\pm h(s)\;.
\end{equation}
Since $\alpha$ is an odd function of $z$, kinematic modes have either even
(quad\-ru\-pole) or odd (dipole) parity, with the following symmetry
conditions at the disc midplane (see, e.g., Ruzmaikin \etal, 1988):
\begin{equation}        \label{symmq}
\deriv{{\cal B}}{z}=0\quad\mbox{and}\quad {\cal A}=0\quad\mbox{at }z=0\quad
\mbox{(quadrupole)}\;,
\end{equation}
or
\begin{equation}        \label{symmd}
{\cal B}=0\quad\mbox{and}\quad\deriv{{\cal A}}{z}=0\quad\mbox{at }z=0\quad
\mbox{(dipole)}\;.
\end{equation}

In order to clarify the nature of the dynamo modes in a thin disc, here we
consider an approximate solution of Eqs~(\ref{localb}) and (\ref{locala})
in the form of expansion in free-decay modes
${\cal B}_n(z)$ and ${\cal A}_n(z)$ obtained for $R_\alpha=R_\omega=0$:
\[
\gamma_n {\cal B}_n=\deriv{^2{\cal B}_n}{z^2}\;,\qquad
        \gamma_n {\cal A}_n=\deriv{^2{\cal A}_n}{z^2}\;,
\]
where $\gamma_n\ (<0)$ is the decay rate of the $n$th mode.  For
the boundary conditions (\ref{localbc}) and (\ref{symmq}) that select
quadrupolar modes, the resulting orthonormal set of basis functions is given
by
\begin{eqnarray*}
\left(\begin{array}{c}{\cal B}_{2n}\\{\cal A}_{2n}\end{array}\right)&=&
\left(\begin{array}{c}\sqrt{2}\cos{\left[\pi(n+\sfrac12){z}/{h}\right]}\\
        0\end{array}\right),\\
\left(\begin{array}{c}{\cal B}_{2n+1}\\{\cal A}_{2n+1}\end{array}\right)&=&
\left(\begin{array}{c}0\\
\sqrt{2}\sin{\left[\pi(n+\sfrac12){z}/{h}\right]}\end{array}\right),\\
\gamma_{2n}=\gamma_{2n+1}&=&-\pi^2(n+\sfrac12)^2\;,\qquad
n=0,1,\ldots\;.
\end{eqnarray*}
The free-decay eigenvalues are all doubly degenerate, and
two vector eigenfunctions, one with odd index and the other with even one,
correspond to each eigenvalue, one with $b_{2n+1}=0$, and the other with
$a_{2n}=0$. The eigenfunctions are normalized to have
$\int_0^h({\cal B}_n^2+{\cal A}_n^2)\,dz=1$.

The solution of Eqs~(\ref{localb}) and (\ref{locala}) is represented as
\[
\left(\begin{array}{c}{\cal B}\\{\cal A}\end{array}\right)\approx
\exp^{\gamma t}\sum_{n=0}^{\infty} c_n
\left(\begin{array}{c}{\cal B}_n\\{\cal A}_n\end{array}\right),
\]
where $c_n$ are constants. We substitute this series into Eqs~(\ref{localb})
and (\ref{locala}), multiply by $(b_k,\ a_k)$ and integrate over $z$ from 0 to
$h$ to obtain an algebraic system of homogeneous equations for $c_k$ whose
solvability condition yields an algebraic equation for $\gamma$.  For our
current purposes, it is sufficient to retain the smallest possible number of
modes, which results in a system of two equations for $c_0$ and $c_1$ and a
quadratic  equation for $\gamma$ whose positive solution is given
by
\begin{equation}
\gamma\approx-\sfrac14\pi^2+\sqrt{W_{01}W_{10}}\;,
                                \label{gapp}
\end{equation}
where
\begin{eqnarray*}
W_{01}&=&-\int_0^h\alpha b_0a_1\,dz=-\sfrac12\quad\mbox{for }
\alpha=\sin\pi z/h\;,\\
W_{10}&=&D\int_0^h b_0a_1\,dz=\frac{2}{\pi}D\;.
\end{eqnarray*}
To assess the accuracy of Eq.~(\ref{gapp}), we note that it yields
$\gamma=0$ for $D=D_\mathrm{cr}=-\pi^5/16\approx-19$, as compared with
the accurate value of $D_\mathrm{cr}=-8$ (Ruzmaikin \etal, 1988). This
solution indicates that the dominant mode is non-oscillatory
($\mathrm{Im}\,\gamma=0$); this is confirmed by other analytical and numerical
solutions of the dynamo equations in thin discs.

A similar solution can be obtained for dipolar modes. The free decay modes of
dipolar symmetry have $\gamma_n=-n^2\pi^2,\ n=1,2,\ldots$, so
that the lowest dipolar mode decays four times faster than the lowest
quadrupolar mode. The reason for that is that the azimuthal field of dipolar
parity has zero not only at $|z|=h$ but also at $z=0$ and so a smaller scale
than the quadrupolar solution. This immediately implies that quadrupolar
modes, with $\mean{B}_\phi(z)=\mean{B}_\phi(-z),\
\mean{B}_s(z)=-\mean{B}_s(-z),\ \mean{B}_z(z)=\mean{B}_z(-z)$, should be
dominant in galactic discs. The dominant symmetry of galactic magnetic fields
is thus expected to be different from that in stars and planets, where dipolar
fields are preferred. This prediction is confirmed by observations (see
Sect.~\ref{QS}).

\subsubsection{The global solution}
The vacuum boundary conditions are often used in analytical and semi-analytical
studies of disc dynamos because of their (relative) simplicity. Most
importantly, they have a local form in the lowest order in $\eps$ --- see
Eq.~(\ref{localbc}). However, this advantage is lost as soon as the next order
in $\eps$ is considered, which is needed in order to obtain a governing
equation for the field distribution along radius, $Q$. To this order,
non-local magnetic connection between different radii has to be included, i.e.,
the fact that magnetic lines leave the disc at some radius, pass through the
surrounding vacuum and return to the disc at another radius. In this section
we discuss the radial dynamo equation, and for this purpose we have to
consider vacuum boundary conditions to the first order in $\eps$.

If the disc is surrounded by vacuum, there are no electric currents outside
the disc, i.e., $\Curl{\mean{\vect{B}}}=\vect{0}$, so that the outer
magnetic field is potential, $\mean{\vect{B}}=-\Grad{\Phi}$. Then axial
symmetry implies that the azimuthal field vanishes outside the disc. Since
magnetic field must be continuous on the disc boundary, this yields the
following boundary condition at the disc surface $z=\pm h(s)$:
\begin{equation}                \label{bcaz}
\mean{B}_\phi|_{z=\pm h}=0\;.
\end{equation}

The vacuum boundary condition for the poloidal field (determined by
$\mean{A}_\phi$) was derived in local Cartesian coordinates by Soward (1978).
Priklonsky \etal\ (2000) rederived it in cylindrical geometry in the form
\begin{equation}        \label{bcpol}
\deriv{\mean{A}_\phi}{z}
-\frac{\eps}{s}{\cal L}\left({\mean{A}_\phi}\right)
=0\quad\mbox{at }z=\pm h(s)\;,
\end{equation}
where the integral operator ${\cal L}(\mean{A}_\phi)$ is defined as
\[
{\cal L}\left(\mean{A}_\phi\right)=\int_0^\infty W(s,s,')
\deriv{}{s'}\left(\frac{1}{s'}\,\deriv{}{s'}s'\mean{A}_\phi\right)\,ds'
\]
with the kernel
\[
W(s,s')=ss'\int_0^\infty J_1(ks)J_1(ks')\,dk\;,
\]
where  $J_1(x)$ is the Bessel function.
Willis \etal\ (2003) obtained another, equivalent form of the integral operator
involving Green's function of the Neumann problem for the Laplace equation.

The integral part of the boundary condition (\ref{bcpol})can be transferred
into a non-local term in the equation for $Q$ which then becomes an
integro-differential equation of the form (Priklonsky \etal, 2000)
\begin{equation}        \label{eqQ}
[\Gamma -\gamma(s)]\,q(s)=\eps p(s){\cal L}\left\{q(s)\right\}\;,
\end{equation}
where
\[
q(s)=Q(s){\cal A}(h;s)\;,
\qquad
p(s)=\frac{{\cal A}(h,s){\cal A}_*(h,s)}{\langle{\bf X,X_*}\rangle}\;,
\qquad
{\bf X}=\left(\begin{array}{l}{\cal B}(z;s)\\{\cal A}(z;s)\end{array}\right).
\]
Here $\bf X$ is the eigenvector of the lowest-order boundary value problem
discussed in Sect.~\ref{local}, the asterisk denotes the eigenvector of its
adjoint problem, and
\[
\langle{\bf X,X_*}\rangle=\int_0^h {\bf X}\cdot{\bf X_*}\,dz\;.
\]
The solution of Eq.~(\ref{eqQ}) subject to the boundary conditions
\[
q(0)=0\,\qquad\mbox{and} \qquad q\to 0\quad\mbox{as}\quad s\to\infty
\]
provides yet another eigenvalue problem, for which the eigenvalue is the
global  growth rate $\Gamma$ and the eigenfunction is $q(s)$
which determines the radial profile of the global eigenfunction $Q$. As shown
by Willis \etal\ (2003), the effect of the integral term in Eq.~(\ref{eqQ})
can be described as enhanced radial diffusion.

Equation (\ref{eqQ}) is complicated enough as to provoke an irresistible
desire to simplify it. Such a simplification, employed by Baryshnikova \etal\
(1987) (see also Ruzmaikin \etal, 1988) consists of neglecting the term
containing $\eps$ in the boundary condition (\ref{bcpol}). This makes the
boundary condition local and leads to the following equation for $Q(s)$:
\begin{equation}                         \label{eqQl}
[\Gamma -\gamma(s)] Q = \eps^2
\deriv{}{s}{}\left(\frac{1}{s}\deriv{}{s}{} sQ\right),
\end{equation}
similar to Eq.~(\ref{eqQ}), but with the integral term replaced by
the diffusion operator. Formally, Eq.~(\ref{eqQl}) can be obtained from
Eq.~(\ref{eqQ}) by replacing the integral kernel by the delta-function,
$W(s,s')\to\delta(s-s')$. In other words, this simplification neglects any
nonlocal coupling between different parts of the disc via the halo, but
includes the local diffusive coupling within  the disc. We note in this
connection that the kernel $W(s,s')$ is indeed singular, although the
singularity is only logarithmic in reality, $W(s,s')\sim\ln|s-s'|$.

The above simplification greatly facilitates the analysis of the global dynamo
solutions and all applications of the thin-disc asymptotics to galaxies and
accretion discs neglect the nonlocal effects. Equation~(\ref{eqQl}) can be
readily solved using a variety of analytical and numerical techniques
(Ruzmaikin \etal, 1988), but some features of the solution are lost together
with nonlocal effects. The most important failure is that the asymptotic
scaling of the solution with $\eps$ is affected, with the radial scale
becoming $\eps^{-1/2}h_0$ instead of the correct value $\eps^{-1/3}h_0$.
However, the difference is hardly significant numerically for the realistic
values $\eps\sim10^{-1}$--$10^{-2}$. We note that the thin-disc asymptotics
are reasonably accurate for $\eps\la10^{-1}$ (Baryshnikova \etal, 1987;
Willis \etal, 2003).

Another effect of the nonlocal effects is that solutions of Eq.~(\ref{eqQ})
possess algebraic tails far away from the dynamo active region, $q\sim
s^{-4}$, whereas solutions of Eq.~(\ref{eqQl}) have exponential tails typical
of the diffusion equation. This affects the speed of propagation of magnetic
fronts during the kinematic growth of the magnetic field: with the nonlocal
effects, the fronts propagate exponentially, whereas the local radial
diffusion alone results in a linear propagation.

These topics are discussed in detail by Willis \etal\ (2003) who compare
numerical solutions of Eqs~(\ref{eqQ}) and (\ref{eqQl}). Whether or not the
nonlocal effects can be neglected depends on the goals of the analysis. There
are several reasons why this simplification appears to be justified. The
neglect of nonlocal effects does not seem to affect significantly any
observable quantities, whereas the parameters of spiral galaxies and of their
magnetic fields are known with a rather limited accuracy anyway. Moreover,
the halos of spiral galaxies can be described as vacuum only in a very
approximate sense, and the finite conductivity of the halo will weaken the
nonlocal effects.

\subsubsection{Non-axisymmetric, nonlinear and numerical solutions}\label{NAX}
The above asymptotic theory can readily be extended to non-axisymmetric
solutions. This generalization is discussed by Krasheninnikova \etal\ (1989)
and Ruzmaikin \etal\ (1988). Starchenko \& Shukurov (1989) developed WKB
asymptotic solutions of the mean-field galactic dynamo equations valid for
$|D|\gg1$. A similar asymptotic regime for one-dimensional dynamo equations
(\ref{localb}) and (\ref{locala}) is discussed in \S9.IV of Zeldovich \etal\
(1983).

Another useful approximate approach, known as the `no-$z$' approximation, was
suggested by Subramanian \& Mestel (1993). In this approximation, derivatives
across the disc in Eqs.~(\ref{meanB}) and (\ref{meanA}) or their
three-dimensional analogues are replaced by division by the disc
semi-thickness, $\partial/\partial z\to 1/h$, and the resulting equations in
$s$ and $\phi$ are solved, e.g., by the WKB method or numerically. This
approach appears to be rather crude at first sight, but it is quite efficient
because the structure of the magnetic field across a thin disc is quite
simple, at least for the lowest mode. A refinement of the approximation to
improve its accuracy is discussed by Phillips (2001). Mestel and Subramanian
(1991) and Subramanian \& Mestel (1993) apply these solutions to study the
effects of spiral arms on galactic magnetic fields. This approximation was
also extensively used in numerical simulations of galactic dynamos (Moss 1995;
see Moss \etal, 2001 for an example).

Nonlinear asymptotics of Eqs~(\ref{localb}) and (\ref{locala}) for
$|D|\gg1$ are discussed by Kvasz \etal\ (1992), where it is
supposed that the nonlinearity affects significantly magnetic field
distribution across the disc, and to the lowest approximation the steady state
of the dynamo is established locally. This, however, may not be
the case. The radial coupling is significant already at the kinematic stage
where it results in the establishment of a global eigenfunction as described
by Eq.~(\ref{eqQ}) or Eq.~(\ref{eqQl}). Nonlinear effects are more likely to
affect the global eigenfunction, and so have to affect the radial equation.
Poezd \etal\ (1993) have derived a nonlinear version of Eq.~(\ref{eqQl})
assuming the standard form of $\alpha$-quenching with the
$\alpha$-coefficient modified by magnetic field as
\begin{equation}                \label{que}
\widetilde{\alpha}=\frac{\alpha}{1+\mean{B}^2/B_0^2}\;,
\end{equation}
where $B_0$ is a suitably chosen saturation level most often identified with
a state where magnetic and turbulent kinetic energy densities are of the same
order of magnitude. As a result, the magnetic field can grow when $\mean{B}\ll
B_0$, but then the growth slows down as the quenched dynamo number obtained with
$\widetilde{\alpha}$ approaches its critical value $D_\mathrm{cr}$, and the
field growth saturates at $\mean{B}\approx B_0$. In terms of the thin-disc
asymptotic model, this implies that $\gamma(s)$ in Eqs~(\ref{eqQ}) and
(\ref{eqQl}) ought to be replaced by $\gamma(s)(1-Q^2/B_0^2)$, so that the
nonlinear version of Eq.~(\ref{eqQl}) with the nonlinearity (\ref{que}) has
been derived in the form
\begin{equation}                               \label{eqQnl}
   \deriv{Q}{t}=\gamma(s)\left(1-\frac{Q^2}{B_0^2}\right)Q +
\eps^2\deriv{}{s}{}\left[\frac{1}{s}\deriv{}{s}{} sQ(s)\right],
\end{equation}
provided the local solution has been normalized in such a way that $Q$
is a field strength averaged across the disc at a given radius. The
derivation of this equation by averaging the governing equations across the
disc can be found in Poezd \etal\ (1993). This equation and its
nonaxisymmetric version have been extensively applied to galactic dynamos (see
Beck \etal, 1996, and references therein).

The detailed physical mechanism of the saturation of the dynamo action is
still unclear. Cattaneo \etal\ (1996) suggest that the saturation is associated
with the suppression of the Lagrangian chaos of the gas flow by the magnetic
field. This mechanism, attractive in the context of convective systems (where
the flow becomes random due to intrinsic reasons, e.g., instabilities), can
hardly be effective in galaxies where the flow is random because of the
randomness of its driving force (the supernova explosions).

Most numerical solutions of galactic dynamo equations that extend beyond the
thin-disc approximation rely on the `embedded disc' approach (Ste\-pin\-ski \&
Levy, 1988; Elstner \etal, 1990). Instead of using complicated boundary
conditions at the disc surface, this approach considers a disc embedded into a
halo whose size is large enough as to make unimportant boundary conditions
posed at the remote halo boundary. Since turbulent magnetic diffusivity in
galactic halos is larger than in the disc (Sokoloff \& Shukurov, 1990; Poezd
\etal, 1993), meaningful embedded disc models are compatible with thin-disc
asymptotic solutions obtained with vacuum boundary conditions and confirm the
asymptotic results. The embedded disc approach was also used to study
dynamo-active galactic halos (Brandenburg \etal, 1992, 1993, 1995; Elstner
\etal, 1995). Further extensions of disc dynamo models include the effects of
magnetic buoyancy (Moss \etal, 1999), accretion flows (Moss \etal, 2000) and
external magnetic fields (Moss \& Shukurov, 2001, 2004).

An implication of the nonlinear model for the thin-disc dynamo is that the
local solution is unaffected by nonlinear effects whose main r\^ole is to
modify the radial field structure. An important consequence of this is that it
can be reasonably expected that the pitch angle of magnetic lines,
$p=\arctan{\mean{B}_s/\mean{B}_\phi}$, is weakly affected by  nonlinear
effects, and so represents an important feature of the solution that can be
directly compared with observations (Baryshnikova \etal, 1987). This
expectation seems to be confirmed by observations (Sect.~\ref{MPA}).
Nevertheless, the modification of the magnetic pitch angle by nonlinear
effects has never been studied in detail, which seems to be a regrettable
omission.

\subsubsection{Dynamo control parameters in spiral galaxies}\label{DCP}
A remarkable feature of spiral galaxies is that they are (almost) transparent
to electromagnetic waves over a broad range of frequencies, so the kinematics
of the ISM is rather well understood, and therefore all parameters essential
for dynamo action are well restricted by observations. This leaves less room
for doubt and less freedom for speculation than in the case of other natural
dynamos. Another advantage is that observations of polarized radio emission at
a linear resolution of 1--3\,kpc (typical of the modern observation of nearby
galaxies) reveal exactly that field which is modelled by the mean-field dynamo
theory (given volume and ensemble averages are identical).

The mean-field dynamo is controlled by two dimensionless parameters
quantifying the differential rotation and the so-called $\alpha$-effect, as
defined in Eq.~(\ref{Dnumbers}). Using Eqs.~(\ref{etat}) and (\ref{alKrause})
and assuming a flat rotation curve, $\Omega=\UU_0/s$, we obtain the following
estimates for the Solar vicinity of the Milky Way:
\[
R_\omega\approx-3\frac{\UU_0}{\uu_0}\,\frac{h_0^2}{\ell s}\approx-20\;,
\qquad
R_\alpha\approx3\frac{\UU_0}{\uu_0}\,\frac{\ell}{s}\approx1\;,
\]
where $\UU_0=s_0\Omega_0$ is the typical rotational velocity.
Since $|R_\omega|\gg R_\alpha$, differential rotation dominates in the
production of the azimuthal magnetic field (i.e., the $\alpha\omega$-dynamo
approximation is well applicable), and the dynamo action is essentially
controlled by a single parameter, the {\em dynamo number\/}
\begin{equation}                 \label{D}
D=R_\alpha R_\omega\sim 10\frac{h_0^2}{\uu_0^2}\,
s\Omega\,\frac{\partial\Omega}{\partial s}
\approx-10\left(\frac{\UU_0 h_0}{\uu_0 s}\right)^2 \approx-20\;,
\end{equation}
where the numerical estimate refers to the Solar vicinity. Thus, $|D|$ does
exceed the critical value for the lowest, non-oscillatory quadrupole dynamo
mode, which then can be expected to dominate in the main parts of spiral
galaxies. It is often useful to consider the local dynamo number $D(s)$, a
function of galactocentric radius $s$, obtained when the $s$-dependent, local
values of the relevant parameters are used in Eq.~\ref{Dnumbers} instead of
the characteristic ones.

The local regeneration (e-folding) rate of the regular magnetic field
$\gamma$ is related to the magnetic diffusion time along the smallest
dimension of the gas layer and to the dynamo number (if $|R_\omega|\gg
R_\alpha$). Using the perturbation solution of Sect.~\ref{local}, the
following expression (written in dimensional form) can be used as a rough
estimate:
\begin{equation}                             \label{gamma}
\gamma\sim \frac{\turb{\eta}}{h^2}
        \left(\sqrt{|D|}-\sqrt{|D_\mathrm{cr}|}\right),
\quad\mbox{for } |D|\ga D_{\rm cr}\;,
\end{equation}
where $D_\mathrm{cr}\approx-6$ can be adopted from the more accurate numerical
solution and numerical factor of order unity has been omitted.
This  yields the local e-folding time $\gamma^{-1}\approx4\times10^8\yr$ for
the Solar neighbourhood. When the radial diffusion is included, i.e.,
Eq.~(\ref{eqQ}) is solved, the growth rate decreases, yielding a global
e-folding time of $\Gamma^{-1}\sim10^9\yr$ near the Sun.
Thus, the large-scale magnetic field near the Sun can be amplified by a
factor of about $10^4$ during the galactic lifetime, $10^{10}\yr$, and the
Galactic seed field had to be rather strong, about $10^{-10}\G$. The
fluctuation dynamo can produce such a statistical residual magnetic field at
the scale of the leading eigenfunction either in the young galaxy (\S{VII.13}
in Ruzmaikin \etal, 1988; Widrow, 2002) or in the protogalaxy (Kulsrud \etal,
1997).

The above growth rate, estimated for the Solar neighbourhood of the Milky Way,
is often erroneously adopted as a value typical of spiral galaxies in
general. It is then important to note that the regeneration rate is
significantly larger in the inner Galaxy (the local dynamo number rapidly
grows as $s$ becomes smaller, $D(s)\propto G\Omega$ --- see
Fig.~\ref{rotcurve}) and in other galaxies. For example, Baryshnikova \etal\
(1987) estimate the global growth time of the leading axisymmetric mode in the
galaxy M51 as $5\times10^7\yr$.

Gaseous discs of spiral galaxies are flared, ie., $h\propto s+\mbox{const}$ at
$s\ga10\kpc$, whereas $\uu_0$ only slightly varies with $s$. For a flat
rotation curve, $\Omega\propto s^{-1}$, Eq.~(\ref{D}) then shows that the
local dynamo number does not vary much with galactocentric radius $s$ and
remains supercritical, $|D(s)|\geq |D_\mathrm{cr}|$ out to a large radius.
It is therefore not surprising that regular magnetic fields have been detected
in all galaxies where observations have sufficient sensitivity and resolution
(Wielebinski \& Krause, 1993; Beck \etal, 1996; Beck, 2000, 2001).

A standard estimate of the steady-state strength of magnetic field
produced by the mean-field dynamo follows from the balance of the Lorentz
force due to the large-scale magnetic field and the Coriolis force that causes
deviations from mirror symmetry (Ruzmaikin \etal, 1988; Shukurov, 1998):
\begin{eqnarray}
\mean{B}&\approx&\left[4\pi\rho\uu_0\Omega\ell
  \left(\left|\frac{D}{D_{\rm cr}}\right|-1\right)\right]^{1/2} \label{B}\\
&\approx&2\mkG
        \left(\left|\frac{D}{D_{\rm cr}}\right|-1\right)^{1/2}
        \left(\frac{n}{1\cmcube}\right)^{1/2}
        \left(\frac{\uu_0}{10\kms}\right)^{1/2},\nonumber
\end{eqnarray}
where $\rho\simeq1.7\times10^{-24}\gcmcube$ is the density of
interstellar gas and $n$ its number density, $n=\rho/m_\mathrm{H}$ with
$m_\mathrm{H}$ the proton mass. This estimate yields values that are in good
agreement with observations, but its applicability has to be reconsidered in
view of the current controversy about the nonlinear behaviour of mean-field
dynamos (see Sect.~\ref{hel}).

It is now clear what information is needed to construct a useful dynamo model
for a specific galaxy: its rotation curve, the scale height of the gas layer,
the turbulent scale and speed, and the gas density. All these parameters are
observable, even though their observational estimates may be incomplete or
controversial. One of successes of the mean-field dynamo theory is its
application to spiral galaxies, where even simplest, quasi-kinematic models
presented above are able to reproduce all salient features of the observed
fields, both in terms of generic properties and for specific galaxies
(Ruzmaikin \etal, 1988). We discuss this in Sect.~\ref{OEDASG}.

Recent observational progress has allowed to explore the effects of galactic
spiral patterns on magnetic fields (Beck, 2000). The corresponding dynamo
models require the knowledge of the arm-interarm contrast in all the relevant
variables (Shukurov \& Sokoloff, 1998; Shukurov, 1998; Shukurov \etal, 2004).

\subsection{The fluctuation dynamo and small-scale magnetic fields}\label{FD}
Similarly to mean-field dynamos, the theory of the fluctuation dynamo is well
understood in the kinematic regime, but nonlinear effects remain
controversial. In this section we present results obtained with kinematic
 models of the fluctuation dynamo and those derived with simplified
nonlinearity. The pioneering kinematic model of the fluctuation dynamo was
developed by Kazantsev (1967), and many more recent developments are based on
it. Detailed reviews of the theory and references can be found in \S{8.IV} of
Zeldovich \etal\ (1983), Ch.~9 of Zeldovich \etal\ (1990) and in Brandenburg
\& Subramanian (2004).

The growth time of the random magnetic field in a random velocity field of a
scale $\ell$ is as short as the eddy turnover time, $\ell/\uu_0\sim10^7\yr$ in
the warm phase for $\ell=0.1\kpc$. The magnetic field produced by the dynamo
action is a statistical ensemble of magnetic flux ropes whose length is of the
order of the flow correlation length, $\ell\simeq0.05$--$0.1\kpc$. The rope
thickness is of the order of the resistive scale, $\ell\Rem^{-1/2}$, in a
single-scale velocity field, where $\Rem$ is the magnetic Reynolds number. A
phenomenological model of dynamo in Kolmogorov turbulence yields the rope
thickness of $\ell\Rem^{-3/4}$ (Subramanian, 1998). The dynamo action can
occur provided $\Rem>\Remcr$, where the critical magnetic Reynolds number is
estimated as $\Remcr=30$--100 in simplified models of homogeneous,
incompressible turbulence. Recent studies have revealed the possibility that
small-scale magnetic fields can have peculiar fine structure because the
magnetic dissipation scale in the interstellar gas is much smaller than that
of turbulent motions, i.e., because the magnetic Prandtl number is much larger
than unity (Schekochikhin \etal, 2002).

Subramanian (1999) suggested that a steady state, reached via the back-action
of the magnetic field on the flow, can be established by the reduction of the
effective magnetic Reynolds number down to the value critical for the dynamo
action, an idea similar to the concept of $\alpha$-quenching in the mean-field
theory. Then the thickness of the ropes in the steady state can be estimated
as $\ell\Remcr^{-1/2}$ or $\ell\Remcr^{-3/4}$. Using a model nonlinearity in
the induction equation with incompressible velocity field, Subramanian (1999)
showed that the magnetic field strength within the ropes $b_0$ saturates at
the equipartition level with kinetic energy density,
$b_0^2/8\pi\sim\sfrac12\rho\uu_0^2$. The average magnetic energy density is
estimated as $\overline{b^2}/8\pi\sim\sfrac12\Remcr^{-1}\rho \uu_0^2$,
implying the volume filling factor of the ropes of order
$f_V\sim\Remcr^{-1}\sim0.01$. Correspondingly, the mean magnetic energy
generated by the small-scale dynamo in the steady state is about 1\% of the
turbulent kinetic energy density, in agreement with numerical simulations.

Shukurov \& Berkhuijsen (2003) interpret thin, random filaments of zero
polarized intensity observed in polarization maps of the Milky Way
(known as canals) as a result of Faraday depolarization in the turbulent
interstellar gas. This interpretation has resulted in a tentative estimate of
the Taylor microscale of the interstellar turbulence
\[
\ell_\mathrm{T}=\ell\widetilde{\Rem}^{-1/2}\sim 0.6\p\;,
\]
where $\widetilde{\Rem}$ is the effective magnetic Reynolds number in the
ISM. This yields the following estimate:
\[
\widetilde{\Rem}=\sim10^4\;.
\]
Of course, this is a very tentative estimate, and further analyses of
observations and theoretical developments will be needed to refine it. The
value of $\widetilde{\Rem}$ obtained is significantly larger than $\Remcr$
obtained in idealized models. This might be due to the transonic nature of
interstellar turbulence as the gas compressibility appears to hinder dynamo
action. Kazantsev \etal\ (1985) have shown that the e-folding time of magnetic
field in the acoustic-wave turbulence (i.e., a compressible flow) is as long
as ${\cal M}^4\ell/\uu_0$, where ${\cal M}\ (\ga 1)$ is the Mach number.

Using parameters typical of the warm phase of the ISM, this theory predicts
that the small-scale dynamo would produce magnetic flux ropes of the length
(or the curvature radius) of about $\ell=50$--100\,pc and thickness
5--$10\p$ for $\widetilde{\Rem}=10^2$ and $0.5$--10\,pc for
$\widetilde{\Rem}=10^4$. The field strength within the ropes, if at
equipartition with the turbulent energy, has to be of order $1.5\mkG$ in the
warm phase ($n=0.1\cmcube$, $\uu_0=10\kms$) and $0.5\mkG$ in the hot gas
($n=10^{-3}\cmcube$, $\uu_0=40\kms$). Note that some heuristic models of the
small-scale dynamo admit solutions with magnetic field strength within the
ropes being significantly above the equipartition level, e.g., because the
field configuration locally approaches a force-free one,
$|(\Curl{\bfB})\times{\bfB}|\ll B^2/\ell$, where $\ell$ is the field scale
(Belyanin \etal, 1993).

The small-scale dynamo is not the only mechanism producing random magnetic
fields (e.g., \S4.1 in Beck \etal, 1996, and references therein). Any
mean-field dynamo action producing magnetic fields at scales exceeding the
turbulent scale also generates small-scale magnetic fields.
Similarly to the mean magnetic field,  this component of the turbulent field
presumably has a filling factor close to unity in the warm gas and its
strength is expected to be close to equipartition with the turbulent energy at
all scales. This component of the turbulent magnetic field
may be confined to the warm gas, the site of the mean-field dynamo action, so
magnetic field in the hot phase may have a better pronounced ropy structure.

The overall structure of the interstellar turbulent magnetic field in the warm
gas can be envisaged as a quasi-uniform fluctuating background with one
percent of the volume occupied by flux ropes (filaments) of a length
50--100\,pc containing a well-ordered magnetic field. This basic distribution
would be further complicated by compressibility, shock waves, MHD
instabilities (such as Parker instability), the fine structure at subviscous
scales, etc.

The site of the mean-field dynamo action is plausibly the warm phase rather
than the other phases of the ISM. The warm gas has a large filling factor (so
it can occupy a percolating global region), it is, on average, in a state of
hydrostatic equilibrium, so it is an ideal site for both the small-scale and
mean-field dynamo action. Molecular clouds and dense $\HI$ clouds have too
small a filling factor to be of global importance. Fletcher \& Shukurov (2001)
argue that, globally, molecular clouds can be only weakly coupled to the
magnetic field in the diffuse gas, but Beck (1991) suggests that a significant
part of the large-scale magnetic flux can be anchored in molecular clouds. The
time scale of the small-scale dynamo in the hot phase is
$\ell/\uu_0\simeq10^6\yr$ for $\uu_0=40\kms$ and $\ell=0.04\kpc$ (the width of
the hot, `chimneys' extended vertically in the disc). This can be shorter than
the advection time due to the vertical streaming, $h/\UU_z\simeq10^7\yr$ with
$h=1\kpc$ and $\UU_z=100\kms$. Therefore, the small-scale dynamo action should
be possible in the hot gas. However, the growth time of the mean magnetic
field must be significantly longer than $\ell/\uu_0$, reaching a few hundred
Myr. Thus, the hot gas can hardly contribute significantly to the mean-field
dynamo action in the disc and can drive the dynamo only in the halo (Sokoloff
\& Shukurov, 1990). The main r\^ole of the fountain flow in the disc dynamo is
to enhance magnetic connection between the disc and the halo (see
Sect.~\ref{hel}).

\subsection{Magnetic helicity balance in the galactic disc}    \label{hel}
Conservation of magnetic helicity $\chi =\langle\bfA\cdot\bfB\rangle$ (where
$\bfB=\Curl{\bfA}$) in a perfectly conducting medium has been identified as an
important constraint on mean-field dynamos that plausibly explains the
catastrophic quenching of the $\alpha$-effect discussed elsewhere in this
volume (Blackman \& Field, 2000; Kleeorin \etal, 2000, 2003; Brandenburg \&
Subramanian, 2004). In a closed system, magnetic helicity can only evolve on
the (very long) molecular diffusion time scale; in galaxies, this time scale
by far exceeds the Hubble time. The large-scale galactic magnetic fields have
significant magnetic helicity of the order of $L B_s B_\phi\sim-\sfrac14 L
B^2$, where $L\ga1\kpc$ is the field scale, $B_s/B_\phi=\sin p$ with
$p\approx-15^\circ$ the magnetic pitch angle. Since the initial (seed)
magnetic field was weak, and so had negligible magnetic helicity, the
large-scale magnetic helicity in a closed system must be balanced by the
small-scale helicity of the opposite sign,
$\simeq\ell_\mathrm{h}\mean{b^2}$, where $\ell_\mathrm{h}$ is an
appropriate dominant scale of magnetic helicity. This immediately results in
an upper limit on the steady-state mean magnetic field (Brandenburg \&
Subramanian, 2004, and references therein)
\begin{equation}        \label{hc}
\frac{\mean{B}^2}{\mean{b^2}}\la 4\frac{\ell_\mathrm{h}}{L}\approx0.4\;,
\end{equation}
where the numerical value is obtained for $\ell_\mathrm{h}=0.1\kpc$ and
$L=1\kpc$. The result of Vainshtein \& Cattaneo (1992),
$\mean{B}^2/\mean{b^2}\sim\Rem^{-1/2}$ is recovered for
$\ell_\mathrm{h}\sim L\Rem^{-1/2}$. The observed relative strength of the mean
field in spiral galaxies is given by $\mean{B}^2/\mean{b^2}\sim0.5$.
The upper limit on the strength of the mean magnetic field (\ref{hc}) appears
to be much lower than the observed field only if $\ell_\mathrm{h}\ll0.1\kpc$.
For $\ell_\mathrm{h}=0.1\kpc$, the observed field strength is compatible with
magnetic helicity conservation. What is perhaps more worrying, is that the
mean magnetic fields can only grow at the long molecular diffusion time scale
to reach this strength.

Blackman \& Field (2000) and Kleeorin \etal\ (2000) suggested that the losses
of the small-scale magnetic helicity through the boundaries of the dynamo
region play the key r\^ole in the mean-field dynamo action. This is an
appealing idea, especially because the mean-field dynamos rely on magnetic
flux loss through the boundaries (\S9.II in Zeldovich \etal, 1983; \S{VII.5} in
Ruzmaikin \etal, 1988). A similar situation occurs with the magnetic moment,
which is a conserved quantity, and it only grows in a dynamo system of a
finite size because the dynamo just redistributes it expelling magnetic moment
out from the dynamo active region (Moffatt, 1978). However, these are the mean
magnetic flux and moment that need to be transferred through the boundaries.
Transport by turbulent magnetic diffusion is sufficient for these purposes.
The new aspect of the magnetic helicity balance is that healthy mean-field
dynamo action requires asymmetry between the transports of the magnetic
helicities of the large- and small-scale magnetic fields.

A useful framework to assess the effects of magnetic helicity flow through the
boundaries of the dynamo region was proposed by Brandenburg \etal\ (2002) who
have presented the balance equation of magnetic helicity in the form
\begin{equation}
\label{balance} \frac{d\chi_B}{dt}+\frac{d\chi_b}{dt}
=-2\eta \chi_J-2\eta \chi_j-Q_B-Q_b\;,
\end{equation}
where
$\chi_B=\bfA\cdot\bfB$ and $\chi_b=\langle\bfa\cdot\bfb\rangle$
are the magnetic helicities of the
mean and random magnetic fields, respectively, $\eta$ is the
molecular magnetic diffusivity, $\chi_J=\bfJ\cdot\bfB$ and
$\chi_j=\langle\bfj\cdot\bfb\rangle$ are the current
helicities (with $\bfJ=\Curl\bfB$ the current density).
The first two terms on the right-hand side of
Eq.~(\ref{balance}) are responsible for the Ohmic losses whereas the last two
terms represent the boundary losses. For illustrative purposes and following
Brandenburg \etal\ (2002), we adopt the following assumptions. (i)~The
magnetic fields are fully helical, so $M_B=\sfrac12 k_B |\chi_B|$ and
$M_b=\sfrac12 k_b |\chi_b|$, where $M_B$ and $M_b$ are the average energy
densities of the mean and random magnetic fields and $k_B$ and $k_b$ are their
wave numbers, respectively. Furthermore, $\chi_J=k_B^2\chi_B$ and
$\chi_j=k_b^2\chi_b$. (ii)~The mean and random magnetic fields have widely
separated scales, $k_B\ll k_b$. (iii)~Approximate equipartition is maintained
between the mean and random magnetic fields, $M_B\approx M_b$. Then
\[
\left|\frac{\chi_B}{\chi_b}\right|
=\frac{k_b}{k_B}\,\frac{M_B}{M_b}
=\frac{k_b^2}{k_B^2}\,\frac{\chi_J}{\chi_j}\;,
\]
and so $|\chi_B|\gg|\chi_b|$ and $|\chi_J|\ll|\chi_j|$.
Assuming for definiteness that $\chi_B,\chi_J>0$, we have
$\chi_b,\chi_j<0$, and Eq.~(\ref{balance}) can be approximated by
\begin{equation}        \label{MB}
\frac{dM_B}{dt}
=2\eta k_b k_B M_b+\sfrac12 k_B (Q_B+Q_b)\;.
\end{equation}

It is important to note that the effective advection velocities for the
large-scale and small-scale magnetic fields are {\it
not\/} equal to each other. Both small-scale and large-scale magnetic fields
are advected from the disc by the galactic fountain flow. With a typical
vertical velocity of order $\UU_z=100$--$200\kms$ and the surface covering
factor of the hot gas $f=0.2$--0.3, the effective vertical advection speed is
$f\UU_z=30$--$70\kms$. However, the large-scale magnetic field is subject to
turbulent pumping (turbulent diamagnetism). Given that the turbulent
magnetic diffusivity in the disc and the halo are
$\turb{\eta}^{(\mathrm{d})}=10^{26}\cm^2\s^{-1}$ and
$\turb{\eta}^{(\mathrm{h})}=2\times10^{27}\cm^2\s^{-1}$ (Poezd \etal, 1993),
respectively, and that the transition layer between the disc and the halo has
a thickness of $\Delta z=1\kpc$, the resulting advection speed is
$\UU_\mathrm{d}=-\sfrac12\nabla\turb{\eta}\approx-(2$--$3)\kms$. Thus, the
vertical advection velocities of the large-scale and small-scale magnetic
fields are $f\UU_z+\UU_\mathrm{d}$ and $f\UU_z$, respectively.

Now we can estimate the magnetic helicity fluxes through the disc surface as
\[
Q_B=-\left(\UU_B+\frac{1}{k_B\tau_\eta}\right)M_B\;,
        \qquad Q_b=\UU_b M_b\;,
\]
where $\tau_\eta=1/(4\turb{\eta}k_B^2)$ is the time scale of the (turbulent)
diffusive transport of the mean magnetic field through the boundary, and
$\UU_B$ and $\UU_b$ are effective advection velocities for the large-scale
and small-scale magnetic helicities, respectively. The latter can be estimated
from the following arguments. Consider advection of magnetic field through
the disc surface $z=h$ by a flow with a speed $\UU$,
$\partial\mean{B}^2/\partial t=-\UU\partial\mean{B}^2/\partial z$. Assuming
for simplicity that $\UU$ is independent of $z$, we obtain by integration over
$z$: $2h\dot{M}_B=-2\UU\mean{B}^2(h)$, where
$M_B=(2h)^{-1}\int_{-h}^h\mean{B}^2\,dz$. With $M_B=-k_B\chi_B/2$, this shows
that advection of magnetic field at a speed $\UU$ produces the large-scale
helicity loss at a rate $\dot{\chi}_B\equiv Q_B=(2\UU/k_Bh)\mean{B}^2(h)$.
Here $\mean{B}(h)$ is the large-scale field strength at the disc surface,
which is given by $\mean{B}^2(h)\equiv\xi M_B$, where $\xi<1$ because the
large-scale magnetic field at the surface must be weaker than that deep in the
disc. For example, $\xi\ll1$ for vacuum boundary conditions where
$\mean{B}_\phi(h)=0$. Thus,
\[
\UU_B=\frac{2\xi}{k_B h}(f\UU_z+\UU_\mathrm{d})\;.
\]
Unlike the large-scale magnetic field, the small-scale magnetic fields are not
necessarily weaker at the disc surface, so similar arguments yield
\[
\UU_b=\frac{2}{k_b h}f\UU_z\neq\UU_B\;.
\]

Thus, there are several reasons for the magnetic helicity fluxes through the
disc surface to be different at small and large scales: most importantly, the
large-scale magnetic field at the surface can be much smaller than that deep
in the disc ($\xi\ll1$) and, in addition, turbulent diamagnetism introduces
further difference ($\UU_\mathrm{d}\neq0$).

Equation (\ref{MB}) has the following
solution satisfying the initial condition $M_B(0)=0$:
\begin{equation}	\label{sol}
\frac{M_B}{M_b}=
\frac{4\eta k_b+U_b}{4\turb{\eta} k_B+U_B}
\left\{1-
\mathrm{exp}\,\left[-\sfrac12\left(\frac{1}{\tau_B}+k_B U_B\right)t\right]
\right\}.
\end{equation}
For $t\ll\tau_B$, this solution captures the exponential growth of
the mean magnetic field at a time scale $2\tau_B$, $M_B\propto t/2\tau_B$.
For $U_B=U_b=0$, we obtain $M_B/M_b\approx\eta
k_b/\turb{\eta}k_B\sim\Rem^{-1}$ --- this corresponds to the catastrophic
quenching of the $\alpha$-effect associated with approximate magnetic helicity
conservation in a medium with (weak) Ohmic losses alone. However, for
$U_b\gg4\eta k_b$ (a condition safely satisfied for any realistically small
$\eta$) and $U_B\gg4\turb{\eta}k_B\approx8\kms$, we obtain

\begin{equation}        \label{sat}
\left.\frac{M_B}{M_b}\right|_{t\to\infty}=\frac{U_b}{U_B}\sim
\frac{k_B}{\xi k_b}
\sim\frac{1}{10\xi}\;,
\end{equation}
where we recall that $\xi<1$ and neglect $\UU_\mathrm{d}$. Thus, states with
$M_B\approx M_b$ cannot be excluded, and this equipartition state is reached
at the time scale of order $\tau_B\sim4\times10^8\yr$.

These arguments suggest that the growth rate of the mean magnetic
field is limited from above by the flux of the mean magnetic helicity
through the boundary of the dynamo region, whereas the upper limit
for its steady state strength is controlled by the
rate at which the small-scale magnetic helicity is transferred through the
boundaries, Eq.~(\ref{sat}). Another limit on the mean field strength arises
from the balance of the Lorentz and Coriolis forces in the disc, Eq.~(\ref{B}).
The steady-state
strength of the mean magnetic field is the minimum of the two values.
These arguments suggest that the restrictions on the mean-field dynamo action
from magnetic helicity conservation can be removed as soon as one allows for
the disc-halo connection and fountain flows in spiral galaxies. Of course,
these heuristic arguments have to be confirmed by quantitative analysis.

\begin{figure}
\begin{center}
\includegraphics[width=0.46\textwidth]{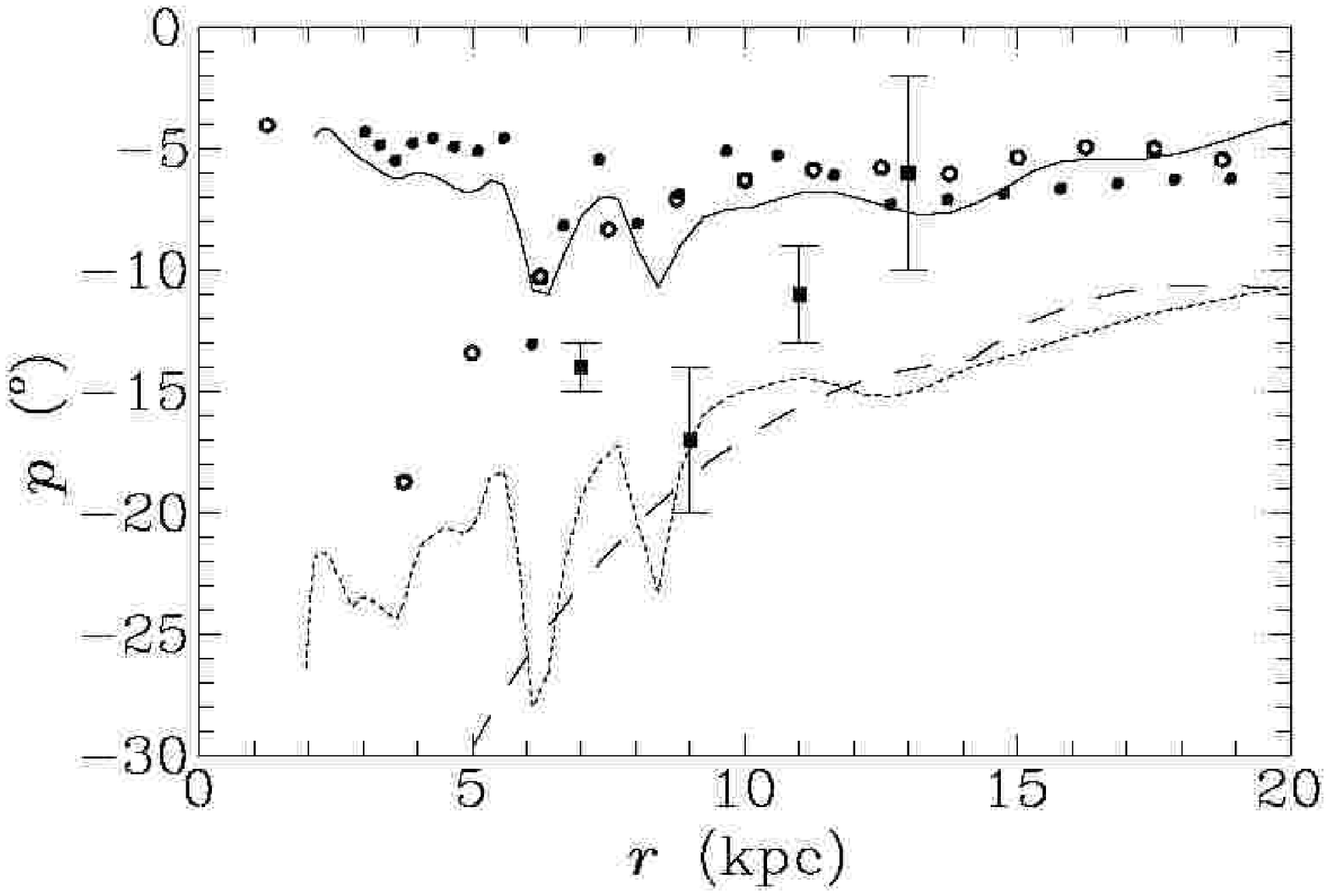}
\includegraphics[width=0.46\textwidth]{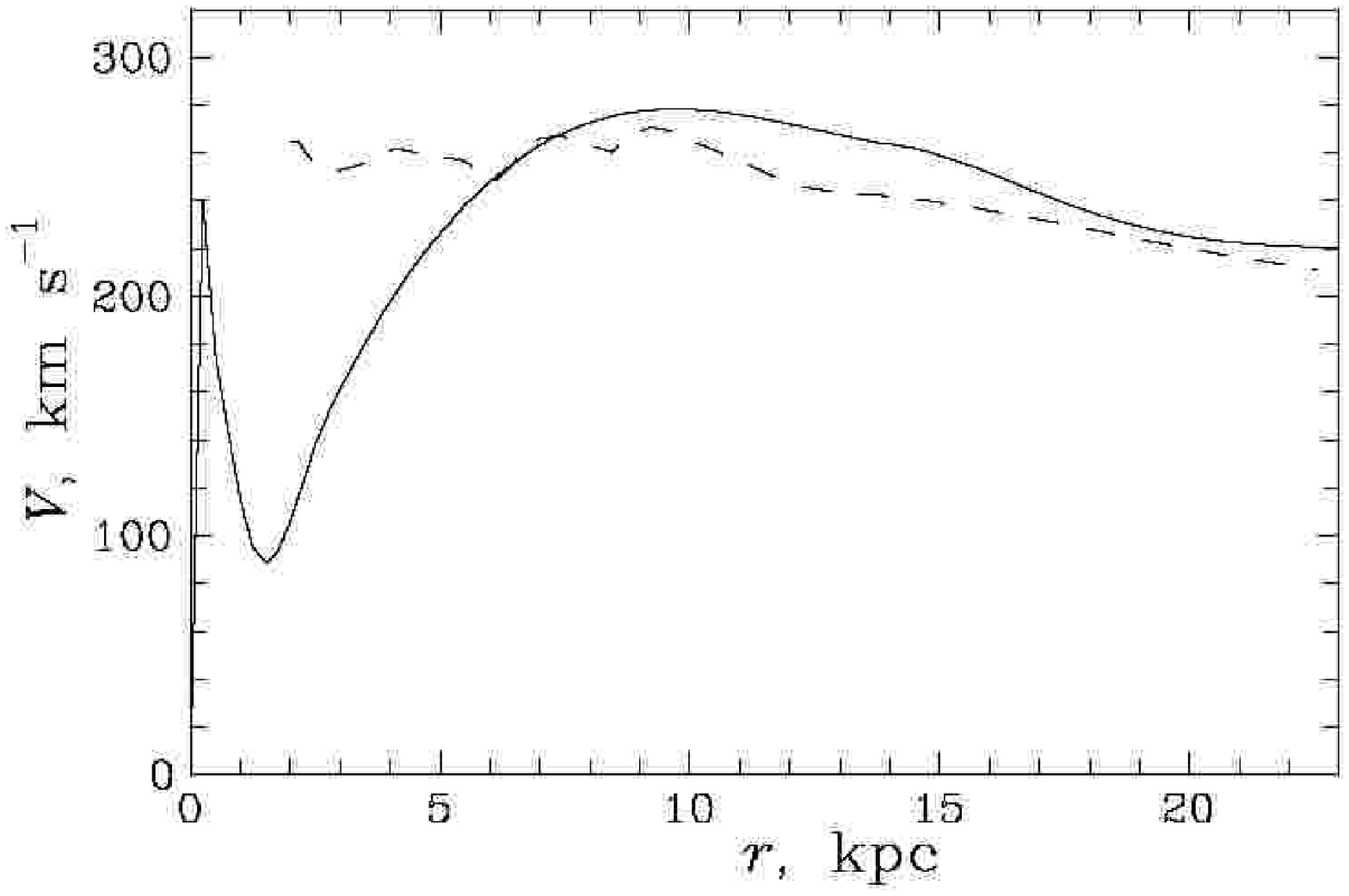}
\end{center}
\caption[]{\label{M31pitch}{\em Left panel:\/} The pitch angle of magnetic
field in the galaxy M31 as obtained from radio polarization observations
(squares with error bars) (Fletcher \etal, 2004), from Eq.~(\protect\ref{p})
using the rotation curve of Deharveng \& Pellet (1975) and Haud (1981)
(dashed) and Braun (1991) (dotted), and from  Eq.~(\protect\ref{pn})
with $D_{\rm cr}=1$ using the same rotation curves (open and filled
circles, respectively); $h(r)$ is twice the $\HI$
scale height of Braun (1991). Results from a nonlinear dynamo model
for M31 (Beck \etal, 1998) are shown with solid line. {\em Right
panel:\/} The rotation curve of M31 from Deharveng \& Pellet (1975)
and Haud (1981) (solid) and from Braun (1991) (dashed).}
\end{figure}

\section{Observational evidence for the origin of\\ galactic magnetic fields}
                                                        \label{OEDASG}
\subsection{Magnetic pitch angle}\label{MPA}
Regular magnetic fields observed in spiral galaxies have field lines
in the form of a spiral with a pitch angle in the range
$p=-(10^\circ$--$30^\circ)$, with negative values indicating a
trailing spiral (e.g., Beck \etal, 1996). As discussed in Sect.~\ref{NAX},
the value of the pitch angle is a useful diagnostic of the mechanism
maintaining the magnetic field.

Consider the simplest from of mean-field dynamo equations (\ref{localb}) and
(\ref{locala}) appropriate for a thin galactic disc, but now written in terms
of dimensional variables for $\mean{B}_s$ and $\mean{B}_\phi$:
\begin{equation}                         \label{dyn}
\deriv{\mean{B}_s}{t}=-\deriv{}{z}(\alpha \mean{B}_\phi)
        +\turb{\eta}\deriv{^2\mean{B}_s}{z^2}\;,
\qquad
\deriv{\mean{B}_\phi}{t}=G\mean{B}_s+\turb{\eta}\deriv{^2\mean{B}_\phi}{z^2}\;.
\end{equation}
Any regular magnetic field maintained by the dynamo must have a non-zero
pitch angle:  for $\mean{B}_s\equiv0$ (a purely azimuthal magnetic field),
equation for $\mean{B}_\phi$ in (\ref{dyn}) reduces to a diffusion equation
$\partial{\mean{B}}_\phi/\partial{t}=\turb{\eta}\partial^2\mean{B}_\phi/\partial{z^2}$
which only has decaying solutions,
$\mean{B}_\phi\propto\mathrm{exp}(-\turb{\eta} t/h^2)$. The same applies to a
purely radial magnetic field.

Consider exponentially growing solutions,
$\mean{B}_{s,\phi}={\cal B}_{s,\phi}\mathrm{exp}(\gamma t)$, and replace
$\partial/\partial z$ by $1/h$ and $\partial^2/\partial z^2$ by $-1/h^2$ (as
in the `no-$z$' approximation) to obtain from Eqs.~(\ref{dyn}) two algebraic
equations,
\[
\left(\gamma+\turb{\eta}/h^2\right){\cal B}_s+\alpha {\cal B}_\phi/h=0\,,\quad
-G{\cal B}_s+\left(\gamma+\turb{\eta}/h^2\right){\cal B}_\phi=0\,,
\]
which have non-trivial solutions only if the determinant vanishes,
which yields $(\gamma+\turb{\eta}/h^2)^2\simeq-\alpha G/h$, and
Eq.~(\ref{gamma}) follows with $D_{\rm cr}=1$. The resulting
estimate of the magnetic pitch angle is given by
\begin{equation}
\tan p
=\frac{{\cal B}_s}{{\cal B}_\phi}
\approx-\sqrt{\frac{\alpha}{-Gh}}
=-\sqrt{\frac{R_\alpha}{|R_\omega|}}
\sim-\frac{\ell}{h}\left|\deriv{\ln\Omega}{\ln s}\right|^{-1/2}\;.
                                        \label{p}
\end{equation}
For $\ell/h\simeq1/4$ and a flat rotation
curve, $\partial{\ln\Omega}/\partial{\ln s}=-1$, we obtain $p\simeq-15^\circ$,
and this is the middle of the range observed in spiral galaxies.
More elaborate treatments discussed by Ruzmaikin \etal\ (1988b) confirm this
estimate of $p$ and yield a more accurate value of $D_{\rm cr}$. For example,
the perturbation solution of Sect.~\ref{local} yields
\begin{equation}        \label{ppert}
p\approx-\sfrac12\pi^{3/2}\sqrt{\frac{R_\alpha}{|R_\omega|}}\;.
\end{equation}

If the steady state is established by reducing $R_\alpha$ to its
critical value as to obtain $R_\alpha R_\omega=D_{\rm cr}$, then
the pitch angle in the nonlinear steady state becomes
\begin{equation}
\tan p\approx-\sfrac12\pi^{3/2}\frac{\sqrt{|D_{\rm cr}|}}{|R_\omega|}\;.
                                \label{pn}
\end{equation}

The magnetic pitch angle in M31 determined from observations and dynamo
theory is shown in Fig.~\ref{M31pitch}. Although the model curves show
noticeable differences from the observed pitch angles, the general agreement
is encouraging. The situation is typical: magnetic pitch angles of spiral
galaxies are in a good agreement with predictions of dynamo theory (Beck
\etal, 1996).

This picture does not explain why the pitch angles of galactic
magnetic fields are invariably close (though not equal) to those of
the spiral pattern in the parent galaxy. A plausible explanation is
that magnetic pitch angles are further affected by streaming motions
associated with the spiral pattern to make the match almost perfect (Moss,
1998). We note, however, that the pitch angle of the large-scale magnetic
field near the Sun differs significantly from that of the local (Orion) arm;
it is not clear whether this misalignment is of a local or global nature.

As shown by Moss \etal\ (2000), magnetic pitch angle can be affected
by an axisymmetric radial inflow (as well as outflow):
\[
\tan p
\approx-\sfrac12\pi^{3/2}\sqrt{\frac{R_\alpha}{|R_\omega|}}
\left(1-\sfrac12 {\cal R}\sqrt{\frac{\pi}{-D}}\,\right),
\qquad {\cal R}=\frac{h^2}{2\turb{\eta}}
\left(\frac{u_s}{s}-\frac{\partial u_s}{\partial s}\right),
\]
which is useful to compare with Eqs~(\ref{p}) and (\ref{ppert}). This effect is
important if $u_s\ga2\turb{\eta}/h\linebreak[0]\simeq1\kms$ (cf.\
Sect.~\ref{RMS}).

Twisting of a horizontal primordial magnetic field by galactic
differential rotation leads to a tightly wound magnetic structure with
magnetic field direction alternating with radius at a progressively
smaller scale $\Delta s\sim s_0/|G|t$ with $\tan p\sim-(|G|t)^{-1}$,
where $s_0\sim10\kpc$ is the scale of variation in $\Omega$ (see \S3.3
in Moffatt, 1978; Kulsrud, 1999 for a detailed discussion).
The winding-up proceeds until a time
$t_0\sim5\times10^9\yr$ such that $|G|t_0\sim|C_\omega|^{1/2},$
where $C_\omega=Gs_0^2/\turb{\eta}=R_\omega s_0^2/h^2\sim10^3$--$10^4$.  At
later times, the alternating magnetic field rapidly decays because of
diffusion and reconnection.  The resulting maximum magnetic field
strength achieved at $t_0$ is given by
\begin{equation}
B_{\rm max}\sim B_0 |C_\omega|^{1/2}\;, \label{Bprim}
\end{equation}
where $B_0$ is the external magnetic field; the magnetic field
reverses at a small radial scale $\Delta s\sim
s_0|C_\omega|^{-1/2}\sim0.1\kpc$.  The magnetic pitch angle at $t_0$ is
of the order of $|p|\sim |C_\omega|^{-1/2}\la1^\circ$, i.e., much
smaller than the observed one. This picture cannot be reconciled with
observations (cf.\ Kulsrud, 1999). It can be argued that streaming
motions could make magnetic lines more open and parallel to the
galactic spiral arms. However, then magnetic field will reverse on a
small scale not only along radius, but also along azimuth. Such
magnetic structures are quite different from what is observed.  The
moderate magnetic pitch angles observed in spiral galaxies are a
direct indication that the regular magnetic field is not frozen into
the interstellar gas and has to be maintained by the dynamo (Beck,
2000).

\begin{figure}\label{MWay}
\includegraphics[width=0.88\textwidth,height=0.25\textheight]{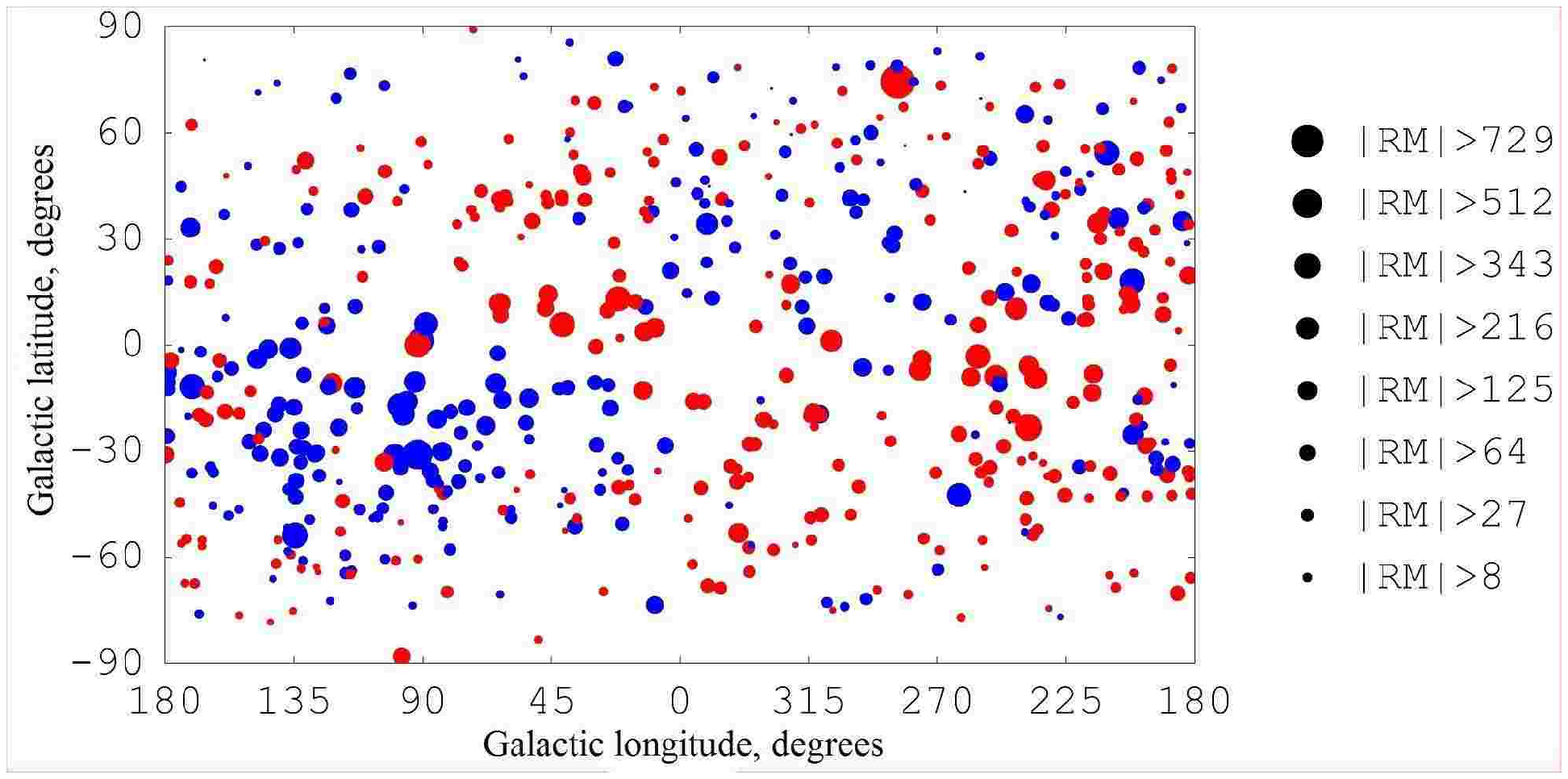}
\includegraphics[width=0.8\textwidth,height=0.25\textheight]{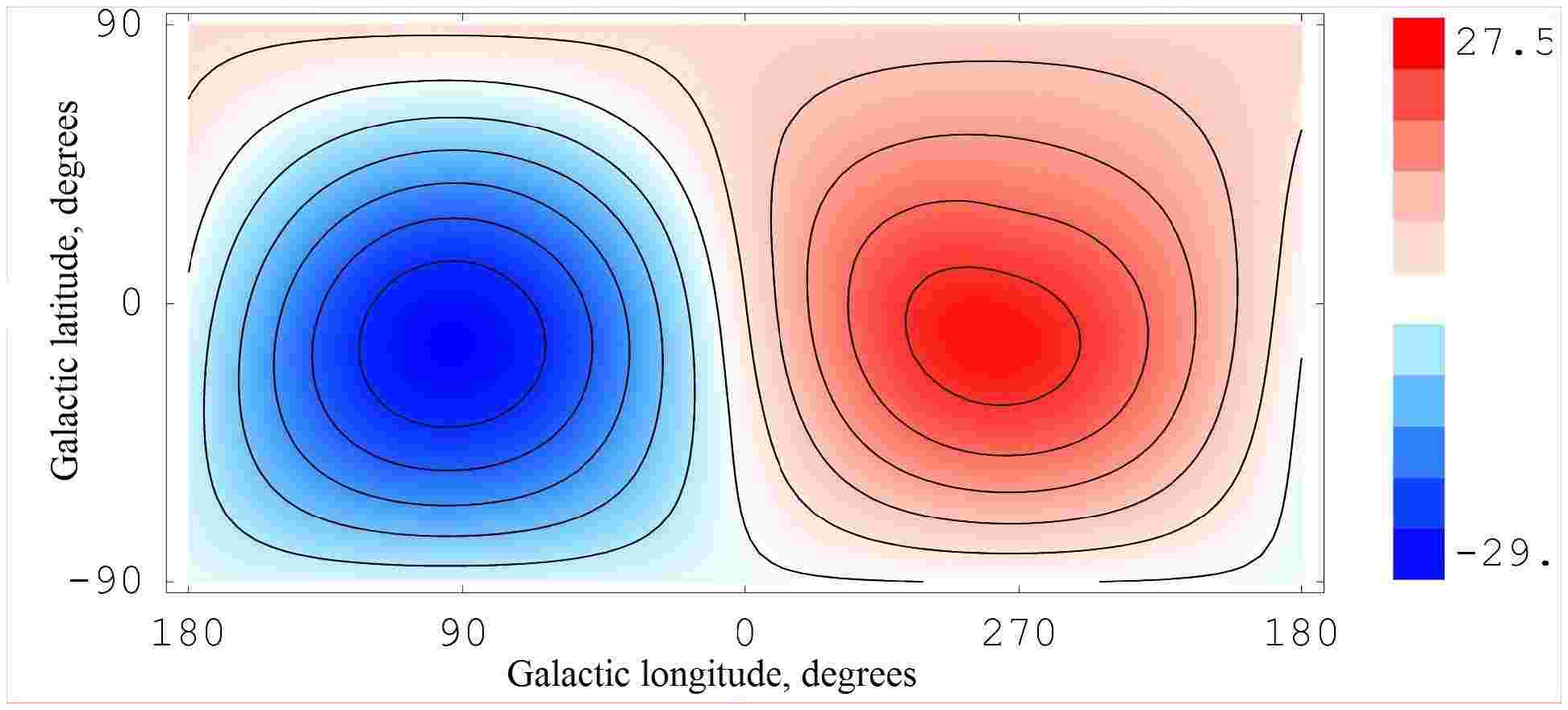}
\includegraphics[width=0.8\textwidth,height=0.25\textheight]{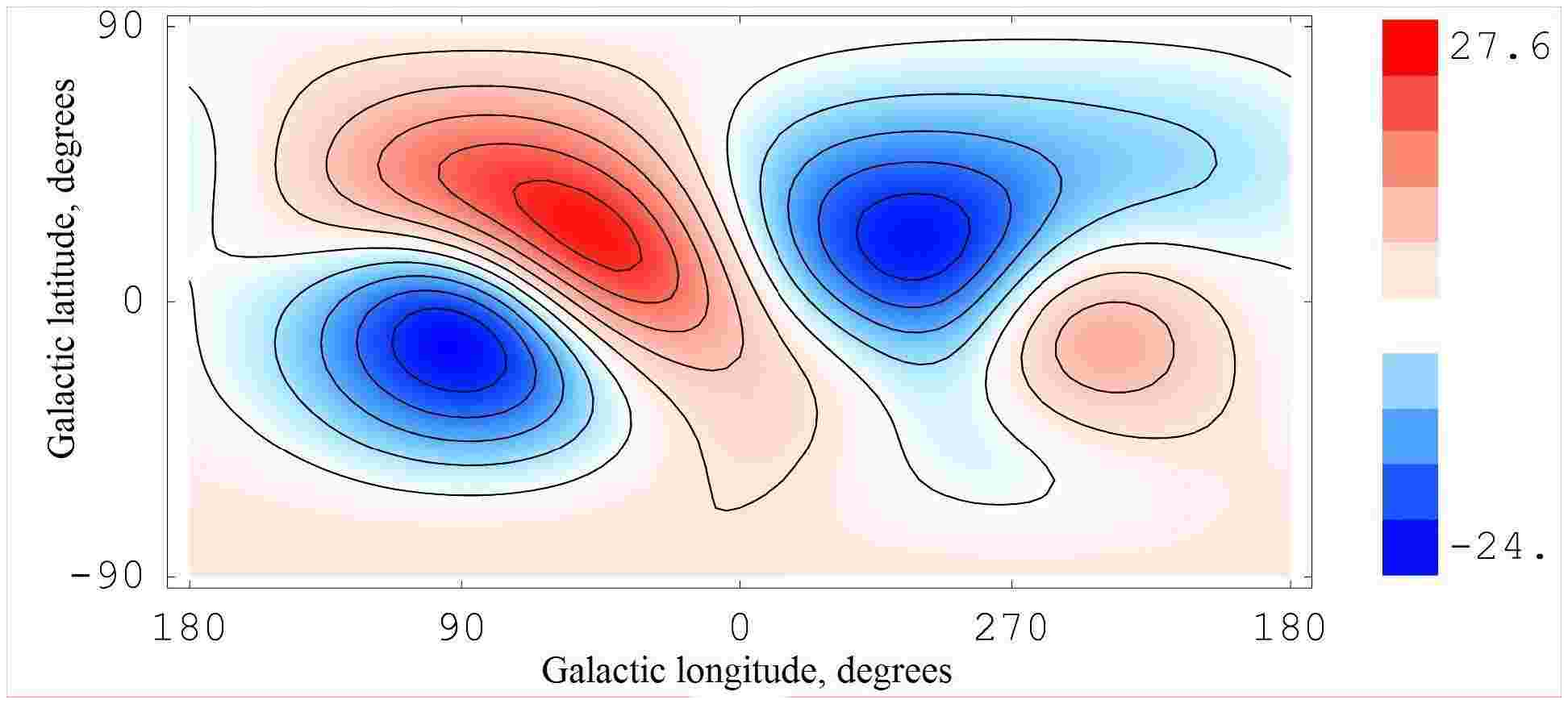}
\caption[]{
{\em Upper panel:\/} Faraday rotation measures of 551 extragalactic radio
sources from the catalogue of Simard-Normandin \& Kronberg (1980) shown in the
$(l,b)$-plane, where $(l,b)$ are the Galactic longitude and latitude in a
reference frame centred at the Sun with the Galactic center in the direction
$l=0$ and Galactic midplane at $b=0$. Positive (negative) $\RM$s are shown
with red (blue) circles whose radius indicates $|\RM|$ (rad\m$^{-2}$) as shown
to the right of the panel. The lower two panels show the wavelet transform of
these data at scales $76^\circ$ (middle panel) and $35^\circ$ (lower panel)
(Frick \etal, 2001). The transform at $76^\circ$ has been obtained with the
region of the Radio Loop~I removed (this radio feature is a nearby supernova
remnant). The wavelet transform at the scale $35^\circ$ is dominated by local
magneto-ionic features.}
\end{figure}

\subsection{The even (quadrupole) symmetry of magnetic field in the Milky Way}
                                                                \label{QS}
One of the most convincing arguments in favour of the galactic dynamo
theory comes from the symmetry of the observed regular magnetic field
with respect to the Galactic equator in the Milky Way. The direction of the
magnetic field is determined from Faraday rotation measures of the cosmic
sources of polarized emission, pulsars and extragalactic radio sources. Since
the Galactic magnetic field has a significant random component and
extragalactic radio sources can have their own (intrinsic) Faraday rotation,
any meaningful conclusions about the Galactic magnetic field must rely on
statistically significant samples of Faraday rotation measures. Even though
the quadrupole symmetry of the galactic magnetic fields has been widely
accepted as a firmly established fact since mid-1970's, its objective
observational verification has been obtained only recently. The main problem
here is that it is difficult to separate local (small-scale) and global
magnetic structures in the observed picture. However, wavelet analysis of the
Faraday rotation measures of extragalactic radio sources has definitely
confirmed that the horizontal components of the local regular magnetic field
have even parity being similarly directed on both sides of the midplane (Frick
\etal, 2001, see Fig.~\ref{MWay}).

The quadrupole symmetry is naturally explained by dynamo theory where
even parity is strongly favoured against odd parity because the even field has
twice larger scale in the vertical coordinate (see Sect.~\ref{local}).

Primordial magnetic field twisted by differential rotation can have
even vertical symmetry if it is parallel to the disc plane. However,
then the field is rapidly destroyed by twisting and reconnection as
described in Sect.~\ref{MPA}. If, otherwise, the primordial field is
parallel to the rotation axis and amplified by the vertical
rotational shear $\partial\Omega/\partial z$ (which, however, is insignificant
within galactic discs, $|z|\leq h$), it can avoid catastrophic decay
(\S3.11 in Moffatt, 1978), but then it will have odd parity in $z$,
which is ruled out by the observed parity of the Milky Way field.

The derivation of the regular magnetic field of the Milky Way from
Faraday rotation measures of pulsars and extragalactic radio sources,
$\RM$, is complicated by the contribution of local magnetic
perturbations, so it is difficult to decide which features of the RM
sky are due to the regular magnetic field and which are produced by
localized magneto-ionic perturbations (e.g., supernova remnants).
Therefore, the same observational data have lead different authors to
different conclusions (see Frick \etal, 2001, for a recent review).
Odd parity of the Galactic magnetic field has been suggested by
Andreassian (1980, 1982) and, for the inner Galaxy, by Han \etal\
(1997). Quantitative methods of analysis (as opposed to the `naked-eye' fitting
of more or less arbitrarily selected models) are especially appropriate in
this case.

Unfortunately, it is difficult to determine the parity of magnetic
field in external galaxies. In galaxies seen edge-on, the disc is
depolarized, whereas Faraday rotation in the halo is weak. Beck \etal\ (1994)
found weak evidence of even magnetic parity in the lower halo of NGC~253. The
arrangement of polarization planes in the halo of NGC~4631 (Beck, 2000) is
very suggestive of odd parity, but this does not exclude even parity in the
disc. In galaxies inclined to the line of sight, the amount of Faraday
rotation produced by an odd (antisymmetric) magnetic field differs from zero
because Faraday rotation and emission occur in the same volume; as a result,
emission originating at the far half of the galactic layer will have small or
zero net rotation, whereas emission from the near half will have significant
rotation produced by the unidirectional magnetic field in that half.
Therefore, Faraday rotation measures produced by even and odd magnetic
structures of the same strength only differ by a factor of two (Krause \etal,
1989a; Sokoloff \etal, 1998) and it is difficult to distinguish between the
two possibilities.

An interesting method to determine the parity of magnetic field in an
external galaxy has been suggested by Han \etal\ (1998). These
authors note that the contribution of the galaxy to the $\RM$ of a
background radio source will be equal to the intrinsic $\RM$ of the
galaxy if the magnetic field has even parity. For odd parity, the
galaxy will not contribute to the $\RM$ of a background source,
whereas any intrinsic $\RM$ will remain. The implementation of the
method requires either a statistically significant sample of background
sources or a single extended background source.

\subsection{The azimuthal structure}
Non-axisymmetric magnetic fields in a differentially rotating object
are subject to twisting and enhanced dissipation as described in
Sect.~\ref{MPA}.  The dynamo can compensate for the losses, but
axisymmetric magnetic fields are still easier to maintain (R\"adler,
1986). A few lowest non-axisymmetric modes with azimuthal wave
numbers
\begin{equation}
m\la\frac{s_0}{h}|R_\omega|^{-1/4}\approx2                \label{mmm}
\end{equation}
can be maintained in thin galactic disks where $h\ll s_0$ (\S{VII.8} in
Ruzmaikin \etal, 1988). The WKB solution of the galactic
$\alpha\omega$-dynamo equations by Starchenko \& Shukurov (1989) shows that
the bisymmetric mode ($m=1$) can grow provided
\[
\frac{\uu}{\ell\Omega}\left(\frac{h}{\ell}\right)^2
\left|\frac{d\ln\Omega}{d\ln s}\right|\la25\;,
\]
which seems to be the case in some galaxies. These results indicate that it is
natural to expect significant deviations from axial symmetry in magnetic
fields of many spiral galaxies. However, the {\em dominance\/} of
non-axisymmetric modes in most galaxies would be difficult to explain because
the axisymmetric mode has the largest growth rate under typical conditions.

Early interpretations of Faraday rotation in spiral galaxies were in striking
contrast with this picture, indicating strong dominance of bisymmetric
magnetic structures $(m=1)$, $\bfB\propto\mathrm{exp}{(i\phi)}$ with $\phi$ the
azimuthal angle (Sofue \etal, 1986), and this was considered to be a severe
difficulty of the dynamo theory and an evidence of the primordial origin of
galactic magnetic fields. It was suggested by Ruzmaikin \etal\ (1986) (see
also Sawa \& Fujimoto, 1986; Baryshnikova \etal, 1987) that the bisymmetric
magnetic structures can be interpreted as the $m=1$ dynamo mode. However,
despite effort, dynamo models could not explain the apparent widespread
dominance of bisymmetric magnetic structures. Paradoxically, what seemed to be
a difficulty of the dynamo theory has turned out to be its advantage as
observations with better sensitivity and resolution and better interpretations
have led to a dramatic revision of the observational picture. The present-day
understanding is that modestly distorted axisymmetric magnetic structures
occur in most galaxies, wherein the dominant axisymmetric mode is mixed with
weaker higher azimuthal modes (Beck \etal, 1996; Beck, 2000). Among nearby
galaxies, only M81 remains a candidate for a dominant bisymmetric magnetic
structure, but the data are old and this result needs to be reconsidered
(Krause \etal, 1989b); the interesting case of M51 is discussed below.
Deviations from precise axial symmetry can result from the spiral pattern,
asymmetry of the parent galaxy, etc.  Dominant bisymmetric magnetic fields can
be maintained by the dynamo action near the corotation radius due to a linear
resonance with the spiral pattern (Mestel \& Subramanian, 1991; Subramanian \&
Mestel, 1993; Moss, 1996) or nonlinear trapping of the field by the spiral
pattern (Bykov \etal, 1997).

Twisting of a horizontal magnetic field by differential rotation
generally produces a bisymmetric magnetic field, $m=1$. Twisting of a
horizontal primordial magnetic field can also produce an axisymmetric
configuration near the galactic centre if the initial state is asymmetric
(Sofue \etal, 1986; Nordlund \& R\"ognvaldsson, 2002), with a maximum of the
primordial field displaced from the disc's rotation axis where the gas density
is normally maximum. Thus, the maximum of the primordial field required by
this scenario has to occur at a different position than the maximum in the gas
density. This can only occur if the primordial field is not frozen into the
gas --- otherwise the field strength scales as a positive power of gas
density. The fact that magnetic fields in most spiral galaxies are nearly
axisymmetric within large radius (in fact, in the whole galaxy) would require
that this strong asymmetry in the initial state occurs systematically for all
the galaxies, which would be difficult to explain.

\begin{figure}
\centerline{
\includegraphics[width=0.549\textwidth]{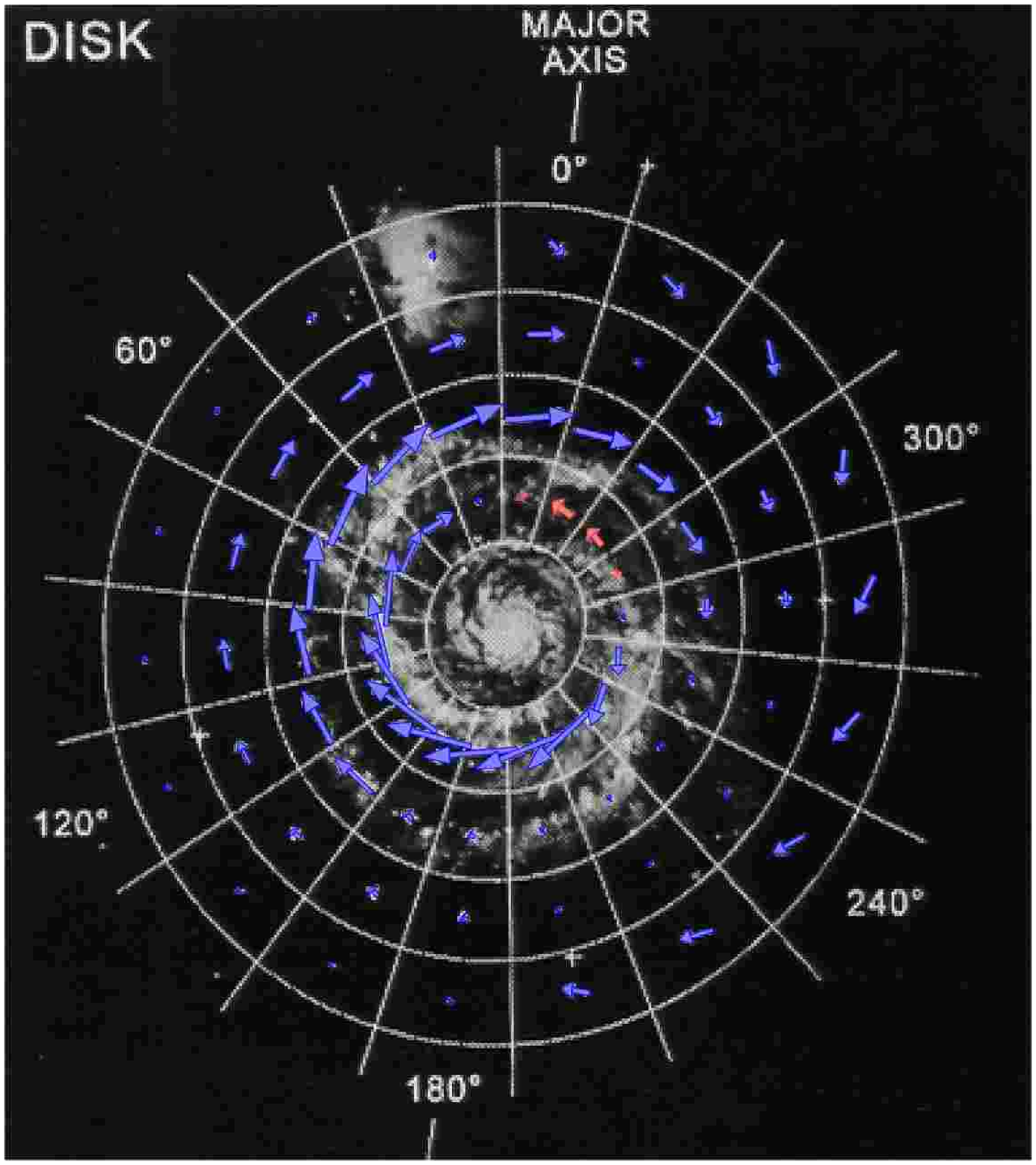}
 \raisebox{2.4cm}{\includegraphics[width=0.449\textwidth]{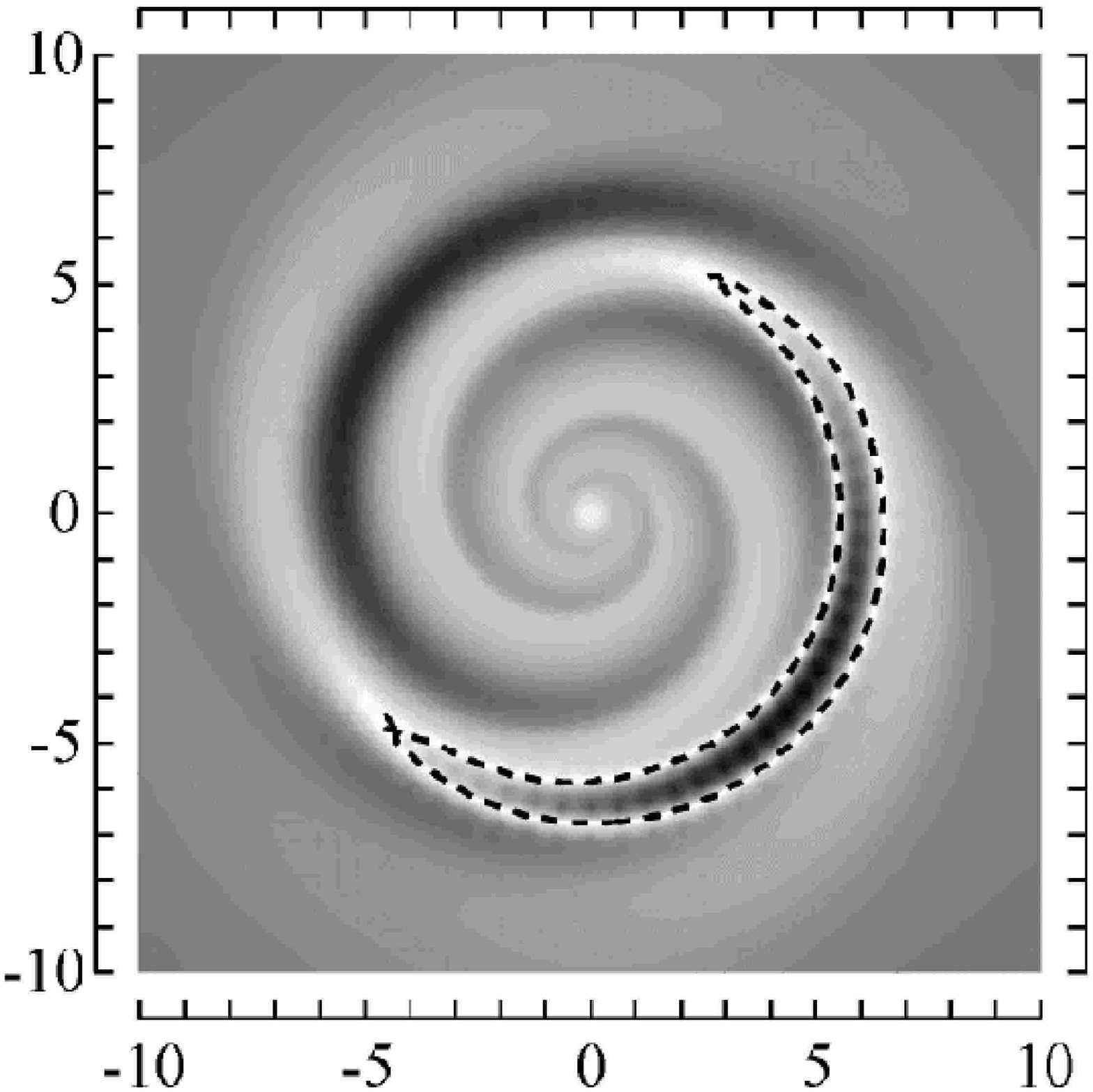}}
}
\vspace*{0.2cm}
\includegraphics[width=0.549\textwidth]{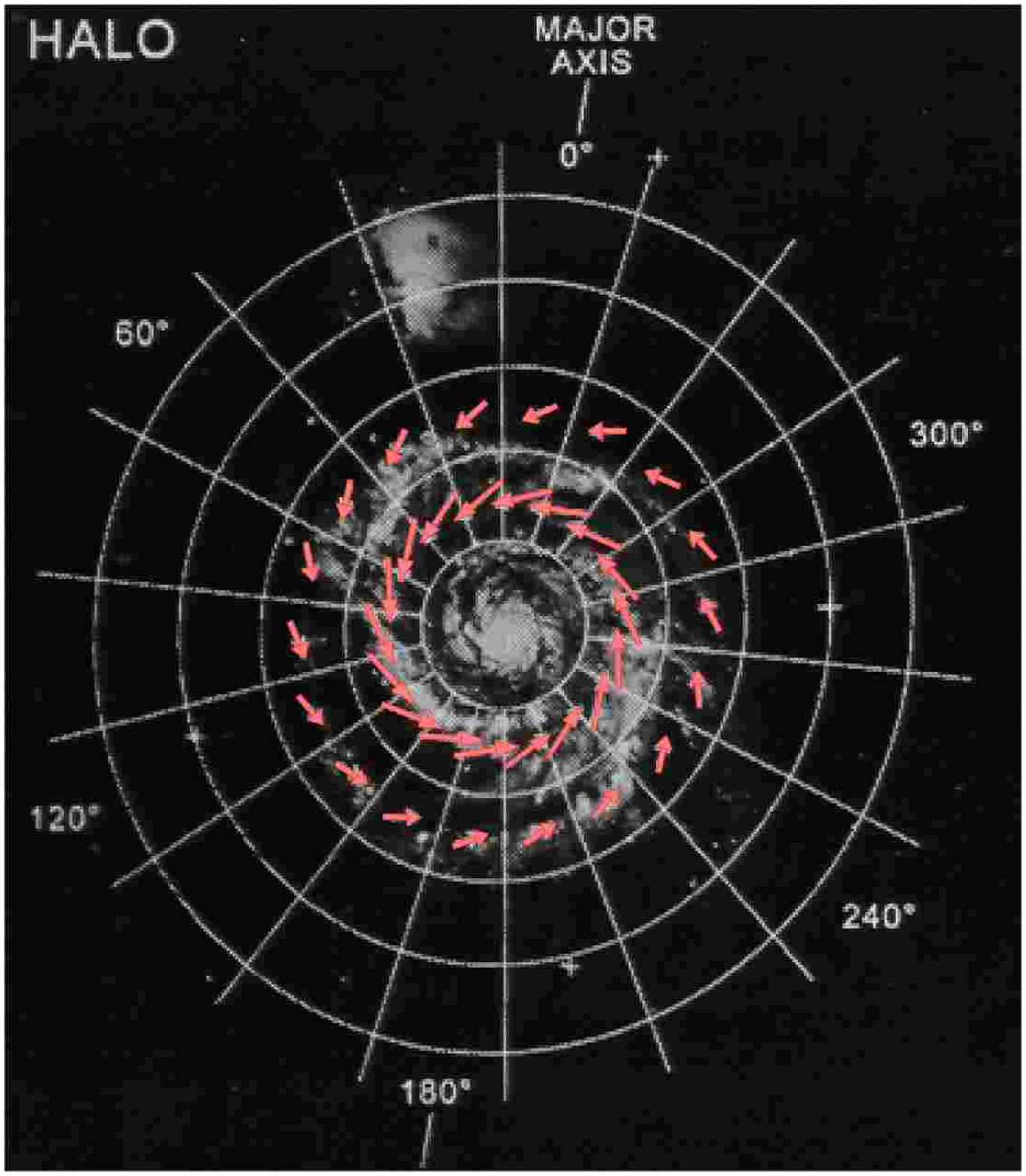}
\raisebox{5.5cm}[0.2\textwidth]{
\begin{minipage}{0.42\textwidth}
\caption[]{\label{M51}
{\em Left panels:\/} The global magnetic structure of the galaxy
M51, in the disc (upper left) and halo (bottom). Arrows show the direction and
strength of the regular magnetic field on a polar grid shown superimposed on
the optical image (Berkhuijsen \etal, 1997).  The grid radii are 3, 6, 9, 12
and 15\,kpc. {\em Upper right panel:\/} Magnetic field strength from the
dynamo model for the disc of M51 (Bykov \etal, 1997) is shown with
shades of grey (darker shade means stronger field). Magnetic field
is reversed within the zero-level contour shown dashed; scale is given in kpc.
The magnetic structure rotates rigidly together with the spiral pattern
visible in the shades of grey.
}
\end{minipage}
}
\end{figure}

\subsection{A composite magnetic structure in M51 and
magnetic reversals in the Milky Way}                        \label{cmsM51}
A striking example of a complicated magnetic structure that can
hardly be explained by any mechanism other than the dynamo has been revealed
in the galaxy M51 by Berkhuijsen \etal\ (1997). These authors used radio
polarization observations of the galaxy at wavelengths 2.8,
6.2, 18.0 and $20.5\cm$ (smoothed to a resolution of $3.5\kpc$). The disc of
this galaxy is not transparent to polarized radio emission at the two longer
wavelengths. Therefore, it was possible to determine the magnetic field
structure separately in two regions along the line the sight, which can be
identified with the disc and halo of the galaxy. As shown in Fig.~\ref{M51},
the regular magnetic fields in the disc is reversed in a region about 3 by
8\,kpc in size extended along azimuth at galactocentric radii $s=3$--6\,kpc
and azimuthal angles $300^\circ$--0 (shown with red arrows). A significant
deviation from axial symmetry in the disc has been detected out to $s=9\kpc$
(in the azimuth range 160--$260^\circ$), although it is too weak to result in
a magnetic field reversal. The field reversal occurs around the corotation
radius in M51, $s\approx6\kpc$ (i.e., the radius where the angular velocity of
the spiral pattern is equal to that of the gas).

A nonlinear dynamo model for M51 was developed by Bykov \etal\ (1997) who used
the rotation curve of M51, with the pitch angle of the spiral arms
$-15^\circ$ and corotation radius 6\,kpc. Figure~\ref{M51}
shows one of their solutions where a region with reversed magnetic
field persists in the disc near the corotation radius of the spiral
pattern. Near the corotation, a non-axisymmetric (bisymmetric) magnetic field
can be trapped by the spiral pattern and maintained over the galactic
lifetime. The effect is favoured by a smaller pitch angle of the spiral arms,
thinner gaseous disc, weaker rotational shear and stronger spiral pattern.
This nonaxisymmetric structure is arguably similar to the structure observed
in M51.

The regular magnetic field in the halo of M51 has a
structure very different from that in the disc --- the halo field is
nearly axisymmetric and even directed oppositely to that in the disc in most
of the galaxy. An external magnetic field should have a rather peculiar form
to be twisted into such a configuration!

Distinct azimuthal magnetic structures in the disc and the halo can
be readily explained by dynamo theory as non-axisymmetric magnetic
fields can be maintained only in the thin disc but not in the
quasi-spherical halo where $h\simeq s_0$ and $|R_\omega|\gg1$ in
Eq.~(\ref{mmm}).  Moreover, dynamo action in the disc and the halo can
proceed almost independently of each other producing distinctly
directed magnetic fields (Sokoloff \& Shukurov, 1990).

Another case of a regular magnetic field with unusual
structure is the Milky Way where magnetic field reversals
are observed along the galactocentric radius in the inner Galaxy between the
Orion and Sagittarius arms at $s\approx7.9\kpc$
and, possibly, in the outer Galaxy between the Orion and Perseus arms at
$s\approx10.5\kpc$ (\S3.8.2 in Beck \etal, 1996, and Frick \etal, 2001);
see, however, Brown \& Taylor 2001). The reversals were first
interpreted as an indication of a global bisymmetric magnetic structure (Sofue
\& Fujimoto, 1983), but it has been shown that dynamo-generated axisymmetric
magnetic field can have reversals at the appropriate scale (Ruzmaikin \etal,
1985; Poezd \etal, 1993). Both interpretations presume that the reversals are
of a global nature, i.e., they extend over the whole Galaxy to all azimuthal
angles (or radii in the case of the bisymmetric structure). This leads to a
question why reversals at this radial scale are not observed in any other
galaxy (Beck, 2000). Poezd \etal\ (1993) argue that the lifetime of the
reversals is sensitive to subtle features of the rotation curve and the
geometry of the ionized gas layer (see also Belyanin \etal, 1994) and
demonstrate that they are more probable to survive in the Milky Way than in,
e.g., M31.

However, the observational evidence of the reversals is restricted to
a relatively small neighbourhood of the Sun, of at most 3--5\,kpc
along azimuth. It is therefore quite possible that the reversals
are local and arise from a magnetic structure similar to that in the
disc of M51 as shown in Fig.~\ref{M51}. The reversed field in the Solar
neighbourhood has the same radial extent of 2--3\,kpc as in M51 and also
occurs near the corotation radius. This possibility has not yet been
explored; its observational verification would require careful
analysis of pulsar Faraday rotation measures.

\subsection{The radial magnetic structure in M31}\label{RMS}
An important clue to the origin of galactic magnetic field is provided by the
magnetic ring in M31 (Beck, 1982), predicted by dynamo theory (Ruzmaikin \&
Shukurov, 1981). Both the large-scale magnetic field and the gas density in
this galaxy have a maximum in the same annulus $8\la s\la12\kpc$, with the
apparent enhancement in the magnetic field strength by about 30\% (Fletcher
\etal, 2004). The kinematic dynamo model of Ruzmaikin \& Shukurov (1981) was
based on the double-peaked rotation curve of shown in Fig.~\ref{M31pitch},
where rotational shear is strongly reduced at $s=2$--6\,kpc. As a result,
$R_\omega$ is small and even positive in this radial range, so
$|D|<|D_\mathrm{cr}|$ and the dynamo cannot maintain any regular magnetic
field at  $s=2$--6\,kpc.

An attractive aspect of this theory is that both magnetic and gas
rings are attributed to the same feature of the rotation curve.
Angular momentum transport by viscous stress leads to matter inflow
at a rate
$\dot M=2\pi\Sigma\turb{\nu}\partial{\ln\Omega}/\partial{\ln s}
\sim0.1\,M_\odot\yr^{-1}$, where $\turb{\nu}\sim\turb{\eta}$ is the
turbulent viscosity, resulting in the radial inflow at a speed $u_s=\dot
M/2\pi s\Sigma$ with $\Sigma$ the gas surface density.  In the nearly-rigidly
rotating parts, $u_s$ is reduced and matter piles up outside this region
producing gas ring. Gravitational torques from spiral arms can further enhance
the inflow (see Moss \etal, 2000, for a discussion), so the total radial
velocity is expected to be $u_s\sim1\kms$ at $s=10\kpc$.

The double-peaked rotation curve of M31 is consistent with the
existence of  both magnetic and gas rings. The situation is different
with the more recent rotation curve of Braun (1991) which does
not have a double-peaked shape (Fig.~\ref{M31pitch}). The difference between
the two rotation curves arises mainly from the fact that Braun allows for
significant displacements of spiral arm segments from the galactic
midplane: this results in a revision of the segments' galactocentric
distances for regions away from the major axis. We note that the CO
velocity field at the major axis (Loinard \etal, 1995) is compatible
with a double-peaked rotation curve.


With Braun's rotation curve, the magnetic field can concentrate into a ring
mainly because the gas is in the ring and $B\propto\rho^{1/2}$ as shown in
Eq.~(\ref{B}).  The dynamo model of Moss \etal\ (1998) based on the rotation
curve of Braun (1991) has difficulties in reproducing a magnetic ring as well
pronounced as implied by the observed amount of Faraday rotation. This has
lead to an idea that magnetic field can be significant at $s=2$--$6\kpc$ in
M31. This has prompted Han \etal\ (1998) to search for magnetic fields at
$s=2$--6\,kpc that could have escaped detection because of reduced density of
cosmic ray electrons at those radii. These authors have found that two out of
three background polarized radio sources seen through that region of M31 have
Faraday rotation measures compatible with the results of Moss \etal\ (1998).
They further conclude that this indicates an even symmetry of the regular
magnetic field. This is encouraging, but a statistically representative sample
of background sources has to be used to reach definite conclusions because of
their unknown intrinsic $\RM$.

With a double-peaked rotation curve, a primordial magnetic field with a
uniform radial component could have been twisted to produce a
magnetic ring by virtue of Eq.~(\ref{Bprim}). In this case the
primordial and dynamo theories have similar problems and
possibilities regarding the magnetic ring in M31.

Lou \& Fan (2000) attribute the magnetic ring in M31 to an
axisymmetric mode of MHD density waves. Because of the axial symmetry
of the wave, the magnetic field in the ring must be purely azimuthal,
$p=0$, in contrast to the observed structure with a significant
pitch angle (Fig.~\ref{M31pitch}). Furthermore, the ring can hardly
represent a wave packet as envisaged in this theory because then its
group velocity must be comparable to the Alfv\'en speed of $30\kms$ (Lou
\& Fan, 1998) and so the ring should be travelling at this speed
along radius to traverse 30\,kpc in $10^9\yr$, a distance much larger than the
ring radius. The implication would be that the ring is a transient with a
short lifetime of order $3\times10^8\yr$. And, of course, the theory cannot
explain the origin of an azimuthal magnetic field required to launch the wave
packet.

\subsection{Strength of the regular magnetic field}\label{SRMF}
Interstellar regular magnetic fields are observed to be close to the energy
equipartition with interstellar turbulence. This directly indicates
that the regular magnetic field is coupled to the turbulent gas
motions. (Note that $\ell\Omega$ does not differ much from the turbulent
velocity $\uu_0$ in Eq.~(\ref{B}).)  To appreciate the importance of this
conclusion, consider primordial magnetic field twisted by
differential rotation. Its maximum strength given by
Eq.~(\ref{Bprim}) as $B_{\rm max}\sim10^2B_0$ is controlled by the
strength of the primordial field $B_0$, and so this theory, if
applicable, would result in stringent constraints on extragalactic
magnetic fields.

\begin{figure}
\begin{center}
\includegraphics[width=0.49\textwidth]{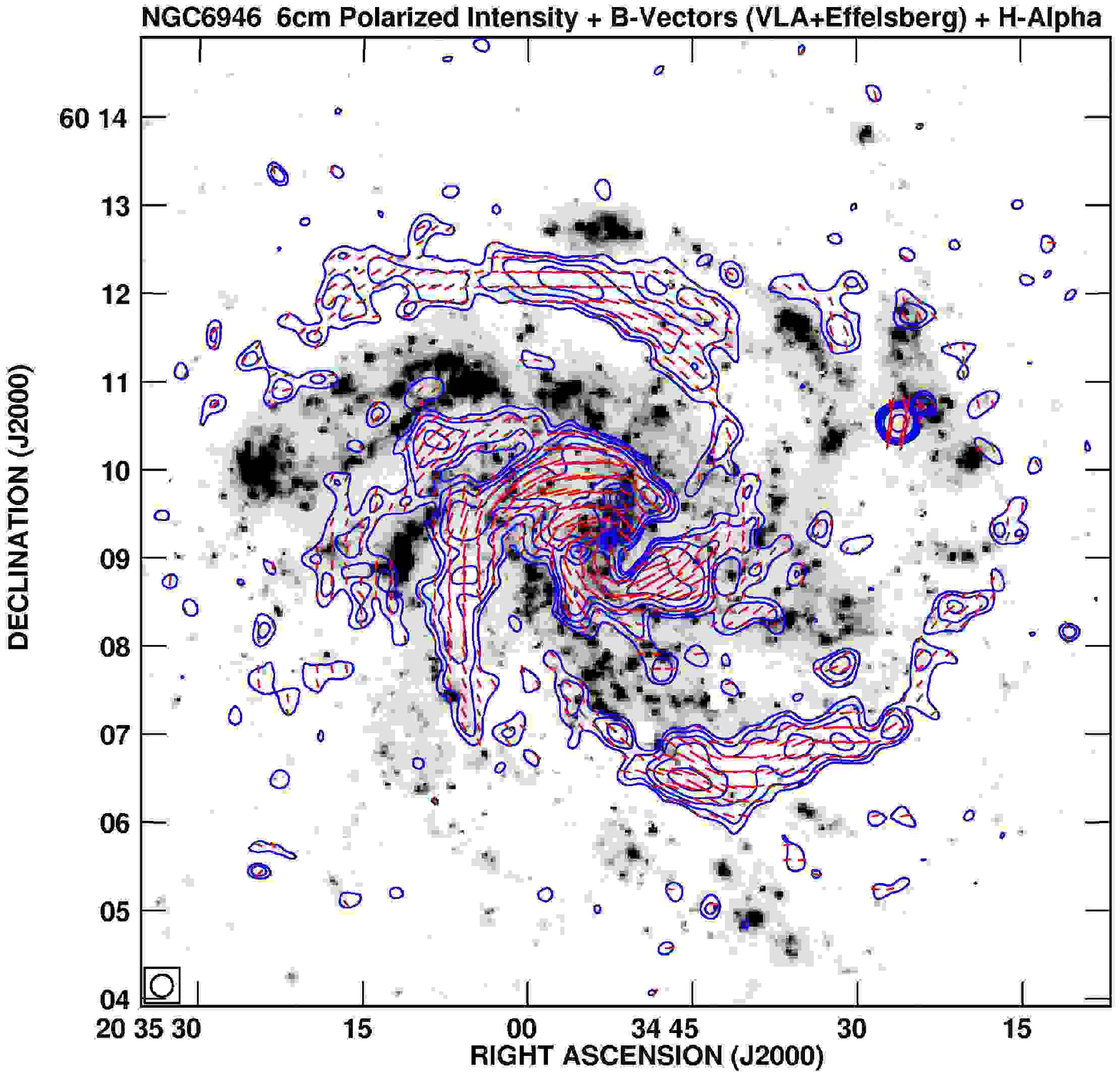}
\includegraphics[width=0.49\textwidth]{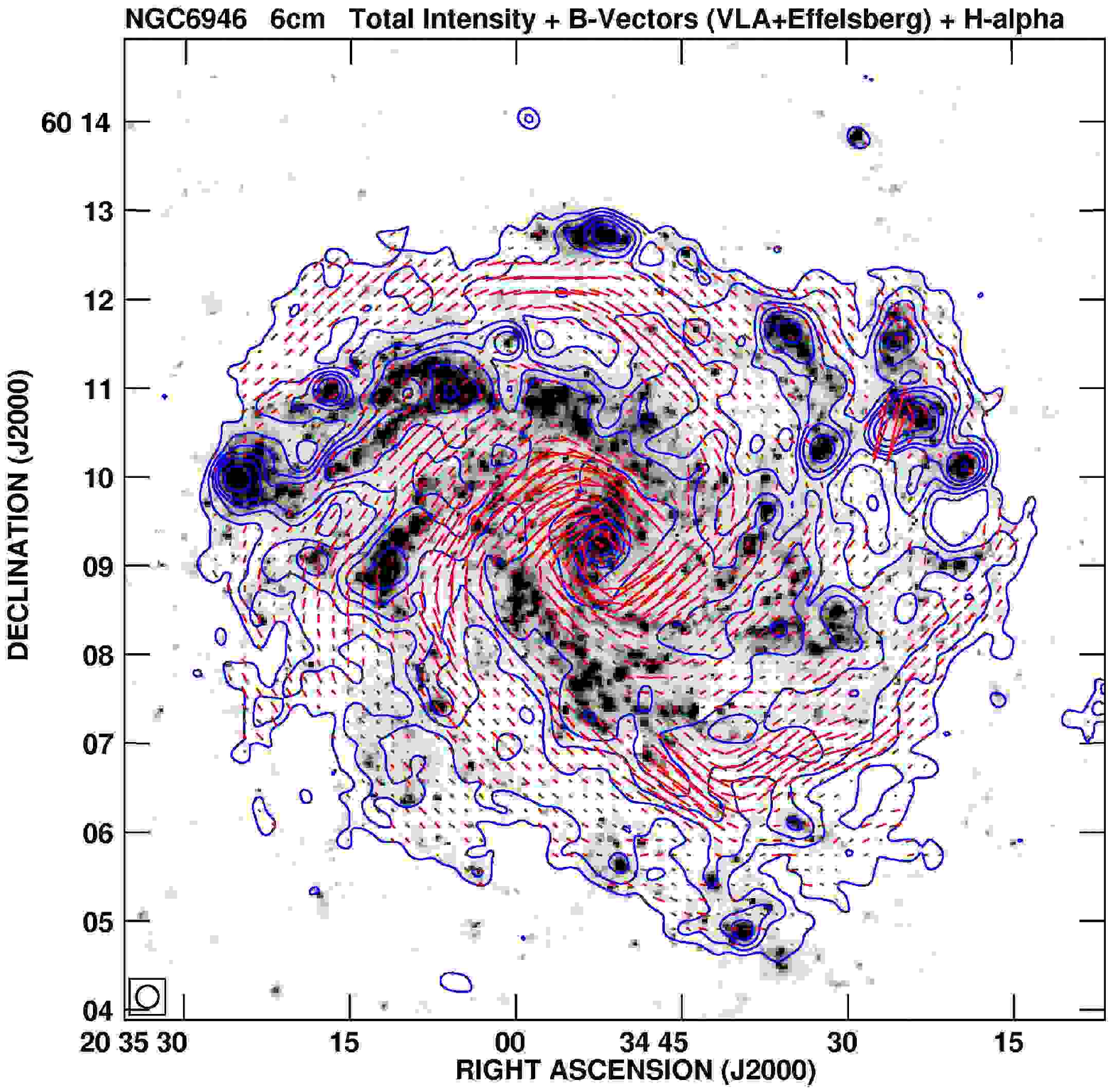}
\end{center}
\caption[]{\label{n6946}{\em Left panel:\/} Magnetic arms in the galaxy
NGC~6946: polarized intensity at the wavelength $\lambda=6\cm$ (blue contours),
a tracer of the large-scale magnetic field $\mean{B}$, superimposed on the
galactic image in the H$\alpha$ spectral line of ionized hydrogen (grey scale).
Red dashes indicate the orientation of the $E$-vector of
the polarized emission (parallel to the direction of intrinsic magnetic field
if Faraday rotation is negligible), with length proportional to the fractional
polarization --- see Eq.~(\protect\ref{polar}). The spiral arms visualized by
H$\alpha$ are the sites where gas density is maximum. The large-scale magnetic
field is evidently stronger between the arms where gas density is lower.
{\em Right panel:\/} As in the left panel, but now for the total synchrotron
intensity, a tracer of the total magnetic field $B^2=\mean{B}^2+\mean{b^2}$.
The total field is enhanced in the gaseous arms. Given that the large-scale
field concentrates between the arms, this means that the random field is
significantly stronger in the arms, a distribution very different from that of
the large-scale field. The size of the beam in the radio maps is shown in the
bottom left of each frame. (Images courtesy of R.~Beck, MPIfR, Bonn.)}
\end{figure}

A striking evidence of the nontrivial behaviour of the large-scale galactic
magnetic field is the so-called magnetic arms, first discovered by Beck \&
Hoernes (1996) in the nearby galaxy NGC~6946. We show in Fig.~\ref{n6946} a
map of polarized radio emission from this galaxy (a tracer of the large-scale
magnetic field strength) superimposed on the map in the H$\alpha$ spectral line
(a tracer of ionized gas). It is evident that magnetic field is stronger {\em
between\/} the spiral arms of this galaxy, i.e., where the gas density (both
total and ionized) is lower. This behaviour is just opposite to what is
expected of a frozen-in magnetic field that scales with a power of gas
density. The phenomenon of magnetic arms confirms in a spectacular manner that
the large-scale magnetic field is not frozen into the interstellar gas, and
therefore cannot be primordial. Shukurov (1998), Moss (1998) and Rohde \etal\
(1999) suggest an explanation of the magnetic arms in terms of the mean-field
dynamo theory. In brief, they argue that dynamo number can be larger between
the gaseous spiral arms, resulting in stronger dynamo action.

The theory of MHD density waves relates magnetic field excess $\Delta\mean{B}$
in spiral arms to the enhancement in stellar density,
$\Delta\mean{B}/\langle\mean{B}\rangle=\Delta\Sigma/\langle\Sigma\rangle$ (Lou
\& Fan, 1998), where $\Sigma$ is the stellar surface density, $\Delta\Sigma$
is its excess in the spiral arms, and angular brackets denote azimuthal
averaging. Arm intensities in magnetic field and stellar surface density in
NGC~6946 have been estimated by Frick \etal\ (2000) who applied wavelet
transform techniques to radio polarization maps and to the galaxy image in
broadband red light, a tracer of stellar mass density. Their results indicate
that the mean relative intensity of magnetic spiral arms remains rather
constant with galactocentric radius at a level of 0.3--0.6. On the contrary,
the relative strength of the stellar arms systematically grows with radius
from very small values in the inner galaxy to 0.3--0.7 at $s=5$--6\,kpc, and
then decreases to remain at a level of 0.1--0.3 out to $s=12\kpc$. The
distinct magnitudes and radial trends in the strengths of magnetic and stellar
arms in NGC~6946 do not seem to support the idea that the magnetic arms are
due to MHD density waves.

\section{Elliptical galaxies}\label{Elliptic}
Elliptical galaxies do not rotate fast enough, so they are ellipsoidal systems
without prominent disc components. The stellar population of elliptical
galaxies is old and the interstellar gas is dilute (Mathews \& Brighenti,
2003). Therefore, both relativistic and thermal electrons have low density,
and any synchrotron emission and Faraday rotation can only be weak.
Nevertheless, there are several lines of evidence, albeit mostly indirect,
suggesting significant magnetic fields in ellipticals (Moss \& Shukurov, 1996;
Mathews \& Brighenti, 1997). The magnetic field should be random, producing
unpolarized synchrotron emission and fluctuating Faraday rotation.  The
r.m.s.\ Faraday rotation measure attributable to the ISM of the ellipticals is
$\langle\mbox{RM}\rangle=5$--100\,rad\,m$^{-2}$.

\subsection{Turbulent interstellar gas in elliptical galaxies}
Interstellar gas in elliptical galaxies is observed via its X-ray
emission. Type~I supernovae
(SNe) (and also stellar winds and random motions of stars) heat the
gas to the observed temperatures $T\simeq10^7\K$.  It is natural to expect
that a fraction $\delta$ of the energy is converted into
turbulent motions of the gas. The turbulent scale $\ell\approx400\,$pc is
given by the diameter of a SN as it reaches pressure balance with the ambient
medium whose typical density is $n\approx10^{-3}\cmcube$.  The balance
between energy injection and dissipation rates yields a turbulent velocity of
$\uu_0\simeq20\kms$ for $\delta=0.1$, assuming the energy
dissipation time $\tau\simeq \ell/\uu_0$ as for the Kolmogorov turbulence. This
estimate of $\uu_0$ is compatible with the constraint
$\uu_0\la50\kms$ resulting from the observed X-ray luminosity.
Another driver of turbulence is the random motions of stars. These generate
random vortical motions at a smaller scale and velocity, $\ell_*\simeq3\p$
and $\uu_*\simeq3\kms$, respectively (Moss \& Shukurov, 1996).

The driving force produced by an expanding quasi-spherical SN remnant is
potential. The above estimates assume that the motions driven by the SNe are
vortical, so $\tau=\ell/\uu_0$ applies. In spiral galaxies, the potential
(acoustic) motions are efficiently converted into vortical turbulence mainly
due to the inhomogeneity of the ISM. The ISM in elliptical galaxies is hot
and, presumably, rather homogeneous at kpc scales. Therefore, Moss \& Shukurov
(1996) suggested that SNe will drive sound-wave turbulence whose correlation
time $\tau$ is $(\uu_0/\cs)^{-2}\ell/\cs\sim3\times10^7\yr$
rather than $\ell/\uu_0$, where $\cs\approx300\kms$ is the
speed of sound. However, Mathews \& Brighenti (1997) noted that sound waves
quickly dissipate, and so cannot form a pervasive turbulent velocity field.
The nature of turbulence in elliptical galaxies needs to be studied further.

\subsection{The fluctuation dynamo in elliptical galaxies}
As in most astrophysical objects, $\Rem$ in elliptical galaxies by far
exceeds 100, so fluctuation dynamo action in ellipticals is quite
plausible (see Sect.~\ref{FD}). The e-folding time of the random field in a
vortical random flow is of the order of the eddy turnover time,
$\tau=2\times10^7\yr$. The magnetic field is concentrated into flux ropes
whose length and thickness are of order $\ell\simeq400\,$pc and
$\ell\Remcr^{-1/2}\simeq40\p$. In the ropes, magnetic
field is plausibly in equipartition with the turbulent kinetic energy,
$b\simeq0.3\mkG$.

Moss \& Shukurov (1996) discuss a two-stage dynamo action by smaller
scale vortical turbulence driven by random motions of the stars and,
at larger scales, by the acoustic turbulence. However, the very
existence of the acoustic turbulence in elliptical galaxies is
questionable (Mathews \& Brighenti, 1997).

Faraday rotation measure produced within a single turbulent cell with
the above parameters is $\RM_0\simeq0.81b\nel\ell$, so the net Faraday
rotation from an ensemble of turbulent cells, observed at a
resolution $D$ such that $D\gg\ell$, is given by
$\langle\RM\rangle\sim\RM_0\sqrt{{N}/{N_{D}}} \sim
\RM_0{\sqrt{L\ell}}/{D}\,,$ where $N=L/\ell$ is the number of cells along
the path length $L$ and $N_{D}=(D/\ell)^2$ is the number of cells in the
resolution element. Thus, $\RM_0\simeq0.1\radm$ and
$\langle\RM\rangle\sim1\radm$. Faraday rotation can be stronger in the central
regions where $\langle\RM\rangle\sim5\radm$ at a distance 8\,kpc from the
galactic centre. These estimates agree fairly with the available observations.

Magnetic field generation in elliptical galaxies was discussed by Lesch \&
Bender (1990), but they considered a mean-field dynamo that
needs overall rotation which is not present in elliptical galaxies. The
fluctuation dynamo in elliptical (radio) galaxies was simulated by De~Young
(1980), but these simulations apparently had $\Rem<\Remcr$ as they resemble
transient amplification of magnetic field by velocity shear rather than
genuine dynamo action.

\section{Conclusions}
The observational picture of galactic magnetic fields is compatible with the
mean-field dynamo theory in its simplest, quasi-kinematic form. It is
important to note that there is not much freedom in varying parameters of
galactic dynamos as observations constrain them fairly tightly. Therefore, this
agreement is not a result of a free-hand parameter adjustments. Moreover,
galactic dynamo theory has demonstrated its predictive power. For example, it
has been clear to dynamo theorists since the early 1970's that the partially
ionized Galactic disc has the scale height 0.4--0.5\kpc, i.e., significantly
larger than that of the neutral hydrogen layer (see \S{VI.2} in Ruzmaikin
\etal, 1988, for a review), but the existence of this component of the
interstellar medium was accepted by a broader astrophysical community only 15
years later (Lockman, 1984).

The agreement of the mean-field dynamo theory with observations discussed in
Sect.~\ref{OEDASG} cannot be considered as a proof of its correctness ---
history of physics is familiar with concepts, such as the ether, that have
proved to be irrelevant despite their perfect agreement with numerous
experimental facts, before a single experiment refuted them. Nevertheless, the
spectacular success of the dynamo theory when applied to galaxies warrants its
careful treatment when compared with other theories. In particular, any rival
theory has to be able to explain {\em at least\/} the same set of
observational data as the dynamo theory. In this sense, there are no fair
rivals to the dynamo theory.

The main current difficulty of the galactic dynamo theory is our lack
of understanding of its nonlinear form. It is important to avoid an unjustified
extension to real galaxies of results obtained for highly idealized
systems. In particular, the disc of a spiral galaxy is {\em not\/} a closed
system. The significance of the disc-halo connection for the mean-field
galactic dynamos, touched upon in Sect.~\ref{hel}, should be carefully
investigated.

Another outstanding problem in theory and observations of galactic magnetic
fields is the effect of the multi-phase structure of the interstellar medium
on the magnetic field. The effect of magnetic fields on the multi-phase
structure is also expected to be very significant, and also poorly understood.
The strength of the magnetic field in the hot phase is not known. The detailed
nature of the balance between cosmic rays and magnetic field has to be
clarified: it is unclear whether this balance is maintained pointwise (at
each location) or only on average (e.g., at scales exceeding the diffusion
scale of cosmic ray particles). The answer to this question is essential for
the interpretation of the synchrotron emission from spiral galaxies. Progress
in this direction will eventually make theory of galactic magnetic fields an
integral part of galactic astrophysics.


\input{Ch7_refs}
\end{document}

%% file: dyn_book.tex
\newcommand{\turb}[1]{#1_{\rm T}}

\def\AAp{\em Astron.\ Astrophys.}
\def\AJ{\em Astron.~J.}
\def\ApJ{\em Astrophys.~J.}
\def\ApJL{\em Astrophys.~J., Lett.}
\def\ApJSS{\em Astrophys.~J., Suppl.\ Ser.}
\def\ARFM{\em Annu.\ Rev.\ Fluid Mech.}

\def\Arep{\em Astron.\ Rep.}
\def\GAFD{\em Geophys. Astrophys. Fluid. Dyn.}
\def\MNRAS{\em Mon.\ Not.\ R.~Astron.\ Soc.}
\def\Nat{\em Nature\/}
\def\PASP{\em Publ.\ Astron.\ Soc.\ Pac.}
\def\VA{\em Vistas Astron.}
\def\IAUC{\em IAU Circ.}

\def\SA{\em Sov.\ Astron.}
\def\NC{\em Nuovo Cimento\/}
\def\NPB{{\em Nucl.\ Phys.} B}
\def\PLB{{\em Phys.\ Lett.}  B}
\def\PRL{\em Phys.\ Rev.\ Lett.}
\def\PRD{{\em Phys.\ Rev.} D}

\def\ZPC{{\em Z.\ Phys.} C}

\newcommand{\etal}{{\it et al.}}

\newcommand{\Rem}{\mbox{Rm}}

\newcommand{\Ros}{\mbox{Ro}}

\newcommand{\bfa}{{\bf a}}
\newcommand{\bfb}{{\bf b}}

\newcommand{\bfj}{{\bf j}}

\newcommand{\bfA}{{\bf A}}
\newcommand{\bfB}{{\bf B}}

\newcommand{\bfJ}{{\bf J}}

\def\exp{{\rm e}}

\def\bfnabla{\mbox{\boldmath $\nabla$}}

\def\mean{\overline}

\newcommand{\Grad}{\bfnabla }
\newcommand{\Curl}{\bfnabla \times}

\def\eps{\mbox{$\varepsilon$}}
\def\epsilon{\varepsilon}

\newcommand{\bt}{\begin{tabular}{@{}p{2cm}p{14.5cm}}}

\def\references{\section*{References}
    \bgroup\parindent=1cm\parskip=1mm\small
    \def\refpar{\par\hangindent=1em\hangafter=1}\reference\ \vskip-10mm}
\def\endreferences{\refpar\egroup}

\def\reference{\relax\refpar \ \hskip-1.1cm} 


%% file: Ch7_gal.bbl
\begin{references}

\def\Journal#1#2#3#4{ {\it #1}, {\bf #2}, #3 (#4)}

\def\AAp{\it Astron.\ Astrophys.}
\def\AJ{\it Astron.~J.}
\def\AN{\it Astron.\ Nachr.}
\def\ApJ{\it Astrophys.~J.}
\def\ApJL{\it Astrophys.~J.\ Lett.}
\def\ApJSS{\it Astrophys.~J.\ Suppl.\ Ser.}
\def\ApSS{\it Astrophys.\ Space Sci.}
\def\ARFM{\it Ann.\ Rev.\ Fluid Mech.}
\def\ARAA{\it Ann.\ Rev.\ Astron.\ Astrophys.}
\def\Arep{\it Astron.\ Rep.}
\def\AZh{\it Astron.\ Zh.}
\def\GAFD{\it Geophys.\ Astrophys.\ Fluid Dyn.}
\def\IAUC{\it IAU Circ.}
\def\JETP{\it Zh.\ Teor.\ Eksper. Fiz.}
\def\MNRAS{\it Mon.\ Not.\ R.\ Astron.\ Soc.}
\def\Nat{\it Nature\/}
\def\PASP{\it Publ.\ Astron.\ Soc.\ Pac.}
\def\NC{\it Nuovo Cimento\/}
\def\NPB{{\it Nucl.\ Phys.} B}
\def\PASJ{\it Publ.\ Astron.\ Soc.\ Japan\/}
\def\PLB{{\it Phys.\ Lett.}  B}
\def\PRD{{\it Phys.\ Rev.} D}
\def\PRep{\it Phys.\ Rep.}
\def\PRL{\it Phys.\ Rev.\ Lett.}
\def\RMP{\it Rev.\ Mod.\ Phys.}
\def\SA{\it Sov.\ Astron.}
\def\SovJETP{\it Sov.\ Phys.\ JETP}
\def\VA{\it Vistas Astron.}
\def\ZPC{{\it Z.\ Phys.} C}


\reference
Andreassian R.~R., On the structure of the Galactic magnetic field.
\Journal{Astrofizika}{16}{707--713}{1980}.

\reference
Andreassian R.~R., A study of the magnetic field of the Galaxy.
\Journal{Astrofizika}{18}{255--262}{1982}.

\reference
Armstrong J.~W., Rickett B.~J.\ \& Spangler S.~R., Electron density power
spectrum in the local interstellar medium.
\Journal{\ApJ}{443}{209--221}{1995}.

\reference
Baryshnikova Yu., Ruzmaikin A., Sokoloff D.\ \& Shukurov A.,
Generation of large-scale magnetic fields in spiral galaxies.
\Journal{\AAp}{177}{27--41}{1987}.

\reference
Beck R., The magnetic field in M31. \Journal{\AAp}{106}{121--132}{1982}.

\reference
Beck R., Magnetic fields and interstellar gas clouds in the spiral galaxy
NGC~6946. \Journal{\AAp}{251}{15--26}{1991}. Erratum:
\Journal{\AAp}{258}{605}{1992}.

\reference
Beck R., Magnetic fields in normal galaxies. \Journal{Phil.\ Trans.\
R.\ Soc.\ Lond.}{A358}{777--796}{2000}.

\reference
Beck R., Galactic and extragalactic magnetic fields.
\Journal{Sp.\ Sci.\ Rev.}{99}{243--260}{2001}.

\reference
Beck R.\ \& Hoernes P., Magnetic spiral arms in the galaxy NGC~6946.
\Journal{\Nat}{379}{47--49}{1996}.

\reference
Beck R., Carilli C.~L., Holdaway M.~A.\ \& Klein U., Multifrequency
observations of the radio continuum emission from NGC 253. I. Magnetic fields
and rotation measures in the bar and halo.
\Journal{\AAp}{292}{409--424}{1994}.

\reference
Beck R., Brandenburg A., Moss D., Shukurov A.\ \& Sokoloff D.,  Galactic
magnetism: recent developments and perspectives.
\Journal{\ARAA}{34}{155--206}{1996}.

\reference
Beck R., Berkhuijsen E.~M., Moss D., Shukurov A.\ \& Sokoloff D., The
nature of the magnetic belt in M31. \Journal{\AAp}{335}{500--509}{1998}.

\reference
Beck R., Shukurov A., Sokoloff D.\ \& Wielebinski R., Systematic bias in
interstellar magnetic field estimates. \Journal{\AAp}{411}{99--107}{2003}
(astro-ph/0307330).

\reference
Belyanin M., Sokoloff D., Shukurov A.,  Simple models of nonlinear
         fluctuation dynamo. \Journal{\GAFD}{68}{237--261}{1993}.

\reference
Belyanin M.~P., Sokoloff D.~D.\ \& Shukurov A.~M., Asymptotic steady-state
solutions to the nonlinear hydromagnetic dynamo equations. \Journal{Russ.\
J.\ Math.\ Phys.}{\bf 2}{149--174}{1994}.

\reference
Berkhuijsen E.~M., Horellou C., Krause M., Neininger N., Poezd A.~D.,
Shu\-ku\-rov A.\ \& Sokoloff D.~D.,  Magnetic fields in the disk and
halo of M51. \Journal{\AAp}{318}{700--720}{1997}.

\reference
Blackman, E.~G., \& Field, G.~B., Coronal activity from dynamos in
astrophysical rotators. \Journal{\MNRAS}{318}{724--732}{2000}.

\reference
Boulares A.\ \& Cox D.~P., Galactic hydrostatic equilibrium with magnetic
tension and cosmic-ray diffusion. \Journal{\ApJ}{365}{544--558}{1990}.

\reference
Brandenburg A.\ \& Subramanian K., Astrophysical magnetic fields and
nonlinear dynamo theory. {\PRep}, in press ({2004}).

\reference
Brandenburg A., Donner K.-J., Moss D., Shukurov, A., Sokoloff D.~D.\ \&
Tuominen I., Dynamos in discs and halos of galaxies.
\Journal{\AAp}{259}{453--461}{1992}.

\reference
Brandenburg A., Donner K.-J., Moss D., Shukurov, A., Sokoloff D.~D.\ \&
Tuominen I., Vertical magnetic fields above the discs of spiral galaxies.
\Journal{\AAp}{271}{26--50}{1993}.

\reference
Brandenburg A., Moss D.\ \& Shukurov A., Galactic fountains as
magnetic pumps. \Journal{\MNRAS}{276}{651--662}{1995}.

\reference
Brandenburg, A., Dobler, W.\ \& Subramanian, K., Magnetic helicity in stellar
dynamos: new numerical experiments. \Journal{\AN}{323}{99--122}{2002}.

\reference
Braun R., The distribution and kinematics of neutral gas in M31.
\Journal{\ApJ}{372}{54--66}{1991}.

\reference
Brown J. C.\ \& Taylor A.~R., The structure of the magnetic field in the outer
galaxy from rotation measure observations through the
disk. \Journal{\ApJL}{563}{L31--L34}{2001}.

\reference
Bykov A., Popov V., Shukurov A., Sokoloff D., Anomalous
persistence of bisymmetric magnetic structures in spiral galaxies.
\Journal{\MNRAS}{292}{1--10}{1997}.

\reference
Cattaneo F., Hughes D.~W.\ \& Kim, E.-J., Suppression of chaos in a simplified
nonlinear dynamo model. \Journal{\PRL}{76}{2057--2060}{1996}.

\reference
Clemens D.~P., Massachusetts--Stony Brook Galactic plane CO survey: the
Galactic disk rotation curve. \Journal{\ApJ}{295}{422--436}{1985}.

\reference
Cox D.~P., The diffuse interstellar medium. In {\em The Interstellar Medium in
Galaxies,} eds H.~A.~Thronson \& J.~M.~Shull, Kluwer, Dordrecht, pp.\ 181--200
(1990).

\reference
Deharveng J.~M.\ \& Pellet A., \'Etude cin\'ematique et dynamique de M31 \`a
partir de l'observation des r\'egions d'\'emission.
\Journal{\AAp}{38}{15--28}{1975}.

\reference
De Young D., Turbulent generation of magnetic fields in extragalactic radio
sour\-ces. \Journal{\ApJ}{241}{81--97}{1980}.

\reference
Dopita M.~A.\ \& Sutherland R.~S., {\em Astrophysics of the Diffuse Universe,}
Springer, Berlin (2003).

\reference
Elstner D., Meinel R.\ \& R\"udiger G., Galactic dynamo models without sharp
boundaries. \Journal{\GAFD}{50}{85--94}{1990}.

\reference
Elstner D., Golla G., R\"udiger G.\ \& Wielebinski R., Galactic halo magnetic
fields due to a `spiky' wind. \Journal{\AAp}{297}{77--82}{1995}.


\reference
Fletcher A.\ \& Shukurov A., Hydrostatic equilibrium in a magnetized,
warped Galactic disc. \Journal{\MNRAS}{325}{312--320}{2001}.

\reference
Fletcher A., Berkhuijsen E.~M., Beck R.\ \& Shukurov A., The magnetic
field of M31 from multi-wavelength radio polarization observations, {\em
Astron.\ Astrophys.}, in press (2004) (astro-ph/0310258).

\reference
Fraternali F., Oosterloo T., Boomsma R., Swaters R.\ \& Sancisi R.,
High velocity gas in the halos of spiral galaxies. In {\em Recycling
Intergalactic and Interstellar Matter,} IAU Symp.\ 217, eds P.-A.~Duc,
J.~Braine \& E.~Brinks, ASP (2003) (astro-ph/0310799).

\reference
Frick P., Beck R., Shukurov A., Sokoloff D., Ehle~M., Kamphuis~J., Radio and
optical spiral patterns in the galaxy NGC~6946.
\Journal{\MNRAS}{318}{925--937}{2000}.

\reference
Frick P., Stepanov R., Shukurov A., Sokoloff D., Structures in the rotation
measure sky.  \Journal{\MNRAS}{325}{649--664}{2001} (astro-ph/0012459).

\reference
Han J.~L., Manchester R.~N., Berkhuijsen E.~M.\ \& Beck R., Antisymmetric
rotation measures in our Galaxy: evidence for an A0 dynamo.
\Journal{\AAp}{322}{98--102}{1997}.

\reference
Han J.~L., Beck R.\ \& Berkhuijsen E.~M., New clues to the magnetic field
structure of M31. \Journal{\AAp}{335}{1117--1123}{1998}.

\reference
Haud U., Gas kinematics in M31. \Journal{\ApSS}{76}{477--490}{1981}.

\reference
Kahn F.~D., Brett L., Magnetic reconnection in the disc and halo.
\Journal{\MNRAS}{263}{37--48}{1993}.

\reference
Kazantsev A.~P., Enhancement of a magnetic field by a conducting fluid.
\Journal{\JETP}{53}{1806--1813}{1967}; English translation:
\Journal{\SovJETP}{26}{1031--1034}{1968}.

\reference
Kazantsev A.~P., Ruzmaikin A.~A.\ \& Sokoloff D.~D., Magnetic field transport
in an acoustic turbulence. \Journal{\SovJETP}{61}{285--292}{1985}.

\reference
Kleeorin N., Moss D., Rogachevskii I.\ \& Sokoloff D., Helicity balance and
steady-state strength of the dynamo generated galactic magnetic field.
\Journal{\AAp}{361}{L5--L8}{2000}.

\reference
Kleeorin N., Moss D., Rogachevskii I.\ \& Sokoloff D., Nonlinear magnetic
diffusion and magnetic helicity transport in galactic dynamos.
\Journal{\AAp}{400}{9--18}{2003}.

\reference
Korpi M.~J., Brandenburg A., Shukurov A., Tuominen I.\ \& Nordlund \AA., A
supernova-regulated interstellar medium: simulations of the turbulent
         multiphase medium. \Journal{\ApJL}{514}{L99--L102}{1999a}.

\reference
Korpi M.~J., Brandenburg A., Shukurov A.\ \& Tuominen I., Evolution of a
superbubble in a turbulent, multi-phased and magnetized ISM.
\Journal{\AAp}{350}{230--239}{1999b}.

\reference
Krasheninnikova Yu.~S., Ruzmaikin A.~A,, Sokoloff D.~D.\ \& Shukurov A.,
    Configuration  of  large-scale magnetic  fields in spiral galaxies.
    \Journal{\AAp}{213}{19--28}{1989}.

\reference
Krause M., Beck R.\ \& Hummel E., The magnetic field structures in two nearby
spiral galaxies. I. The axisymmetric spiral magnetic field in IC342.
\Journal{\AAp}{217}{1--17}{1989a}.

\reference
Krause M., Beck R.\ \& Hummel E., The magnetic field structures in two nearby
spiral galaxies. II. The bisymmetric spiral magnetic field in M81.
\Journal{\AAp}{217}{17--30}{1989b}.

\reference
Kvasz L., Sokoloff D.~D.\ \& Shukurov A.,  A steady state of the disk dynamo.
\Journal{\GAFD}{65}{231--244}{1992}.

\reference
Kulsrud R.~M., A critical review of galactic dynamos.
\Journal{\ARAA}{37}{37--64}{1999}.

\reference
Kulsrud R.~M., Cen R., Ostriker J.~P.\ \& Ryu D., The protogalactic origin for
cosmic magnetic fields. \Journal{\ApJ}{480}{481--491}{1997}.

\reference
Lesch H.\ \& Bender R., Magnetic fields in elliptical galaxies.
\Journal{\AAp}{233}{417--421}{1990}.

\reference
Lockman F.~J., The $\HI$ halo in the inner galaxy.
\Journal{\ApJ}{283}{90--97}{1984}.

\reference
Loinard L., Allen R.~J., Lequeux J., An unbiased survey for CO emission in the
inner disk of the Andromeda galaxy. \Journal{\AAp}{301}{68--74}{1995}.

\reference
Lou Y.-Q., Fan A., Coupled galactic density-wave modes in a composite system
of thin stellar and gaseous discs. \Journal{\MNRAS}{297}{84--100}{1998}.

\reference
Lou Y.-Q., Fan A., Large-scale magnetohydrodynamic density-wave structures in
the Andromeda nebula. \Journal{\MNRAS}{315}{646--654}{2000}.

\reference
Lozinskaya T.~A., {\em Supernovae and Stellar Winds in the
Interstellar Medium,} AIP, New York (1992).

\reference
Mathews W.~G.\ \& Brighenti F., Self-generated magnetic fields in galactic
cooling flows . \Journal{\ApJ}{488}{595--605}{1997}.

\reference
Mathews W.~G.\ \& Brighenti F., Hot gas in and around elliptical galaxies.
\Journal{\ARAA}{41}{191--239}{2003}.

\reference
Mestel L.\ \& Subramanian K., Galactic dynamos and density wave theory.
\Journal{\MNRAS}{248}{677--687}{1991}.

\reference
McKee C.~F.\ \& Ostriker J.~P., A theory of the interstellar medium: three
components regulated by supernova explosions in inhomogeneous substrate.
\Journal{\ApJ}{217}{148--169}{1977}.

\reference
Moffatt H.~K.,  {\em Magnetic Field Generation in Electrically
Conducting Fluids,} Cambridge Univ.\ Press (1978).

\reference
Moss D., On the generation of bisymmetric magnetic-field structures in spiral
galaxies by tidal interactions. \Journal{\MNRAS}{275}{191--194}{1995}.

\reference
Moss D., Parametric resonance and bisymmetric dynamo solutions in spiral
galaxies. \Journal{\AAp}{308}{381--386}{1996}.

\reference
Moss D., The relation between magnetic and gas arms in spiral galaxies.
\Journal{\MNRAS}{297}{860--866}{1998}.

\reference
Moss D.\ \& Shukurov A., Turbulence and magnetic fields in elliptical
galaxies. \Journal{\MNRAS}{279}{229--239}{1996}.

\reference
Moss  D.\ \& and Shukurov A., Galactic dynamos with captured magnetic flux and
an accretion flow. \Journal{\AAp}{372}{1048--1063}{2001} (astro-ph/0012436).

\reference
Moss  D.\ \& and Shukurov A., Accretion disc dynamos opened up by external
magnetic fields, \Journal{\AAp}{413}{403--414}{2004}.

\reference
Moss D., Shukurov A.\ \& Sokoloff D., Galactic dynamos driven by magnetic
buoyancy.  \Journal{\AAp}{343}{120--131}{1999}.

\reference
Moss D., Shukurov A.\ \& Sokoloff D., Accretion and galactic dynamos.
\Journal{\AAp}{358}{1142--1150}{2000}.

\reference
Moss D., Shukurov A., Sokoloff D., Beck R.\ \& Fletcher A., Magnetic fields in
barred galaxies. II. Dynamo models. \Journal{\AAp}{380}{55--71}{2001}
(astro-ph/0107214).

\reference
Niklas S., {\em Eigenschaften von Spiralgalaxien in hochfrequenten
Radiokontinuum.} PhD Thesis, Max-Planck-Institut f\"ur Radioastronomie, Bonn
(1995).

\reference
Nordlund \AA.\ \& R\"ognvaldsson \"O., Magnetic fields in young galaxies.
\Journal{Highlights Astron.}{12}{706--708}{2002}.

\reference
Parker E.~N., The generation of magnetic fields in astrophysical bodies. II.
The galactic field. \Journal{\ApJ}{163}{252--278}{1971}.

\reference
Phillips A., A comparison of the asymptotic and no-$z$ approximations for
galactic dynamos. \Journal{\GAFD}{94}{135--150}{2001}.

\reference
Poezd A., Shukurov A.\ \& Sokoloff D., Global magnetic patterns in
the Milky Way and the Andromeda nebula. \Journal{\MNRAS}{264}{285--297}{1993}.

\reference
Priklonsky V., Shukurov A., Sokoloff D.\ \& Soward A.,
Non-local effects in the mean-field disc dynamo. I.  An asymptotic expansion.
\Journal{\GAFD}{93}{97--114}{2000} (astro-ph /0309666).

\reference
R\"adler K.-H., On the effect of differential rotation on axisymmetric and
non-axisymmetric magnetic fields of cosmical bodies. In {\em Plasma
Astrophysics}, ESA Publ.\ SP-251, pp.~569--574 (1986).

\reference
Rohde R., Beck R.\ \& Elstner D., Magnetic arms in NGC~6946 generated by a
turbulent dynamo. \Journal{\AAp}{350}{423--433}{1999}.

\reference
Ruzmaikin A.~A.\ \& Shukurov A.~M.,  Magnetic field generation in the
Galactic disk. \Journal{\SA}{25}{553--558}{1981}.

\reference
Ruzmaikin A.~A., Sokoloff D.~D.\ \& Shukurov A.~M., Magnetic field
distribution in spiral galaxies. \Journal{\AAp}{148}{335--343}{1985}.

\reference
Ruzmaikin A.~A., Sokoloff D.~D.\ \& Shukurov A.~M., Magnetic fields in spiral
galaxies. In {\it Plasma Astrophysics}, ESA Publ.\ SP-251, pp.\ 539--544
(1986).

\reference
Ruzmaikin A.~A., Shukurov A.~M.\ \& Sokoloff D.~D., {\em Magnetic
Fields of Galaxies,} Kluwer, Dordrecht (1988).


\reference
Sawa T.\ \& Fujimoto M., Bisymmetric spiral configuration of magnetic fields
in spiral galaxies. I. Local theory. \Journal{\PASJ}{38}{551--566}{1986}.

\reference
Schekochikhin A., Cowley S., Hamnett G.W., Maron J.~L.\ \& McWilliams J.~C.,
A model of nonlinear evolution and saturation of the turbulent MHD dynamo.
\Journal{New J.\ Phys.}{4}{84.1--84.22}{2002} (astro-ph/0207503).

\reference
Segalovitz A., Shane W.~W., de Bruyn A.~G., Polarisation detection at radio
wavelengths in three spiral galaxies, \Journal{\Nat}{264}{222--226}{1976}.

\reference
Shapiro P.~R.\ \& Field G.~B., Consequences of a new hot component of the
interstellar medium. \Journal{\ApJ}{205}{762--765}{1976}.

\reference
Shukurov A., Magnetic spiral arms in galaxies.
\Journal{\MNRAS}{299}{L21--L24}{1998}.

\reference
Shukurov A.\ \& Berkhuijsen E.~M., Faraday ghosts: depolarization canals in
the Galactic radio emission. \Journal{\MNRAS}{342}{496--500}{2003}. Erratum:
\Journal{\MNRAS}{345}{1392}{2003} (astro-ph/0303087).

\reference
Shukurov A.\ \& Sokoloff D., Galactic spiral arms and dynamo control
parameters. \Journal{Studia Geoph.\ et Geod.}{42}{391--396}{1998}.

\reference
Shukurov A., Sarson G.~R., Nordlund \AA., Gudiksen B.\ \& Brandenburg A.,
The effects of spiral arms on the multi-phase ISM. {\ApSS}, in press (2004)
(astro-ph/0212260).

\reference
Simard-Normandin M., Kronberg P.~P., Rotation measures and the galactic
magnetic field. \Journal{\ApJ}{242}{74--94}{1980}.

\reference
Sofue Y., Fujimoto M., A bisymmetric spiral magnetic field and the spiral arms
in our Galaxy. \Journal{\ApJ}{265}{722--729}{1983}.

\reference
Sofue Y.\ \& Rubin V., Rotation curves of spiral galaxies.
\Journal{\ARAA}{39}{137--174}{2001}.

\reference
Sofue Y., Fujimoto M.\ \& Wielebinski R., Global structure of magnetic fields
in spiral galaxies. \Journal{\ARAA}{24}{459--497}{1986}.

\reference
Sokoloff D.~D.\ \& Shukurov A., Regular magnetic  fields  in coronae  of
   spiral  galaxies. \Journal{\Nat}{347}{51--53}{1990}.

\reference
Sokoloff D.~D., Bykov A.~A., Shukurov A., Berkhuijsen E.~M., Beck R.\ \&
Poezd A.~D., Depolarisation and Faraday effects in galaxies and other
extended radio sources. \Journal{\MNRAS}{299}{189--206}{1998}; Erratum:
\Journal{\MNRAS}{303}{207--208}{1999}.

\reference
Soward A.~M., A thin disc model of the Galactic dynamo,
    \Journal{\AN}{299}{25--33}{1978}.

\reference
Soward A.~M., Thin disc $\alpha\omega$-dynamo models.
   I.  Long length scale modes.  \Journal{\GAFD}{64}{163--199}{1992a}.

\reference
Soward A.~M., Thin disc $\alpha\omega$-dynamo models. II.
   Short length scale modes.  \Journal{\GAFD}{64}{201--225}{1992b}.

\reference
Soward A.~M., Thin aspect ratio $\alpha\Omega$-dynamos
   in galactic discs and stellar shells. In {\it Advances in Nonlinear
    Dynamos,} Eds.\  A.~Ferriz-Mas \& M.~N\'u\~nez, Taylor \& Francis,
London, pp.\ 224--268 (2003).

\reference
Starchenko S.~V.\ \& Shukurov A., Observable parameters of  spiral
galaxies  and  galactic magnetic fields. \Journal{\AAp}{214}{47--60}{1989}.

\reference
Stepinski T.~F.\ \& Levy E.~H., Generation of dynamo magnetic fields in
               protoplanetary and other astrophysical accretion disks.
               \Journal{\ApJ}{331}{416--434}{1988}.

\reference
Strong A.~W., Moskalenko I.~V.\ \& Reimer O., Diffuse continuum gamma rays
from the Galaxy. \Journal{\ApJ}{537}{763--784}{2000}; Erratum:
\Journal{\ApJ}{541}{1109}{2000}.

\reference
Subramanian K., Can the turbulent galactic dynamo generate large-scale
magnetic fields? \Journal{\MNRAS}{294}{718--728}{1998}.

\reference
Subramanian K., Unified treatment of small- and large-scale dynamos in helical
turbulence. \Journal{\PRL}{83}{2957--2960}{1999}.

\reference
Subramanian K.\ \& Mestel L., Galactic dynamos and density wave theory -- II.
An alternative treatment for strong non-axisymmetry.
\Journal{\MNRAS}{265}{649--654}{1993}.

\reference
Tenorio-Tagle G.\ \& Bodenheimer P., Large-scale expanding superstructures in
galaxies. \Journal{\ARAA}{26}{145--197}{1988}.


\reference
Vainshtein S.~I.\ \& Ruzmaikin A.~A., Generation of the large-scale galactic
magnetic field. \Journal{\AZh}{48}{902--909}{1971} (English translation in
\Journal{\SA}{15}{714}{1972}).

\reference
Vainshtein S.~I.\ \& Ruzmaikin A.~A., Generation of the large-scale galactic
magnetic field. II. \Journal{\AZh}{49}{449--452}{1972} (English translation in
\Journal{\SA}{16}{365}{1972}).

\reference
Vainshtein, S.~I., \& Cattaneo, F., Nonlinear restrictions on dynamo
action. \Journal{\ApJ}{393}{165--171}{1992}.

\reference
Wakker B.~P.\ \& van Woerden H., High-velocity clouds.
               \Journal{\ARAA}{35}{217--266}{1997}.

\reference
Widrow, L.~M., Origin of galactic and extragalactic magnetic fields.
               \Journal{\RMP}{74}{775--823}{2002}.

\reference
Wielebinski R.\ \& Krause F., Magnetic fields in galaxies.
\Journal{\AAp\ Rev.}{4}{449--485}{1993}.

\reference
Williams R.~M., Chu Y.-H., Dickel J.~R., Beyer R., Petre R., Smith R.~C.\ \&
Milne D.~K., Supernova remnants in the Magellanic Clouds. I. The colliding
remnants DEM L316. \Journal{\ApJ}{480}{618--632}{1997}.

\reference
Willis A.~P., Shukurov A., Soward A.~M.\ \& Sokoloff D., Nonlocal effects in
the mean-field disc dynamo. II. Numerical and asymptotic solutions. {\GAFD},
submitted (2003) (astro-ph/0309667).

\reference
Woltjer L., Remarks on the Galactic magnetic field.  In
{\it Radio Astronomy and the Galactic System,} Proc.\ IAU Symp.\ 31,
ed.\ H.~van~Woerden, Academic Press, London, 1967,  pp.~479--485 (1967).

\reference
Zeldovich Ya.~B., Ruzmaikin A.~A., Sokoloff D.~D., {\em Magnetic Fields in
Astrophysics,} Gordon and Breach, New York (1983).

\reference
Zeldovich Ya.~B., Ruzmaikin A.~A.\ \& Sokoloff D.~D., {\em The Almighty
Chance,} World Sci., Singapore (1990).

\end{references}
